\documentstyle[11pt,psfig]{article}
\newcommand{\lift}{\mu}
\newcommand{\restrict}{{\mathcal{M}}}

\newcommand{\tstep}{{\mathcal T}}

\begin{document}

\title {Equation-Free Multiscale Computation: \\
enabling microscopic simulators to perform system-level tasks}

\author{Ioannis G. Kevrekidis$^{1,2}$,
\thanks{{\it to whom correspondence should be addressed,
yannis@princeton.edu} }  \\
C. William Gear$^{1,3}$,
James M. Hyman$^{4}$,
Panagiotis G. Kevrekidis$^5$, \\
Olof Runborg$^2$, \footnote{Current address: Numerical Analysis and
Computer Science, KTH, 100 44 Stockholm, Sweden.}
and Constantinos Theodoropoulos$^6$ \\
\\
$^1$Department of Chemical Engineering; \\
$^2$Program in Applied and Computational Mathematics \\
Princeton University, Princeton, NJ 08544;
$^3$NEC Research Institute; \\
$^4$CNLS and T-7, Los Alamos National Laboratory
Los Alamos, NM 87545; \\
$^5$Department of Mathematics, University of Massachusetts,
Amherst, MA \\
$^6$Department of Process Integration, UMIST, Manchester M60 1QD, UK.}

\date{}
\maketitle

\begin{abstract}

We present and discuss a
framework for computer-aided multiscale analysis,
which enables models at a ``fine" (microscopic/stochastic) level of description
to perform modeling tasks at a ``coarse" (macroscopic, systems) level.
These macroscopic modeling tasks, yielding information over {\it long} time and
{\it large} space scales, are accomplished through appropriately initialized calls to
the microscopic simulator for only {\it short} times and {\it small} spatial domains.
Traditional modeling approaches first involve the derivation of macroscopic
evolution equations (balances closed through constitutive relations).
An arsenal of analytical and numerical techniques for the efficient solution
of such evolution equations (usually Partial Differential Equations,
PDEs) is then brought to bear on the problem.
Our equation-free (EF) approach, introduced in \cite{PNAS},
when successful, can bypass the derivation of the macroscopic evolution
equations {\it when these equations conceptually exist but are not
available in closed form}.
We discuss how the mathematics-assisted development of a computational superstructure
may enable alternative
descriptions of the problem physics ({\it e.g.} Lattice Boltzmann (LB),
kinetic Monte Carlo (KMC)
or Molecular Dynamics (MD) microscopic simulators, executed over relatively short
time and space scales) to perform systems level tasks (integration over relatively
large time and space scales,``coarse" bifurcation analysis,
optimization,  and control) directly.
In effect, the procedure constitutes a systems identification based,
``closure on demand" computational toolkit, bridging microscopic/stochastic
simulation with traditional continuum scientific computation and numerical analysis.
We illustrate these ``numerical enabling technology" ideas through
examples from chemical kinetics (LB, KMC), rheology (Brownian Dynamics),
homogenization and the computation of ``coarsely self-similar" solutions,
and discuss various features, limitations and potential extensions
of the approach.

\end{abstract}

\newpage

\tableofcontents

\newpage

\section{INTRODUCTION}

The purpose of this work is to analyze
a synergism that exists between ``conventional"
numerical methods/analysis
and microscopic complex systems modeling,
dynamics and control.
Based on this synergism, it is possible (under conditions discussed below)
to perform the computer-assisted analysis of an evolution equation for the coarse,
macroscopic level, closed description of a physical/material phenomenon
{\it without having this closed description explicitly available}.
This is accomplished by extracting from a different, fine level,
microscopic description (e.g. MD, KMC, kinetic theory based Lattice Gas (LG) or
LB-BGK (Bhatnagar, Gross and Krook, \cite{BGK} codes) the
information that ``traditional" numerical procedures would obtain through
direct function evaluation from
the macroscopic evolution equation, had this equation been available.
This type of information includes, for example,
numerical approximations of residuals,
(actions of) Jacobians and Hessians, partial derivatives with respect to 
parameters etc.
Circumventing the derivation of the macroscopic-level description
lies at the heart of the approach.
We collectively refer to this class of methods as ``Equation Free (Multiscale)" Methods
(EF or EFM Methods).
The purpose of this ``mathematics assisted" computational technology is, therefore,
to enable microscopic-level codes to perform system-level analysis directly,
without the need to pass through an intermediate, macroscopic-level,
explicit ``conventional'' evolution equation description of the dynamics.

In general introductory terms, a persistent feature of complex systems is the
emergence of macroscopic, coherent behavior from the interactions of microscopic
``agents" -molecules, cells, individuals in a population- between themselves and with
their environment.
The implication is that macroscopic rules (description of behavior at a high level)
can somehow be deduced from microscopic ones (description of behavior at a much finer
level).
For some problems (like Newtonian fluid mechanics) the successful macroscopic
description (the Navier Stokes equations) predated its microscopic derivation
from kinetic theory.
In many important problems, ranging from ecology to materials science,
and from chemistry to engineering, the physics are known at the microscopic/individual
level, and the closures required to translate them to a high-level macroscopic
description are simply not available.
Severe limitations arise in trying to either find these closures,
or to directly computationally bridge the enormous gap between 
the scale of the available
description and the scale at which the questions of interest are asked and
the answers are required.
These computational limitations constitute a major stumbling block in current
complex system modeling.

In this work, using as a tool the ``{\it coarse time-stepper}" \cite{PNAS}
we discuss the formulation (and in some instances demonstrate the implementation)
of a set of ``mathematics assisted" computational
super-structures (libraries) that can be wrapped around whatever the best
computer model a scientist would come up with for her/his system -
be it a Monte Carlo description of a chemical reaction or a kinetic theory
based model of multiphase flow.
The procedure is intended to be used when a macroscopic description
is {\it conceptually possible yet unavailable in closed form}.
If {\it accurate} macroscopic models {\it are} available in closed form,
one should probably work with them directly.
It is interesting, however, even in this case, 
to explore the performance of the EF methods we
describe here.

We discuss a systematic framework for extracting directly
(and, we hope, as more work is performed, and experience gained, efficiently)
from microscopic simulations the information one would obtain from
macroscopic models, had these macroscopic models been available in closed form.
It is the difficulty in obtaining such closed models that the proposed
approach may circumvent.
In a sense, the approach consists of ``unavailable macroscopic model-motivated"
processing of microscopic dynamical computations - it is a {\it system
identification} based procedure, processing the results of short bursts
of (appropriately initialized) microscopic simulation,
that enables us to create an ``equation-free"
computational framework.

If what we did was to run a full (completely resolved in space and
time) microscopic simulation, and merely extract the macroscopic behavior from it,
the size of the resolved computation would obviously limit us
to small systems over short times.
A key feature of this work is that microscopic simulations are performed only
in small ``elements" in space-time: small space domains (``patches") for short times.
The solution is evolved in each patch, and then macroscopically interpolated across
the patches (which can be thought of as nodes of a coarse mesh) to report the evolution
of the macroscopic fields of interest; consecutive such reports can be used to estimate
the time-derivative of the unavailable equation for the evolution of the coarse fields.
These time derivative estimates can then be used as input to
(a) a large time-step integrator --the projective
integrator \cite{Gear1,GKT}; (b) a zero-finder, to determine steady states, \cite{PNAS};
(c) an Arnoldi type eigensolver, to determine stability; and, more generally,
to macroscopic task codes (controller design software, optimization software etc.).
These points are briefly summarized in Table 1, where ``small space" and ``short time"
refer to the microscopic simulation, while ``large space" and ``long time" refer to
the macroscopic tasks.

\begin{table}[htbp]
\begin{center}
\title{EFM Methods \& Scope}
\vspace{0.3cm}
\begin{tabular}{|l|ccc|}
\hline
Method & Actual & & Enabled \\
\hline
 Legacy timestepper + RPM & repeated (ST) &$\longrightarrow$& S$T_{\infty}$ \\
{\tiny Shroff and Keller, 1993} \cite{S&K} &&&\\
Coarse timestepper + RPM & repeated (S$\tau$) &$\longrightarrow$& S$T_{\infty}$ \\
{\tiny Theodoropoulos et al., 2000} \cite{PNAS} &&&\\
\hspace*{17mm}{\small (coarse bifurcation)}&&&\\
Coarse integration & S$\tau$$\tau$$\tau$$\dots$ &$\longrightarrow$& ST \\
{\tiny Gear and Kevrekidis, 2001, 2002}\cite{Gear1,GKT} &&&\\
Gaptooth scheme & (ss$\dots$s)$\tau$ &$\longrightarrow$& S$\tau$ \\
{\tiny Kevrekidis, AIChE 2000} \cite{AIChE2000} &&&\\
\hline
 gaptooth + coarse integration & (ss$\dots$s)$\tau$$\tau$$\tau$$\dots$ &$\longrightarrow$& ST \\
 $\equiv$ patch dynamics &&&\\
 \hline
 gaptooth + RPM & repeated ((ss$\dots$s)$\tau$$\tau$$\tau$$\dots$) &$\longrightarrow$& S$T_{\infty}$ \\
 $\equiv$ patch bifurcations &&&\\
\hline
\end{tabular}\vspace{0.3cm}\\
\caption{\small EFMs can be catalogued by their spatial -- s(S) small(large) -- 
and their temporal -- $\tau$(T,$T_{\infty}$)
short (intermediate,``infinite") computational scales. The method named on the
left uses ``actual" space-time computations to obtain ``enabled" space-time
results.}
\label{Tab:meth_scop}
\end{center}
\end{table}

The microscopic simulation should use the best current microscopic
description of the physics (and the code implementing it) available; our
procedure will be wrapped around this microscopic simulator calling it as a
subroutine.
As microscopic simulators improve, the overall results will also improve.
If the simulator does not embody correct physics, the EFM ``wrapper" will extract
(hopefully quickly and efficiently) the macroscopic consequences of the incorrect
physics; this can help in realizing what is missing and modifying the microscopic
model or code.

The macroscopic computations will take advantage of the arsenal of tools that have
been developed for continuum evolutionary equation (usually Ordinary and Partial
Differential Equation, ODE and PDE)
descriptions of physical phenomena (e.g. \cite{MATLAB,AUTO,LOCA,gPROMS}).
Standard such tools include discretization and integration techniques,
numerical linear algebra -most
importantly, iterative large scale linear algebra, matrix-free solvers and eigensolvers-,
contraction mappings,
continuation and numerical bifurcation analysis, 
as well as optimization -local and possibly global-,
and controller design computations (e.g. the solution of large scale Riccati
equations and in general optimal control techniques)

The paper is organized as follows.
We start our presentation of EFM methods
with a review of our recent work on the coarse timestepper, and its use
in coarse integration and coarse bifurcation analysis of unavailable macroscopic
evolution equations through microscopic simulators.
We continue with a long conceptual discussion of the building blocks that
make the EFM computational enabling technology possible: system identification,
the numerical analysis of legacy codes, and separation of scales.
This section may be skipped at a first reading of the paper, and the reader
can proceed directly to the new material, returning to the conceptual
discussion later.
We then discuss the ``gaptooth" scheme (first presented in \cite{AIChE2000})
and subsequently show how to combine it with coarse integration (for patch dynamics)
and coarse bifurcation methods (for patch bifurcations).
These steps enable the EF computation of macroscopic behavior over large spatial
domains and long time intervals
through post-processing (``patching together") the results of appropriately
initialized and executed bursts of microscopic simulation over
small spatial domains and short time intervals.
We will then present an anthology of results from ongoing investigations that
illustrate the scope of the EFM methods to analyze various types of problems,
both in terms of the type of inner (microscopic) simulator, but also in terms
of alternative tasks, like control-related tasks, homogenization --more precisely
``effective" equation analysis--, and the computation of coarsely self-similar solutions.
In conclusion we will discuss various features and limitations of the EFM framework, as
well as areas in which we hope the methods may be fruitfully applied.

Before we start, it is worth making certain disclaimers
(and we will come back to this again in the conclusions and discussion sections).
Many of the mathematical and computational tools combined here
(for example, system identification, or inertial
manifold theory) are well established; we only borrow them as necessary.
We hope that the {\it synthesis} we propose here has new tools and insights to offer.
Its main point, the ``on-line" bridging of scales through ``on demand" closure is,
of course, a mainstream avenue in current research, where innovative multiscale
techniques (such as the quasi-continuum method of Philips and coworkers \cite{Phillips,Shenoy},
the kinetic-theory based solvers proposed by Xu and Prendergast \cite{Xu1,Xu2},
the optimal prediction methods of Chorin \cite{Chorin1,Chorin2} and
the novel stochastic integration techniques of Shardlow and Stuart \cite{Stuart})
are constantly being proposed and explored.

This work is written more in the style of a manifesto of principles;
it contains suggestions and speculation,
as well as results and research directions that we believe 
will be fruitful for further study.

\section{THE COARSE TIME-STEPPER}

The main tool that allows the performance of numerical tasks at the
macroscopic level using microscopic/stochastic simulation codes
is the so-called ``coarse timestepper'' discussed in \cite{PNAS}
(see also \cite{Olof,Alexei1,GKT}).
By timestepper, or time-$T$ map,
we mean the solution operator of a differential equation
for a fixed time $T$.
We denote it generically by $\tstep$:
a differential equation solution $u(t,x)$ satisfies
$u(t+T,x)={\mathcal T}u(t,x)$.
The coarse timestepper, denoted $\tstep_c$, implements
the time-$T$ map for the --unavailable in closed form-
equation that governs the macroscopic evolution for the problem of interest.
For this task it only uses
a fine timestepper, $\tstep_f$,
of the detailed microscopic/ stochastic evolution of the system.
The coarse timestepper consists of the following basic
elements (see Fig. \ref{fig:RPM}.a):
\begin{itemize}
\item[(a)] Select the statistics of interest for describing the
coarse behavior of the system and an appropriate representation
for them. For example, in a gas simulation at the particle level,
the statistics would probably be the pressure, density, and
velocity. For our LB-BGK reaction-diffusion
and our KMC lattice gas illustrative problems, the statistics of
interest are the zeroth moments of the
distribution (i.e. the concentrations). We will call this the
macroscopic description, $u$. These choices determine a
restriction operator, $\restrict$, from the microscopic-level
description $U$, to the macroscopic description: $u = \restrict
U$. The maximum likelihood inference based ``field estimator"
of Li Ju {\it et al.} \cite{LiJu1,LiJu2,LiJu3},
and the cumulative probability distributions
in \cite{Gear3}, are instances of this restriction operator.
This operator may involve averaging over microscopic space and/or
microscopic time and/or number of realizations if an ensemble of
simulations has been used.

\item[(b)] Choose an appropriate {\it lifting} operator, $\lift$, from the
macroscopic description, $u$, to a consistent microscopic description, $U$.
By consistent we mean that
$\restrict \lift =I$ , i.e. lifting from the
macroscopic to the microscopic and then restricting (projecting)
down again should have no effect.
For many types of microscopic descriptions, the lifting operator 
constructs distributions conditioned on (one or more of
their) lower moments.
For example, in a gas simulation using pressure etc. as the
macroscopic-level variables, $\lift$ would probably be chosen to make
random particle assignments consistent with the macroscopic
statistics.

\item[(c)] Prescribe a macroscopic initial condition (e.g.
concentration profile)  $u(t=0)$;

\item[(d)] Transform it through lifting to one (or more) fine,
consistent microscopic realizations $U(t=0) = \lift u(t=0)$;

\item[(e)] Evolve these realizations using the microscopic
simulator (the fine timestepper) for the desired short macroscopic
time $T$, generating the value(s) $U(T)=\tstep_f U(t=0)$.

\item[(f)] Obtain the restriction $u(T)=\restrict U(T)$
and define the coarse timestepper as $u(T)=\tstep_c u(t=0)$.
In other words, $\tstep_c = \restrict\tstep_f\lift$.
\end{itemize}

Separation of scales arguments, similar to those used in deriving
explicit closed macroscopic equations, can in principle be used to show
that this coarse timestepper indeed approximates the timestepper of
the macroscopic equation.
The hope is that the coarse solution $u(T)$ (or, more generally, $u(nT)$) at
these discrete points in time ``come from" an evolution equation (a semigroup).
In other words, we hope that a closed evolution equation --a formula for the
time derivative--
exists and closes for $u$, and that its solution (defined for all $t$) agrees
with our coarse-timestepper obtained $u(T)$ (or $u(nT)$).
The coarse time-stepper procedure is a natural one, and it is easy to realize
numerically.
For problems in which several microscopic copies of the same
coarse initial conditions must be evolved, it is also easy to parallelize: 
each ``lifted" copy of
the same coarse initial condition runs independently on a different processor.
(As we will see below, an additional level of parallelism enters when
we decimate a large computational domain to ``patches", each of which can
also be run on a different processor.)
Such procedures have been employed for the construction of averaged equations
from coarse-graining the master equation (e.g. \cite{Prigogine,Balescu,Gaspard})
and, similarly, for the derivation of hydrodynamic descriptions of interacting
particle systems (e.g. \cite{Lebowitz1,Lebowitz2,Chapman,Vincenti,BoonYip}).
In this work we provide, through several illustrations, ``computational
collateral" for this idea, suggesting a program for the proof of existence of
a semigroup for such a procedure (and for its ``gaptooth" and ``patch dynamics"
extensions that will be presented in later Sections).

If the the coarse time-stepper is accurate enough, it is immediately obvious
(and we have demonstrated in the literature) \cite{PNAS,GKT,Olof,Alexei1,Bubbles,
BubblesPRL,Alexei2}
that a computational superstructure like the Recursive Projection Method
of Shroff and Keller (see next section) can be ``wrapped around" it
and enable it to perform the
time-stepper based bifurcation analysis of the (unavailable in
closed form) coarse description of the problem.
Through this ``lift-run-restrict'' procedure, we enable a code doing time
evolution at a fine level of description, to perform bifurcation analysis at
a completely different, coarse level of description.

\section{COARSE BIFURCATION ANALYSIS}

We briefly review here, for completeness, previous results so that they
can be compared later with results obtained through more recent approaches.
In our original papers we used, as the basic illustrative example, a simple,
well known set of nonlinear coupled partial differential equations (the
FitzHugh-Nagumo, FHN) system in one spatial dimension \cite{FHN1, FHN2,PNAS}.

\begin{equation}
            u_t({\bf x},t)=u_{xx}({\bf x},t)+u({\bf x},t)-u^3({\bf
x},t)-v({\bf x},t)
   \label{fhn1}
\end{equation}
\begin{equation}
     v_t({\bf x},t)=\delta v_{xx}({\bf x},t)+\epsilon (u({\bf
x},t)-a_1v({\bf x},t)-a_0)
   \label{fhn2}
\end{equation}

Here $u({\bf x},t)$ and $v({\bf x},t)$ are the local concentrations of the
two participating reactants (the ``activator" and the
``inhibitor"), $\delta$ is a diffusion coefficient and the
remaining constants pertain to the kinetic terms; we can clearly
recognize the terms corresponding to diffusion and those corresponding
to chemical kinetics.
In all the simulations that will be presented in this paper, the
diffusion coefficient is set to $\delta=4.0$ and the values of the kinetic
parameters are chosen to be $a_0=-0.03$ and $a_1=2.0$. Finally,
$\epsilon$ was varied as in some of our simulations it represented the
continuation/bifurcation parameter.
The use of such models for both analysis and design purposes then hinges
on the exploitation of numerical discretization techniques to turn them
into large sets of ordinary differential or differential-algebraic equations.

One of the reasons for choosing this model in \cite{PNAS} was that it is
possible to analyze it through both the closed form PDE, as well as through a
kinetic theory motivated Lattice Boltzmann-BGK code \cite{qian1995}.
The PDE, at the appropriate limit, is a closed equation for the zeroth moments
of the discrete velocity distribution - so it can be considered as a ``coarse"
closed deterministic model; the LB-BGK model is the ``fine" model in this case.
Another reason for choosing this example is that it exhibits a rich variety of
nonlinear dynamic patterns, including multiplicity of spatially structured
steady states (and we will work with sharp, front like ones, both stable and
unstable) as well as spatiotemporal oscillations consisting of the palindromic
movement of sharp concentration fronts (we will also work with these).
The rich nonlinear dynamics and the spatial sharpness of the solutions make
this simple but nontrivial one dimensional example a good testing ground for new tools.

The behavior can be analyzed through
{\it direct simulation} using a code that evolves an initial profile in time
(a time-stepper) \cite{PNAS}.
Using the time-stepper in a ``successive substitution" form,
$u^{n+1}=\tstep u^n$ with $u^n\equiv u(nT)$,  
corresponds to doing on the computer what one would observe in an experiment
for the problem: setting parameter values, setting initial conditions and
evolving forward in time.
Doing on the computer what ``nature does" in a laboratory makes sense, but may
not be the best or fastest way of discovering features like steady states or
bifurcation points for an evolution PDE 
\begin{equation}
            u_t(x,t)= P (\partial_x, x, u(x,t);\lambda))
   \label{tim}
\end{equation}
where $\lambda$ is the bifurcation parameter.
Different algorithms
(contraction mappings, augmented
systems for the detection of bifurcation points) are routinely now constructed
\cite{AUTO,LOCA} that use {\it the same right hand side, (RHS)} 
$P(\partial_x,x,u;\lambda)$,
but in different ways than direct integration.
Indeed, techniques like Newton-Raphson, or multiparameter continuation codes
inspired by numerical bifurcation theory, can be thought of as ``alternative
simulation ensembles": they use the same system equations
(the same RHS formula) but in different ways, that
{\it do not} correspond to evolution in physical time.
The detailed dynamics of these algorithms (e.g. the dynamics of a Newton Raphson
iteration) do not make sense as physical
dynamics; yet {\it the results} (the fixed points, the bifurcation points) are
the same as the corresponding states of the original system - and they have
been found ``much easier" than an experimentalist, or a ``direct simulator",
would find them.

The plan is, therefore, to construct such alternative computational ensembles
that will discover the information we want to obtain from the code ``faster
and easier" than direct simulation.
This is routinely done today, for example through numerical bifurcation theory
algorithms, or through optimization algorithms -- using knowledge about the
mathematical characteristics of the problem we want to solve (e.g. eigenvalue
conditions for a boundary of stability, Karush-Kuhn-Tucker conditions for optimality)
we construct algorithms that wrap additional considerations around
the same right-hand-side and search more efficiently than direct simulation could
search.
In general, such codes (optimization codes, bifurcation codes,
controller design codes) have to be written from scratch, even if they
do use some of the subroutines that the direct simulation code uses.

The ``numerical analysis of legacy codes" discussed later in Section V,
and the associated ``time-stepper based bifurcation analysis" from which our
original inspiration comes \cite{S&K,KellerIMA,J&M1987a,J&M1987b,TBIMA}
combine the two approaches: direct simulation {\it and}
alternative algorithm (``ensemble") construction in what we believe is an
ingenuous ``software engineering" way.
One uses the direct simulator as a timestepper
-a simple subroutine that integrates starting at an initial condition
in time, and reports the result a little later.
The point of algorithms like Shroff and Keller's Recursive Projection Method
(RPM) \cite{S&K} is to extract from the time-stepper
what a fixed point Newton algorithm would need: residuals (that is easy) and
the action of Jacobians (that
is a little more intricate, and uses iterative linear algebra / matrix free ideas).
Using the time-stepper as a tool that allows us to identify the quantities that
we need in order to do numerical analysis of the ``unavailable equation" lies,
as we have already said, at the heart of the overall procedure.

One might, at some level, consider that what we are doing is adaptive control
on the time-stepper using it as an experiment.
The difference though --and it is a crucial difference--
is that the time-stepper can be {\it initialized at will},
while a physical experiment, even in closed loop, is not so easy to initialize,
and evolves as the physical dynamics dictate in real time.

In RPM the user repeatedly calls the timestepper routine for
successive (relatively short)
time steps as well as from nearby initial conditions.
The main assumption is that the system has a clear separation
of time scales: there exist a few eigenvalues in a narrow strip
around the imaginary axis (possibly some unstable ones too);
then comes a {\it spectral gap}, and the rest of the eigenvalues are
``far to the left" in the complex plane.
These eigenvalues arise, for example, in linearizing around a
steady state of the evolution equation Equ.(\ref{tim})
which is obviously also a fixed point
\begin{equation}
            u^*(x)=\tstep u^*(x)
   \label{fixp}
\end{equation}
of the timestepper, $u^{n+1} = \tstep u^n$,
the result of integration of Equ.(\ref{tim}) above with initial conditions
$u(x,t)$ for time  $T$.
This translates to having a few eigenvalues in a strip around the unit
circle, then a gap, and then many eigenvalues in a small disk around zero
for the time-stepper.
Eigenvalues at zero for the linearization around a steady state
of Equ.(\ref{tim}) (at 1 for Equ.(\ref{fixp}))
are special, and can be dealt with.
While so much structure seems like a lot to expect {\it a priori} from
a model (or from the physical process
that the model springs from) the usually dissipative PDEs modeling reaction
and transport (and including diffusion, viscous dissipation, heat conduction
etc.) often do
possess such a separation of time scales.

Similar assumptions, for example, underpin the theory
of Inertial Manifolds and Approximate Inertial Manifolds
{\cite{Temam,Foias,JKTi}, and many singularly perturbed systems that
arise in engineering modeling.
In addition, as we will qualitatively discuss below, even if the problem itself
does not have a separation of time scales, it is possible that the Fokker-Planck
equation for the evolution of the statistics of the problem may have a separation
of time scales.
The EF approach will then be used on a timestepper for that level of
(coarse, averaged) description of the statistics of the
problem \cite{CKT}.

Under the loose assumptions described above, the results of the repeated calls 
to the timestepper can
be used to adaptively approximate a low-dimensional subspace ${\bf \mathcal
P}$ of the linearization
of the system along which time evolution is slowest, possibly slightly
unstable.
One then performs a contraction mapping ({\it e.g.} a Newton iteration based on a
small approximate Jacobian)
in this subspace, aided by the integration itself (Picard iteration for the
timestepper, Equ.(\ref{fixp}))
in its orthogonal complement ${\bf \mathcal Q}$.
This combination of (approximate) Newton iteration in the
low-dimensional, slow subspace, and Picard iteration (integration) to
provide a contraction in
its complement, justifies the classification of such iterative methods as
``Newton-Picard" methods.
Such procedures, and their extensions, are being successfully used to perform
continuation,
stability and bifurcation analysis of dissipative PDEs and low-index PDAEs
close to low-codimension bifurcations
\cite{S&K,vonSosen,Lust,LustThesis,Burr,Erhel,Davidson,Love,Koronaki}.

Substituting the ``coarse time-stepper" (Fig. \ref{fig:RPM}.a) for the
``PDE time-stepper" (Fig. \ref{fig:RPM}.b)
is (almost) all  it takes to turn RPM into a coarse bifurcation code.
The main new component is {\it lifting}: one needs to construct 
microscopic initial states (e.g. distributions)
conditioned on the coarse variables (typically, some of the lower moments of
said distributions).
We have discussed the one-many nature of this procedure in much more detail in
\cite{Alexei1,GKT,Alexei2,Graham}.
{\it In principle}, if there is enough of a separation of time scales,
it should not matter what we initialize the higher moments of the
distributions to; we expect that, if a coarse deterministic equation for the
low moments exists and closes, the high moments will very quickly relax to
functionals of the lower moments.
We refer to this process as the ``healing" of the higher moments,
and it basically produces
``mature" distributions (as opposed to ``fresh" distributions that have the
same low moments but ``inaccurate" or ``improbable" higher moments).
This is, of course, a singularly perturbed problem (the high moments quickly
evolve to an attracting manifold parametrized by the low moments), and the
problem of consistent initialization has been systematically studied in this
context.
Consistent initialization of DAEs is, essentially, the same problem \cite{Petzold};
initializing on slow manifolds in meteorology (what we call ``mature"
initial conditions are referred to as ``bred" initial conditions there,
e.g. see \cite{Toth,Patil}) is also in
essence the same problem, as is initializing as close to an inertial manifold
as possible.
Of course, after a short integration time
the dynamics themselves will ``bring us down" to the manifold; but the random
initialization of the high moments, even though quickly healed, is a source
of noise for the deterministic part of the computation, and
reducing this noise is helpful in the ``outer" computational tasks.
The ``strong" and ``weak" cone conditions in the theory of inertial manifolds
\cite{Foias}
provide a framework for the rate at which trajectories ``off" the manifold approach
it, and several playful analogies to airplanes and landing strips can be made.
This issue has important practical implications for the effectiveness of the
computations, and it has been studied in the statistical mechanical literature for
a long time \cite{Lewis} (see also \cite{Ilya}).

Averaging over an ensemble of consistent (i.e. similarly conditioned) initial
distributions is, of course, also a means to variance reduction.
In the systems and control area a vast literature exists on optimal
estimation and model identification (e.g. \cite{Kalman,Astrom,Ljung}),
and good
filtering is vital when one tries to estimate the numerical derivatives (or
the approximations to the matrix-vector products) necessary to enable 
Newton-Picard type, subspace iteration based computations.
Our standard illustrative problem (the LB-FHN model) is extremely forgiving in 
terms of
variance reduction and consistent initialization; this is illustrated in
Fig.\ref{fig:heal}.
A single simulation copy (a single realization) conditioned on
the zeroth moments, at various levels of coarseness, initialized at local equilibrium
or away from local equilibrium for the high moments, does ``the same thing"
as far as the RPM code is concerned at the level of accuracy in our computations.

Other problems (like the nematic liquid crystal closure \cite{Graham}
as well as the surface
kinetic Monte Carlo examples \cite{Alexei1,Alexei2} we briefly discuss below)
are much more sensitive to ``mature" initial conditions.
Indeed, care has to be taken (i.e. a separate Metropolis Monte
Carlo code written) to ``equilibrate" the consistent initialization \cite{Alexei2};
as we discuss in \cite{Alexei1} and demonstrate in \cite{Graham}, it may become
necessary to include additional moments (which now have gradually become ``slow")
as independent coarse variables.
For surface lattice gases, for example, we may lift with 
densities {\it and} pair probabilities as independent variables, and include
a separate short MC step that imposes desired pair probabilities 
for every initialization (``lifting").

Newton-based and RPM-based bifurcation analysis results (stable and unstable steady
states, limit cycles, computation and continuation of codimension one ``coarse
bifurcation" points) for our FHN test problem, as well as other problems, have appeared
in the past \cite{PNAS,GKT}.
A summary of these results (to be compared to additional, ``small space" gaptooth
results later) is included in Fig.\ref{fig:bifhop}
(bifurcation diagram of the FHN with
respect to the bifurcation parameter $\epsilon$, showing a Hopf bifurcation),
Fig.\ref{fig:uvss}
(showing representative $u$ and $v$ spatially varying steady states for
a value of the parameter), Fig.\ref{fig:eigval}
(showing the leading eigenspectrum just before
and just after the Hopf bifurcation)
and Fig.\ref{fig:eigvec}
(showing the corresponding critical eigenvectors
(both the ``PDE-based" and the ``LB-RPM small Jacobian" identified ones).

Let us stress once more the two-tier structure of the EFM numerics: an inner
numerical algorithm (the LB-BGK time integrator) is called repeatedly, and its results
processed by an outer numerical algorithm (the Newton-Picard-type RPM code).
It is interesting that one can study the numerical analysis
(stability, convergence) aspects of the method
without having to worry about the exact nature of the ``inner" code.
The only element required to parametrize this inner code from the point
of view of the outer one 
is the ``coarse linearization" spectrum of the inner code; this
can be obtained off-line, and using the inner code only \cite{Gear1,Gear2}.
As we will discuss below, this two-tier ``inner-outer" structure makes also
sense in the case of ``legacy code" numerical analysis, when the ``inner"
timestepper is not a microscopic/stochastic code, but rather, a continuum
model solver.

\section{COARSE PROJECTIVE INTEGRATION}

We first present the coarse projective integration (CPI) approach
(whose analysis can be found in \cite{Gear1,Gear3},
and which has been extensively used in \cite{GKT})
in the context of accelerating a ``legacy simulator" using
system identification to estimate time-derivatives.
Consider a problem for which a powerful integrator code
(a time-stepper) has been written with, for example,
a {\it hardwired} time step of $10^{-6}$ (appropriate for the
stiffness of the problem).
Suppose the solution displays smoothness in time, suggesting
that a much larger time step, say $0.1$, could be used.
For a variety of reasons, it may not be feasible to change the time step.
The user may not have access to the source code; in an implicit
integrator the Newton iteration may fail to converge for
larger time steps; or (for some
unknown reason) the code becomes unstable (e.g. unacceptable splitting
errors arise in a split step based code).
The only tool then availabe is the executable with the very small
step that we cannot change.
And now, to complete the story, we need to integrate over long time horizons.

One can, of course, concatenate many calls to the $10^{-6}$ step subroutine
we have available in the traditional successive substitution
use of the timestepper.
Clearly, if the (``lost") right-hand-side was available,
one could take large 
time steps.
The compromise is, of course, to use a few evaluations of the  $10^{-6}$
black-box timestepper to {\it estimate} the right-hand-side
of the unavailable (lost) equation on the slow/inertial manifold
--a system identification task--, and then perform a time step
(e.g. an Euler step).
The time-stepper is the ``inner" integrator; it
is repeatedly called, and its
results processed, by an ``outer" integrator - the ubiquitous two-tier 
structure of the algorithms we propose.

The simplest projective integration method,
the {\it Projective Forward Euler} (PFE) method (see \cite{Gear1})
integrates over $k+1+M$ time steps of size $h$ from $t_n$ to
$t_{n+k+1+M}$ in the following manner:
\begin{enumerate}
\item Use a suitable {\it inner integrator} to integrate for $k$ time steps
from $t_n$ to $t_{n+k}$.  It does not matter in our discussion what
method is used for this inner integrator except that it is stable
and of at least first order.
\item Perform one more inner integration to compute $y_{n+k+1}$ from
$y_{n+k}$.
\item Finally, perform an extrapolation over $M$ steps using
$y_{n+k+1}$ and $y_{n+k}$ to compute $y_{n+k+1+M}$ as
$$
y_{n+k+1+M} = (M+1)y_{n+k+1} - My_{n+k}
$$
\end{enumerate}
This outer procedure is called a ``Forward Euler'' method because it can
also be written as

$$
 y_{n+k+1+M} = y_{n+k+1} + (Mh)y^\prime_{n+k+1}
$$
where the derivative approximation is given by
$$
y^\prime_{n+k+1} = (y_{n+k+1} - y_{n+k})/h
$$

The $k$ steps of the inner integrator are taken so that the
fast components of the solution are sufficiently damped, 
and their growth at the extrapolation step is neutralized.
In fact, each
application of the inner integrator multiplicatively reduces the fast
components, so the error reduction scales with a power of $k$ whereas
the growth in the extrapolation is linear in $M$.

A detailed study of this type of algorithms can be found in \cite{Gear1,Gear2};
while implicit versions of these algorithms are possible, and are partially
discussed there, the heart of the process is an {\it explicit} outer integrator
built around calls to an ``inner" integrator.
As such, the algorithms bear many resemblances to explicit codes for stiff
systems \cite{Lebedev,Medovikov,Verwer};
yet ours have been devised with
multiscale problems in mind, and a microscopic/stochastic inner integrator.
The main enabling step is again the coarse time-stepper through the
``lift-run-restrict" evaluation sequence.
There are some significant differences from explicit RK codes, inlcuding
the applicability in the ``legacy code" or noisy inner integrator
cases, and ease in dynamically selecting step sizes (see \cite{Gear2} for
details).
It would also be interesting to compare our projective integrators with
Krylov/exponential integrators for large systems with separations of time
scales \cite{Lubich,Friesner,Gallop1992,CSP}.

Using the coarse timestepper as the ``inner" integrator, Coarse Projective
Integration (CPI) can be performed.
The procedure is illustrated in Fig.\ref{fig:project}: an initial condition is taken
in ``coarse space" (e.g. a density field) and {\it lifted} to a consistent distribution
in microscopic space (e.g. cells for a chemotaxis problem).
The microscopic code is used to evolve the distribution {\it long enough} for the
higher moments, that have been ``incorrectly" initialized, to heal.
A few more evolution steps are then taken, and the solutions restricted to coarse space (densities).
Successive density profiles are used to estimate the time derivatives of the
(unavailable) density equation (again, an identification step).
The profile of the (coarse) density function can be represented by 
the basis functions
used in any of the  traditional numerical discretization techniques, 
including finite differences, finite volumes,
finite elements and spectral methods.
The time-derivatives for the coefficients of the basis function will be estimated,
and used in the projective integration step; finite difference (nodal), 
finite element
as well as empirical basis function representations have been 
illustrated in \cite{GKT}.

The smooth macroscopic fields can be estimated by low-dimensional
projections of Cumulative Distribution Functions (CDFs) on appropriate
basis functions \cite{Gear3}; alternative maximum-likelihood based methods for estimating
smooth macroscopic fields from distributions include the ``thermodynamic field
estimator" \cite{LiJu2}.
This estimated time derivative is then used by the outer integrator to
perform a {\it projection step} in
density space - a step much longer than the time-step in the molecular code.
What carries the day is the separation of time scales in the space
of moments of the distribution (in that, for example, higher moments get very
quickly slaved to density, the zeroth moment, here).
Of course, even {\it within}
the space of the evolution for the zeroth moment (density profiles) slaving and
smoothness is important, and another level of separation of time scales ensues
(center manifolds and possibly inertial manifolds for the density PDE).
But the important thing \cite{Gear1,GKT,Gear2}
is that the detailed, microscopic dynamics themselves ``kill" (damp)
the coarse strong stable modes enough
during the initial, maturing stages of the integration.
Eventually, for {\it coarse} problems
with gaps in their spectrum, the only effective limitation is the time scales of
the coarse slow modes, and the ``outer" integration steps can be taken (with care)
commensurate to these.

The projective step can be run forwards or (interestingly) backwards in time.  
A reason for running it backwards would be to obtain an estimate of
an earlier point {\it on} the ``slow" or inertial manifold --when 
the integration was started off that manifold.  
If, for example, we used $k$ steps, each of length $h$, of an inner integrator, 
--where $k$ was large enough for the fast components to significantly damp 
and approach the manifold--, took one more
step, and then took a projective step of length $kh$ backwards in
time using the chord from the last two integration steps, we would have an
estimate of the initial values {\it on} the inertial manifold.  
Simple analysis (at least, in the linear case) shows that if the inner 
integrator is approximately correct for the fast component -- i.e., 
it approximates $\exp(h\lambda)$ -- then the
initial deviations from the inertial manifold are damped by a factor
proportional to $kh\lambda\exp(kh\lambda)$.   
For negative $\lambda$ this factor can be made arbitrarily small by increasing $k$.

This ``backwards estimate" can also be used to estimate the {\it time derivatives} 
on the manifold at the ``start".  
In this case, we must do at least 2 additional inner steps, 
and then interpolate a quadratic or higher polynomial
through the last 3 (or more) points.  
Estimating the time derivative at the ``start" by computing the derivative 
of this polynomial can be similarly shown to reduce the impact of any 
off-manifold discrepancies in the initial conditions by a 
comparable factor.
An additional benefit of this ``backwards estimate" is that one can now
construct a subroutine which, given a coarse initial condition,
returns an estimate of the coarse time-derivative {\it at this condition}.
This would allow for the easier incorporation of the equation-free approach,
based on the coarse time-stepper, into existing large-scale, equation based,
scientific computing integration packages.

Note that the coarse projective integration procedure can be ``telescoped" 
recursively over
several tiers of description: not just densities to molecules, but (not
too seriously) densities to molecules to wavefunctions, with the microscopic
integration always performed at the inner level.
One then has an hierarchical structure, where the inner timestepper
does the real work, the ``first outer" time-stepper ``projects" inner time-stepper
results, a ``second-outer" timestepper projects ``first outer" timestepper
results and so forth.
The key is to do enough inner integration (that is where the damping takes place)
so that the iterative outer projections remain stable and accurate;
we have some preliminary
experience with these telescoping projective integrators, and we are currently
studying them further \cite{Gear2}.
It may be interesting (although not immediately useful) to remark that the
stability regions of these telescoping integrators have asymptotically
(at the infinite iteration limit) fractal boundaries.

In summary, and in terms of the description of the coarse time-stepper above,
it is possible (see Fig.\ref{fig:project}), through

\begin{enumerate}
\item repeating steps ({\bf c}) through ({\bf f}) above over several
time steps and several successive $U(t_i)$ as well as their
restrictions $u(T_i)={\cal M}U(T_i)$ 

\item using the chord connecting these successive time-stepper
output points to estimate the right-hand-side of the equations we do
not have, we can then

\item use this derivative in another ``outer" integrator scheme.
\end{enumerate}
to estimate the coarse solution ``farther" into the future.
More generally, one can \cite{GKT}
\begin{enumerate}
\item From an initial value at the microscopic level, $U(t=0)$, run the
microscopic simulator (the fine timestepper) for a
number of simulated time steps, generating the values $U(t_i)$ for $i = 1,
2,\cdots, n$.
\item Obtain the restrictions $u(t_k) = \restrict{U(t_k)}$,
for $k =n, n-s, n-2s,\cdots,
n-qs$ for some integers $s$ and $q$.
\item Let $u(t)$ be the $q$-th order polynomial in time, $t$, through
$u(t_k)$.
\item Evaluate $u(T)$, where $T = t_n + H$ and $H$ is a large time step,
appropriate for the macroscopic-level description.
\item Lift $u(T)$ to get a new consistent
microscopic $U = \lift u(T)$ and use it as a new
starting value for repeating steps 3 to 6.
\end{enumerate}

Steps 1 and 5, in the spirit of the discussion above, may be
performed for an appropriately
chosen ensemble of microscopic initial conditions $U_i$, all consistent with
the same macroscopic condition
$u:(\restrict U_j = u, \forall j)$.
The projective step is performed on the macroscopic description, and is
followed by a new lifting step.

Notice that when a microscopic inner integration is continued beyond the first
restriction (``coarse reporting horizon"), we can skip the lifting
step, since the evolving distribution is already mature, and simply
continue evolving for some additional time before restricting again.
The schematic does not contain such ``intermediate" liftings; the
reporting horizons for the subsequent restriction results can therefore,
in principle, be significantly shorter.
As an aside, we note that there exist cases where
intermediate liftings may be necessary
(see the discussion in \cite{Alexei1} about the ``density of collapses",
and also the later discussion about the effect of boundary conditions
in the gaptooth and patch dynamics calculations).

We have analyzed and used such two-tier (and multi-tier) EFM
projective integrators for what we term
``coarse integration" of the --unavailable in closed form macroscopic
equations-- using microscopic as well as stochastic inner time-steppers.
Figure \ref{fig:lbfem}, taken from \cite{GKT} shows time-series as well
as long-term attractors of such an ``inner LB, outer FEM" one stage
projective forward Euler integrator; because of the nature of the
outer description, we call these two-tier codes ``micro-Galerkin methods".
Using different sets of basis functions for the outer integrator is
discussed in \cite{GKT}.
In particular, we have used empirical basis functions (EOFs) for the
representation of the macroscopic field.
These are also known as KL modes (from the Karhunen-Loeve expansion),
POD modes (from the Proper Orthogonal Decomposition) or simply PCs
(from Principal Component Analysis) see \cite{HOLMES,Sirovich}.
Such basis functions (modes) are useful for the parsimonious representation of macroscopic
fields in complicated geometries, and are extensively used in nonlinear
identification and model reduction (for data compression, identification,
bifurcation, control and optimization tasks), e.g. \cite{Deane,Aubry,StasAICHE,JPCPOD,
Patera}.

\section{A DISCUSSION OF THE INGREDIENTS}

In this section we discuss several elements
(all of them more or less well established) that underpin the
``computational bridge" we want
to construct between microscopic simulation and traditional,
macroscopic scientific computing / numerical analysis.
This discussion can be skipped at a first reading of the paper,
and studied after a first pass of the new algorithms and results sections.

\subsection{System Identification}
The most important element comes from systems theory: system identification lies
at the heart of what we propose.
The point is to consider the microscopic simulator as an experiment, from
which measurements are made.
It is crucial that
this computational experiment can be {\it repeatedly initialized}
almost at will; this is to be contrasted to a physical experiment, which
cannot practically be initialized at will.
Using information obtained from several short dynamic runs of the microscopic
code, runs that have been initialized ``intelligently", possibly by post-processing
results of previous runs, one can essentially identify (estimate) useful elements
of the unavailable continuum model.
This procedure has components of what are called ``just in time"
or ``model on demand" approaches in learning theory \cite{Cybenko,Rivera}.
Whether because of the lifting procedure, because of the noise inherent
in stochastic ``inner" timesteppers, or even from the choice of the restriction
operator, it is clear that accurate estimation of various coarse quantities
and derivatives is crucial for the successful performance of the ``outer"
algorithm.
We rely on theory developed in the systems area \cite{Kalman,Astrom,Ljung}
for the best unbiased estimates of these quantities (see also the review
\cite{Sinha}).
Since we can initialize the simulation at will, however, we also have the option
of selecting appropriate ensembles, so that we can improve
variance reduction for a given computational effort; variance reduction
plays a crucial role in the accurate estimation of coarse values and
coarse derivatives.

Because in our case the missing element is the closure of the macroscopic equations
-and that is what, in effect, is accomplished through the identification procedure-,
we believe that ``closure on demand" is an appropriate description for our
computational framework.
One then ``passes" the on-demand identified coarse model
quantities to the standard continuum scientific computing subroutines.
These proceed to perform the computer assisted
analysis ``without realizing" that the numbers they crunch do not come from
a subroutine evaluating an explicit coarse description, but rather from a
subroutine that identifies these quantities ``on demand" through
short computational bursts of the microscopic description.
Part of the macroscopic operations performed includes the post-processing
of simulation results to construct new, useful initial conditions
for the microscopic timestepper:  for example, a new initial guess for an
outer Newton iteration, an outer optimization algorithm,
or for an outer Arnoldi eigensolver.

\subsection{The ``Numerical Analysis of Legacy Codes"}

The next important element is what, for lack of a better expression, could be called
``the numerical analysis of legacy codes";
this is the context in which we first encountered such ideas.
It is a context
of great practical importance in itself, but at the same time it is truly valuable
since it provides a natural framework for the numerical analysis of what we
propose.
The Recursive Projection Method (Shroff and Keller, 1993 \cite{S&K}, see also
\cite{J&M1987a}) is representative of this context (and of what has come to be called
``time-stepper-based" numerical bifurcation theory \cite{TBIMA}).
Consider that a company or National Laboratory (which, as we understand,
was the case for
the inception of RPM) has a large scale dynamic simulator of a physical process (what we
will call ``the legacy code").
This dynamic simulator requires often tens of man-years of development and validation,
and it embodies the best physics description of the process available.
As an illustration, consider that
this code is used, say, to detect a steady state by transient integration;
it is then conceivable that the acquisition of the desired result
(the steady state) may be significantly accelerated by a Newton-Raphson procedure.
However, rewriting the code as a Newton-Raphson iteration may be practically impossible -
the original
programmers may have left or retired, and rewriting the code from scratch might
be an immense undertaking.
Alternatively,
the evolution code may not involve simple right-hand-side evaluations,
but, for example, split-step iterations, and
a steady solver will be far from a small modification.
The alternative is to write a software wrapper ``around" this code, calling it as
a subroutine.
The ``wrapper" will identify, using ``intelligent" calls to the code, what
a Newton iteration would require; and thus exploit the simulator (the ``time-stepper")
making it the core of a continuation - stability - bifurcation code.

This combination of system identification with numerical analysis is motivated by the
structure of large scale numerical linear algebra techniques, based on
matrix vector product computation.
These matrix-vector products can be approximated
through timestepper calls in a ``matrix free" environment.
This was demonstrated in the work of Shroff and Keller
where, exploiting separation of time scales for the original problem,
it was proved that the overall procedure converges.
The main element in the procedure is its two-tier structure: one has an ``inner"
simulator (the legacy code) and an ``outer" processor (identification plus whatever
we do with the identification results, here a contraction mapping).
It is this two-tier
structure that is important for numerical analysis purposes.
There is an extensive and growing literature on the use of techniques like
RPM in the inner ``legacy timestepper" context
(e.g. \cite{vonSosen,Love,Davidson,Koronaki}), not only for steady states
but also for periodic solutions,
whether autonomous \cite{Lust,LustThesis} or periodically forced
\cite{Siettos_PSA}.

In our research so far we have substituted the ``inner simulator" with the
(appropriately initialized) microscopic system model.
However, one can perform the analysis of the
two-tier process assuming a ``general" inner simulator: for example, a ``perfect"
inner simulator as will be shown for the ``gaptooth scheme" below, or a
``typical" inner integrator as can be found in our work on Projective Integrators
\cite{Gear1,Gear2}.
One can then proceed to analyze the stability and convergence of the procedure
based on nominal representative properties of the inner scheme (like the amplification
factor for an integrator, as in \cite{Gear1}) without worrying at this level about
what the detailed inner scheme will eventually actually be.
The results of the overall process for a particular inner scheme may then be found
through the properties (e.g. the integration amplification  factor) of that scheme
itself.

As we discussed above, RPM is a computational superstructure
that enables a ``short time" legacy code to perform ``infinite time" calculations
(find elements of the $\omega$-limit set of the problem).
What we call ``projective integration" is similarly a computational framework
that enables a ``short time" legacy code to perform ``longer time" calculations.
Along the same lines, the so-called ``gaptooth" scheme is a computational
framework that enables a ``small space, short time" legacy code to perform
``large space, short time" calculations.
The gaptooth-projective integration combination (that we call ``patch dynamics")
allows then the ``small space, short time" legacy codes to perform
``large space, long time" calculations, while the gaptooth-RPM combination
allows them to perform ``large space, infinite time" calculations.
The scope of these schemes is summarized in Table 1.
It is not clear which of these names will withstand the test of time, but for
the moment we will keep using them for brevity.

Over the last few years we have demonstrated that this ``enabling legacy
codes" approach can be used by substituting at the inner level an
appropriately initialized microscopic code at the place of the legacy simulator.
The ``coarse timestepper" in effect, evolves distributions -and thus
moments of distributions- conditioned on (initialized consistently with)
initial values of typically a few of these moments; thus the microscopic
code becomes our main tool for conditional probability estimation \cite{Chorin1,Chorin2}.
Some of the combinations we consider (timestepper and RPM,
projective integration) may make
perfect sense for legacy codes; the rationale of other
combinations (the ``gaptooth scheme")
might appear a little warped in the context of legacy codes,
but much more natural, as we shall see, and even quite powerful,
in the context of microscopic inner simulators.
Shroff and Keller's RPM, the Newton-Picard methods of Lust et al.,
our own Projective Integrator analysis \cite{Gear1,GKT,Gear2} as
well as the glimpses of gaptooth numerical analysis presented
below make, we hope, the point.

\subsection{Separation of Scales}

After system identification, the matrix-free computations that arise in the
numerical analysis of legacy codes bring up the next important element in the
overall procedure: separation of time scales (and the concomitant separation of
space scales and the smoothness it induces on the expected behavior).
As one sees in the numerical analysis of RPM, a gap in the eigenvalue spectrum of
the linearization of the problem one studies is useful for the fast convergence of
the method.
The same appears in the study of projective integrators - and is also very important
in the identification of the slow subspace of the evolution of the ``unavailable
coarse equation".

Separation of time (and sometimes concomitant space) scales arises in several
important ways.
Possibly the most important is in the way it appears in the RPM itself:
as a spectral gap {\it within} the dynamics of the legacy code (or the ``unavailable
PDE").
This is typical of dissipative evolution equations (equations characterized
by diffusion -molecular or turbulent-, viscosity, heat conduction...) where
the spatial operators have dispersion relations that give rise to large,
increasingly negative eigenvalues.
If only a few modes, with eigenvalues close to the imaginary axis (possibly
unstable) are ``slow", then a spectral gap exists; to the left of it follow
the (infinitely many) fast, strongly stable modes, and RPM is shown to converge
``fast".
This dissipative PDE/separation of time scale context is also the realm of
Inertial Manifolds/Forms and Approximate Inertial Manifolds/Forms (IMs, AIMs,
IFs and AIFs respectively, \cite{Foias,Temam}).
It is also the context in which the singularly perturbed control methodologies
by Kokotovic and coworkers \cite{KKOR} have been developed over the years; these
will become important in several ways below.

Time scale separation also arises through {\it irreversibility}
and averaging, as one takes hydrodynamic limits of microscopically
reversible processes (e.g. \cite{Lebowitz1,BoonYip}).
In essentially all current microscopic/multiscale models we
evolve distributions (of atoms, particles, cells, individuals)
and try to analyze (simulate, design, optimize and control)
closed deterministic equations for a few order parameters, usually the first few
moments of these distributions (alternatively, well-chosen phase field
variables).
A typical example would be the distribution of molecules (``kinetic
entities" for a kinetic theory) over physical space, time and velocity space.
It is convenient to consider the Navier-Stokes as an example of explicitly
closed macroscopic ``coarse" equations for fields {\it of the first two moments}
(density and momentum) of this distribution over velocity space - (a kind
of {\it Approximate Inertial Form} closure of the moments hierarchy of the
Boltzmann equation).
We will discuss below the issue of choosing possibly ``better" coarse variables
than these first few moments; let us temporarily think in terms of
writing an (infinite) hierarchy of (coupled) equations for the evolution
of moments of the distribution.

{\it If} coarse, deterministic, macroscopic behavior exists (we fully realize
this is a colloquial, and not a mathematical statement), the implication is that
--for the Boltzmann/NSE example--
one can indeed predict, over coarse space and time scales, the behavior of
the first two moments of the distribution (density and momentum) based {\it only}
on these two moments.
It is the {\it closure} of the infinite moment hierarchy that allows us to write
the Navier-Stokes: they contain a model of the effect of all the higher moments
on the first two (Newton's law of viscosity).
The fact that a {\it useful} deterministic equation exists and closes in terms
of only the first two moments does not imply that the higher moments are negligible -
far from that.
It implies that the higher moments of the distribution get {\it very quickly}
(compared to the time scale over which coarse deterministic predictions are made)
slaved to (become functionals of) the lower moments - density and momentum.

The theory of inertial manifolds provides  good ``thought analogies"
and nomenclature for such a
situation. 
In the same way that one writes an (approximate) inertial form in terms
of the first few eigenmodes of the linear part of the problem, one here writes
the Navier-Stokes in terms of the first few moments of the evolving distribution.
By analogy to ``determining modes", we can call these first few moments
{\it the ``determining moments"} for the coarse evolution of the distribution.

By analogy to the parametrization of the (approximate) inertial manifold as a
function over the determining modes, we can think of the existence of an
``almost always" forward attracting, smooth, invariant manifold
in moments space (in the case of IMs, the manifold is Lipschitz and
exponentially attracting \cite{Foias,Temam}).
When the distribution is initialized ``far away" from it, the
higher moments get quickly attracted to this manifold, and thus
become functionals of the lower (governing) moments.
In that sense, we can think of the Navier-Stokes as some kind of ``approximate
inertial form" of the (non-closed) hierarchy of equations for the evolution
of the moments of the distribution.
Alternatively, this ``slow moment manifold" can be thought of as
embodying a closure in the mode hierarchy in the appropriate space
(see \cite{Berez} for a recent corresponding study for diffusion).

It stands to reason that such a slaving is at play when coarse macroscopic
deterministic behavior is observed: if the higher moments {\it were truly individually
important} then knowing only the first few (in effect, conditioning the distribution
on the first few) would not be enough to have deterministic predictions
later on in time.
It is conceivable that one can come up with examples in statistical mechanics
where closed equations can be obtained for some moments {\it even without} slaving;
in most of the practical physical problems that we want to model, however,
quick dynamic slaving
of the high moments of the distributions, effected by the inner, microscopic
timestepper, lies at the heart of macroscopic coarse determinism.

Perhaps the most typical example of this ``irreversibility induced"
separation of time scales is the derivation of the diffusion equation
from molecular dynamics; under appropriate conditions, and
after an initial few molecular collisions,
the density field effectively satisfies the diffusion equation;
all higher moments of the distribution have been slaved
to the zeroth moment (density) field.
Similarly, microscopic (MD,DSMC) simulators effectively ``see" the Navier Stokes
equations after a brief initial startup period \cite{BoonYip,NicolasDSMC}.
This is probably best illustrated by thinking of a (laminar, isothermal) fluid
mechanical experiment as compared to a solution of the Navier-Stokes equations.
We know that measuring the density and momentum
with some resolution allows us to predict what they will become later in time:
we interpolate (smoothness is important) the measurements,
give them as initial conditions to the Navier Stokes, evolve
the Navier-Stokes and obtain deterministic predictions for the same fields after
some time.
Clearly, the NS are ``successful" closed equations for the expected values of two
moments of the distribution; the actual positions and momenta of each molecule at
the initial time are not necessary to make coarse deterministic predictions.
It is worth reiterating that the microscopic (Newtonian collision) dynamics
are fully reversible, yet the macroscopic coarse dynamics are dissipative.

There is a third way by which separation of time scales enters the problem,
and this is the one that we have the most difficulty articulating;
the discussion is still, we hope, informative.
There exist physical contexts in which it does not make sense
to attempt the derivation of an equation for the
(expected value of a) {\it single} experiment; the system is ``too noisy".
While it is not
possible to have deterministic predictions of the coarse state of a single
experiment, we can still obtain good deterministic predictions for the
{\it probability distribution} of coarse states of many experiments over
said coarse state values (e.g. through Fokker-Planck equations).
One evolves simultaneously several realizations of the problem
with the same {\it coarse} initial conditions, and observes the evolution
of the ensemble; noise comes from the {\it fine} initial conditions and/or
the fine simulator.
It is now the dynamics {\it for the probability distributions} of the coarse results
that may acquire time-scale separation: higher order moments again quickly
become functionals of lower order moments.
In this case, however, these are moments of the evolving probability
distributions of the coarse variables (alternatively, of 
the solution of the Fokker-Planck for the evolving probability distribution
of the coarse variables).

A nice example arises in the rigid rod model of
nematic liquid crystal rheology, \cite{Larson,Armstrong,Oettinger},
see \cite{Russo,Graham}.
One cannot write a deterministic equation for the molecular
orientation - but {\it can} write a good deterministic equation at the
level of the 
single rod orientation distribution.
This equation certainly has exponential attractors, and probably even has an
inertial manifold.
Macroscopic closures for the problem
(that is, approximate inertial forms in terms of order parameters -
which can be identified with low moments of the distribution) 
are well known and quite accurate in certain parameter regimes.
So, inertial manifold ``technology" may be invoked to explore
whether indeed high moments of evolving probability density distributions
get slaved to low moments of these distributions.
In this case, it is the averaging over many realizations that
``is responsible for" the
time scale separation, the spectral gaps and the slaving;
and it is the running and averaging of many
similar experiments that allows us to make deterministic predictions
for the {\it statistics} of the expected coarse state, rather than for the
coarse state itself.
One has to work at a different level of coarse description (statistics of
evolving ensembles) in order to restore coarse determinism.
Since the single realization is not ``ergodic enough" any more,
the modeler changes the problem
by following many (similarly coarsely initialized) realizations.

What we do is to use the microscopic timesteppers to identify, on the
fly, what amounts to the time-$T$ map (the timestepper) and
slow evolution subspace of the unavailable equation,
be it an equation for the expected state of an experiment (like
the velocity field in the Navier-Stokes) or
the expected statistics of an experiment (like, possibly, the ``energy versus
wavenumber" spectrum in a turbulence calculation, as we will discuss below).
We then use scientific computation to analyze/compute this equation as if we had it.
Statistical mechanics and probability theory must therefore
play a vital role in establishing or contesting the validity
of the attempt to extend the``numerical
analysis of legacy codes" into the ``coarse numerical analysis of
microscopic timesteppers".

\subsection{Some Additional Considerations}

It is interesting to consider that, in addition to direct simulation, most of
the computational technology we use to enhance the exploration of our computational
models is based on smoothness and the evaluation of derivatives.
The Newton-Raphson is such
a case, as are other steady solvers \cite{Kelley},
continuation algorithms, parametric sensitivity, the integration of
variational and Riccati equations, or optimization algorithms.
Within these algorithms we need to evaluets Jacobians, Hessians,
derivatives with respect to parameters etc.
If the equations are not available, these derivatives have to be identified from
calls to the ``inner" legacy code, and so numerical derivative estimation becomes
important.
In the context of microscopic/stochastic codes, we need to 
obtain accurate {\it coarse} derivative information directly 
from the timestepper: variance
reduction is absolutely {\it crucial} for the practical usefulness of the ideas
we discuss.
There are several ways to accomplish variance reduction: in addition to large samples
(something we want to avoid if possible for computational reasons) one could
try ``intelligent" (i.e. slightly biased) microscopic simulations, such as Brownian
Configuration Fields \cite{Brule1,Brule2}, see alse \cite{OSCM}.
Systems theory comes once more to the aid of such
computations; indeed, the filtering and estimation techniques that have
been developed for noisy data processing and identification (from the famous
Kalman filtering to the recent interest in synchronization and nonlinear observers
\cite{IEEE}) can be brought to bear on the variance reduction problem.

One of the important issues in the ``coarse observation" of microscopic simulations
is the extraction of smooth, coarse, macroscopic fields (output) from, say, the
detailed locations, velocities, and possibly recent histories of molecules in
an MD simulation (the ``microscopic state").
Such problems, using systems theory methods, are currently being tackled in materials
science computations (see for example the ``field estimator" \cite{LiJu2} or
the use of the Cumulative Distribution Function and filtering in \cite{Gear3}).

As we will see, additional elements of systems theory (in particular, controllers)
will be combined with elements of continuum numerical analysis
(the ``gaptooth" and ``patch dynamics" schemes) and with new ensembles from
microscopic simulation (in the spirit of the Dual Volume Grand Canonical MC or MD
simulations \cite{Heff1,Heff2,Heff3,Heff4}).
The crucial element here is the effective imposition of coarse boundary conditions on
microscopic simulators; and the use of control-motivated methodologies to come
up with the corresponding microscopic ensembles.

\section{THE GAP-TOOTH SCHEME}

The discussion up to this point involved results that have either
been published, are in press or under review; we will now proceed
to discuss new results.
We started with the discussion of previous work because we wanted
to show the framework in which what follows was developed as an integral
step of a general program.
Table 1 starts with what we have discussed so far: we have enabled
``large space, short time" coarse timesteppers (through the
``lift-run-restrict" approach, identification, and connection
with traditional scientific computation) to perform ``large space,
infinite time" tasks (find steady states and bifurcations) as well
as ``large space, long time" tasks (coarse integration).
We will now discuss a different ``enabling technology": enabling
a ``small space, short time" timestepper to perform ``large space,
short time" tasks - what for short we will refer to as ``the gaptooth scheme"
\cite{AIChE2000}.

The main idea is related to the above discussion for enabling
a wonderful (but with hardwired $10^{-6}$ step) black-box legacy timestepper
to perform much longer steps {\it when the solution is smooth enough}.
We now do something similar for enabling a wonderful
(but with hardwired domain size $10^{-6}$) black-box PDE solver
to solve in much larger domains {\it again when the solution is smooth
enough}.
While {\it extrapolation} of temporally smooth solutions was the main element
in coarse projective integration, it is the {\it interpolation} of spatially
smooth solutions that will be the main element in the gaptooth scheme and
later in patch dynamics.

We will motivate the work in this section by presenting what will appear
as a ``pretty dumb" scheme for the numerical solution of the diffusion equation
in one spatial dimension.
We will then discuss how this ``gap-tooth" scheme,
can lead to hybrid, two-tier (e.g. finite difference - molecular
dynamics or finite difference - Monte Carlo) timesteppers.
These hybrid timesteppers will be useful for problems for which
microscopic evolution laws (MC, MD, LB) exist and are implemented
in scientific codes;  but
mean field type
evolution equations for ``coarse" quantities,
(such as concentrations or moments of the particle distribution functions)
even though they conceptually exist,
they are not explicitly available in closed form.
Our purpose is to construct accurate timesteppers for the
evolution equations of such coarse quantities {\it without} explicitly
deriving these equations through
analytical %***************
mean-field or averaging procedures.
We use instead the microscopic rules themselves, %,*****************
{\it but in smaller parts of the domain}
and use computational averaging within the subdomains,
followed by interpolation %**************************
to evaluate (estimate, identify) these coarse fields,
their timesteppers, and the time derivative fields over the entire domain.

We start with a one-dimensional case where ``coarse equations" (like the diffusion
equation for a concentration, the zeroth moment of the single particle
distribution function) exist {\it and we know them in closed form}.
We denote the exact solution (of the coarse equation) by  $u^c(x,t)$.
Consider a finite difference (FD) discretization of the diffusion equation
on the full, one-dimensional domain, on a computational grid with spatial stepsize $H$,
which is capable of accurately approximating the (time-dependent as well as stationary)
true solution of the diffusion equation.
(If we are lucky, we may know this solution analytically).
This is our ``Coarse Mesh"; note that it is fine enough for the numerical
approximation to the solution to be accurate for macroscopic purposes.
We let $u^{FD(n)}_i$ be the
grid function approximation of the exact coarse solution on this
mesh, $u^{FD(n)}_i\approx u^c(iH,nT)$.

Now, centered around each discretization point of the coarse mesh,
consider a small computational domain (a ``patch") of size (here, in
1-D,%d******************
of length) $h \ll H$.
These small boxes are the ``teeth" of the scheme, and are separated by
the ``gaps", each of size $H-h$.

In an explicit finite difference scheme, what happens during a single
time step $T$ in a ``big box" (represented by the concentration values at
the coarse nodal positions) is computed (approximated) using fluxes,
proportional to derivatives evaluated through finite differences.
For reaction-diffusion problems, we also need reaction source-sink terms
evaluated {\it at} the nodal point(s)).

We want to represent what happens (during a time step of the explicit
finite difference scheme, which is assumed small enough for the coarse
finite difference scheme to be stable, i.e. one satisfying the CFL
condition) in the ``big box" (of size $H$) through what
happens in the small box (of size $h$) appropriately weighting the
various terms by box size.
This is {\em another} way of approximating the
exact coarse solution $u^c(x,t)$, and we denote the
corresponding grid function with $u^{c(n)}_i$, so that
$u^{c(n)}_i\approx u^c(iH,nT)$.
{\it The small box in this approximation
plays the role of the nodal point of the coarse finite
difference scheme}.

In each such small box we solve the same transient
governing balance equations as those for the whole domain,
up to the reporting time, $T$ (the time step of the coarse FD scheme).
Let $u_i(y,t)$
denote the solution in the small box $i$ with
appropriate boundary and initial conditions;
the spatial coordinate $y$
is local to each box $0 \leq y \leq h$).
Hence, given a small box size $h$, for each box $i$,
\begin{itemize}
   \item Construct boundary conditions and initial data
         for the small box from
         the coarse grid function $u^{c(n)}_i$.
   \item Compute $u_i(y,t)$ by solving the governing balance equation
         {\em exactly} for time $t\in[0,T]$ in each small box
         $y\in [0,h]\equiv [x_i-h/2,x_i+h/2]$ with these boundary conditions
         and initial data.
 \item Define the discrete approximation at the
       next time level by the small box average of $u_i$ at
       the next time level $t=T$,
$$
  u_{i}^{c(n+1)} = \bar{u}_i(T)\equiv \frac1h
  \int_{0}^{h}u_i(y,T)dy.
$$
Note that this should correspond to the
explicit FD solution $u_{i}^{c(n+1)}\approx u_i^{FD(n+1)}$.
\end{itemize}

For a balance equation (diffusion, reaction diffusion) the
local boundary conditions should be the ``correct"
slopes (determining the fluxes) in and out of each  {\it small} box.
The derivative $s_{i+\frac{1}{2}}$
at the midpoint $x_{i+\frac{1}{2}} = \frac{1}{2}(x_i + x_{i+1})$   between two adjacent ``coarse'' nodes
$x_i$ and $x_{i+1}$ can be approximated by the corresponding centered
finite differences

\begin{equation}
\label{slop}
    s_{i+\frac{1}{2}}=(u_{i+1}^c-u_{i}^c)/H.
\end{equation}

\noindent
Let the left and the right %**********************
slope boundary conditions for each small box
be %,***************
$s^-_i$ and $s^+_i$.
They can then be computed by linear interpolation between
$s_{i-\frac{1}{2}}$ and $s_{i+\frac{1}{2}}$
(see inset of Fig.\ref{fig:oldAIChE_schem}).
Therefore

\begin{eqnarray}
\label{s'1}
s^-_i=\frac{(H+h)s_{i-\frac{1}{2}}+(H-h)s_{i+\frac{1}{2}}}{2H}\\
\label{s'2}
s^+_i=\frac{(H-h)s_{i-\frac{1}{2}}+(H+h)s_{i+\frac{1}{2}}}{2H}.
\end{eqnarray}

We have not yet discussed what to use as an initial
condition for the simulation in the small box with these
boundary conditions.
It is reasonable to expect that the average (over the small box)
of the quantity we are studying will be initially equal to the
``nodal value"  $u_i^{c(n)}$.
{\it Any} initial condition with the correct average ought to do a good
job in the scheme we envision ({\it as we discuss below, it is
possible that we may have to run an} {\bf ensemble} {\it of
such ``macroscopically consistent" initial conditions and average
over their results)}.
If the initial condition does not satisfy the BC
at $t=0$ there will be sharp startup transients close to the
boundary, but in
principle the time-reporting horizon for the simulation in the small box
(corresponding to one time step of the FD scheme for the
``big" box) should be enough for these
transients to smoothen out.
We will return extensively to the discussion of the time-reporting horizon
of the simulation in each small ``tooth".
{\it This is also an important point for
when we get to microscopic simulations: how quickly do ``random" initial
conditions consistent with (conditioned upon) some moment fields, heal -
that is, how quickly do their higher moment fields become functionals of (get slaved
to) the lower, governing moments fields ?).
}

Let us study the gap-tooth scheme for the  linear problem
\begin{equation}
    \label{gov}
    u_{t}=u_{xx}.
\end{equation}
Since the boundary conditions for the small boxes
are slopes, the average is given, and the
profile is assumed to be smooth, it is reasonable (from all possible
initial conditions with the same mean) to choose the simplest polynomial
that satisfies the B.C: a quadratic function of space within the small
box:
$$u(y) = Ay^2 + By + C, \qquad
y \in [0,h] \equiv \left[x_i - \frac{h}{2}, x_i + \frac{h}{2}\right].$$

\vspace{0.1in}

The explicit (coarse) FD solution we assume is given by
the forward Euler method,
\begin{equation}
\label{expl}
u_i^{FD(n+1)}=u_i^{FD(n)}+T \frac{u_{i+1}^{FD(n)}-2u_i^{FD(n)}+u_{i-1}^{FD(n)}}{H^2}.
\end{equation}
The ``gap-tooth" scheme
starts with the coarse $u_i^{c(n)}$ from the previous time step.
For each box $i$,
\begin{itemize}
    \item  Use $u_i^{c(n)}$
 to compute the boundary conditions $s^-_i$
and $s^+_i$ at the beginning and end of each small box respectively,
based on the linear interpolation formula (equ.(\ref{s'1}) and (\ref{s'2}))
with $s_{i-\frac{1}{2}}$ and $s_{i+\frac{1}{2}}$ calculated from equation (\ref{slop}),
\begin{equation}
\label{boxbc}
   \partial_y u_i(0,t) = s_i^-, \qquad
   \partial_y u_i(h,t) = s_i^+, \qquad t>0.
\end{equation}
    \item  Use $u_i^{c(n)}$
to find the initial condition
through the quadratic expression
\begin{equation}
\label{boxic}
u_i(y,0) = Ay^2+By+C,
\end{equation}
with the coefficients
$A$, $B$ and $C$ computed so that the profile matches the BC and the (mean) IC.
The slope at every $y$ is given by
$2Ay+B$ and, from matching the BC's, the correct $A$ and $B$
can be calculated as
\begin{eqnarray}
\label{a1}
    B&=&s^-_i,\\
\label{b1}
    A&=&(s^+_i-s^-_i)/2h.
\end{eqnarray}
Moreover, for consistent lifting,
\begin{equation}
\label{integ}
    \bar{u}_i(0)\equiv
\frac{1}{h}\int_{0}^hu_i(y,0)dy=
\frac{1}{h}\int_{0}^h{(Ay^2+By+C)dy}=u_i^{c(n)}.
\end{equation}
The average value of $u_i(y,t)$ in the box is identified with
the $u^c(x,t)$ value at the corresponding ``coarse'' node
$u_i^{c(n)}$ (which is also supposed to correspond to the coarse FD
value $u_i^{FD(n)}$).
We can compute $C$ from Eq. (\ref{integ}) as
\begin{equation}
\label{c1}
    C=u_i^{c(n)}-\frac{1}{h}\int_{0}^h{(Ay^2+By)dx}.
\end{equation}
Note that $A$, $B$ and $C$ are linear in $u_i^{c(n)}$ by
equations (\ref{a1}), (\ref{b1}) and (\ref{c1}).

    \item Integrate the equation in the small box for time
$T$, corresponding to the time step of the coarse FD scheme.
During this
time the boundary conditions remain constant and equal with their values
at the beginning of the step. Hence, solve
$$
   \partial_t u_i = \partial_{yy}u_i, \qquad y\in (0,h),\quad t\in (0,T],
$$
with initial data (\ref{boxic}) subject to boundary
conditions (\ref{boxbc}).
While in general the solution would need to be approximated numerically
(e.g. through a {\it fine} FD scheme, {\it within} each box,
and with well-chosen
fine time-step, or --- for the diffusion equation --- through a reasonable
truncation of the solution obtained through separation of variables)
here we are lucky: for the diffusion equation,
the particular initial condition we chose for each
box (a parabolic profile satisfying the mean and the BC) has an
{\it explicit} solution:
    $$u_i(y,t) = Ay^2 + By + C + 2At,$$.

\item  Compute the discrete approximation at
time $T$ by
averaging in space over the small box solution $u_i(y,t)$ at
the final time $T$.
We obtain
\begin{equation}
    u_i^{c(n+1)}=\bar{u}_i(T) = \frac{1}{h}\int_{0}^h(Ay^2+By+C+2AT)dy.
\end{equation}
By using (\ref{b1}), (\ref{s'1}) and (\ref{s'2})
we thus have
\begin{equation}
\label{alm}
    u_i^{c(n+1)}= u_i^{c(n)} +  T \frac{s^+_i-s^-_i}{h}
= u_i^{c(n)} +  T \frac{s_{i+\frac{1}{2}}-s_{i-\frac{1}{2}}}{H}.
\end{equation}
Finally, from (\ref{slop})
we obtain the new coarse mesh quantities:
\begin{equation}
\label{fin}
    u_i^{c(n+1)}=u_i^{c(n)}+  T
\frac{u_{i+1}^{c(n)}-2u_i^{c(n)}+u_{i-1}^{c(n)}}{H^2}.
\end{equation}
\end{itemize}

The procedure then repeats itself: we interpolate to obtain a new
coarse field, i.e. use the $u_i^{c(n+1)}$ to compute the new slopes $s_i^+$ and
$s_i^-$ and so on.
The right-hand side of the above equation is {\it exactly} the same as the
right-hand side
of equation (\ref{expl}); therefore, this hybrid scheme {\it reduces 
exactly to the
corresponding
coarse FD discretization}.

In this example we used a nodal representation of the macroscopic field
as well as linear interpolation of coarse slopes to determine the
patch boundary conditions.
These features are shared by the FD scheme that was used to motivate the
approach.
We refer to this as the {\it ``nodal--Exact"} gap-tooth scheme.
The first component of the
name (nodal) refers to our chosen representation of the macroscopic field
which guides the number and location of coarse nodes as well as the
patch boundary condition procedure.
The second component (here ``exact") is the way by which
we solve the equation in each small box.

In the simple diffusion example, the truncation error
for ``nodal--Exact'' is $O(H^2+T)$, independent of $h$.
To analyze the accurcay of the scheme, consider first the limiting
case when $h=H$ i.e., when there are no gaps between the teeth.
The scheme will then have  truncation error is $O(H^{p_0}+T^{p_1})$.
The values of $p_0$ and $p_1$ depend on the order with which
the boundary conditions are interpolated in space
and extrapolated in time from the data $u_i^{c(n)}$.
Furthermore, an asymptotic analysis reveals
that the difference between ``nodal--Exact''
with $h<H$ and with $h=H$ in one timestep is
$O(T(H^{q}-h^{q}))$, for some $q$  (for  $h,H \ll 1$).
Thus the total
truncation error for ``nodal--Exact'' is bounded by
\begin{equation}\label{FDExacttrunc}
O\left(H^{p_0}+T^{p_1}+|H^{q}-h^{q}|\right).
\end{equation}
When the ``nodal--Exact'' scheme described above is applied to
a reaction-diffusion equation, $u_t=u_{xx}+f(u)$,
we will have $p_0=q=2$ and $p_1=1$. Since $p_0\leq q$, we note
that in this case the order of accuracy is still asymptotically
independent of $h$.

If we solve problems where the solution within each small box
is not explicitly available, we also need to compute it
numerically using, for example,
a finite difference scheme on fine mesh,
with $N_f$ ($f$ for {\it fine}) nodes
within each small box and time step $T_f$.
This would be the ``nodal--FD" scheme.
In this case, the truncation
error of the FD scheme
will be $O((h/N_f)^{r_0}+T_f^{r_1})$, with $r_0$ and $r_1$
depending on the scheme.
Assuming stability,
this is also the
error for one coarse time step $T$ ``within" each small box.
The total truncation error for ``nodal--FD'' is then
\begin{equation}\label{FDFDtrunc}
 O\left(H^{p_0}+\left(\frac{h}{N_f}\right)^{r_0}
        +T^{p_1} + T_f^{r_1}+|H^{q}-h^{q}|\right).
\end{equation}

In addition to the consistency of the schemes as expressed by
(\ref{FDExacttrunc}) and (\ref{FDFDtrunc}), we also need to
consider their stability at both the coarse
level and at the fine level independently.
Stability conditions will
typically restrict the possible sizes of $T$, $H$ and $T_f$, $N_f$.
Optimal $h/H$ as well as $T_f$ and $N_f$ will
appropriately balance errors and computational cost under these constraints.

This example was chosen to illustrate an EFM approach in situations
where the macroscopic balance equations  are {\bf not} explicitly available.
This ``two-tier", inner-outer structure of the solver is a persistent feature
in EFM computations.
In general, the ``inner" solver will not come from a
finer description of the {\it same} representation of the physics (i.e.
the diffusion equation) but from a different level representation of the
physics, i.e. random walkers or molecular dynamics.

In these situations the above procedure is modified so that

\begin{itemize}
    \item The initial condition in each small box is somehow ``lifted" to
    a {\bf microscopic} initial condition consistent with the
    coarse profile (or an appropriate ensemble of such consistent,
    microscopic initial conditions --distributions conditioned on
    a few low moments--); boundary conditions for each small box are
	computed from the coarse field.
    \item Microscopic versions of the BC are applied to the small box
    problems.  ({\it New ensembles for doing, say, MC under various
    boundary conditions will need to be invented, see for example
    the work of Heffelfinger et al. on Dual-Volume Grand
    Canonical Molecular Dynamics or -Monte Carlo as an example of such
    ensembles, \cite{Heff1,Heff2,Heff3,Heff4}}).
    \item The small box problems are propagated in time using
    microscopic evolution rules (MD, MC, LB) with, in principle,
    constant boundary conditions over the reporting horizon
    (corresponding to one coarse time step).
    \item The results at the end of the ``coarse" time interval
    are averaged within each small box, subsequently interpolated
        in space between these
        averaged values. This gives the {\it coarse field timestepper}
	evolution for one step. The restriction operator maps the
    fine to the coarse description levels.
\end{itemize}
    The procedure is then repeated.

It is conceivable that we may have to average over several realizations
-possibly over several initial conditions- of the simulations in each
small box
to get good variance reduction
(i.e., ``good" final expected values for the coarse quantity of interest
in this ``tooth"; good enough for the interpolation in coarse space that follows).

It is important, as was discussed above, that the time-reporting horizon
for the microscopic simulation is long enough for the higher order moments
(incorrectly, ``off manifold" initialized when microscopically lifting) 
to have time to ``heal",
i.e. to become functionals of the governing moments, the ones for which
we believe that an evolution equation exists and closes.
A comparable ``healing" process should occur in terms of the spatial interpolation-
the time integration should be long enough for interpolation errors
in the higher spatial frequency
components of the coarse field to get slaved to lower spatial frequency components.
Similar thoughts (through a ``center manifold" argument) for dissipative PDEs, at
a single level of physics description, have been discussed by Roberts \cite{Roberts}.

While this scheme seems computationally intensive (solving in each small
box, averaging and interpolating is more expensive than
the corresponding ``straight" coarse FD solution), it is aimed
towards the simulation of {\it microscopic systems} on large computational domains.
An extra advantage is that the small box computations for the reporting
time can be performed in parallel (a separate processor for each ``box" or
patch)
significantly reducing the total computational time.
What was described here was templated on an {\it explicit} outer solver; in
the spirit of implicit projective integrators, we expect that implicit 
(predictor-corrector) versions of this procedure can also be constructed.

Clearly, the ``gaptooth" scheme can also be combined with what are traditionally called
``hybrid codes" \cite{Shenoy,Garcia,Abraham,HadjPat}: codes that combine a continuum
with a microscopic description at
different parts of the domain, matching them at ``computational interfaces".
If in parts of the domain an available coarse description is valid, we evolve
the mesh quantities there with the coarse equation ``outer" scheme directly.
In the regions were the explicit macroscopic description fails, we use ``teeth"
(boxes, patches in higher dimensions) and the ``lift-run-restrict" approach.
It appears as if the entire domain has been evolved with the ``outer scheme".
But the
numbers that this outer scheme crunches come in part from simple function evaluations
(in parts of the domain where the coarse equation is valid) and in part from
the appropriately initialized and evolved microscopic timestepper (in the regions
of the domain where the coarse description is inaccurate and we need to revert to
the microscopic physics).
Many current hybrid applications have only one ``tooth":
the {\it entire} portion of the domain where the macroscopic equation fails
is evolved microscopically, and matched to the continuum.
If this portion of the domain is large, however, the gaptooth approach
may reduce the extent of physical space over which microscopic simulations
must be evolved.

One can also envision (although the details need to be worked out) that
multitiered {\it outer} solvers (e.g. multigrid solvers), can be combined with
the inner microscopic small-box timestepper.
As a matter of fact, in the spirit of telescopic projective integrators,
different levels of physics descriptions can be used at different
levels of a composite, multitier solver.
In established multigrid approaches one works with different coarseness levels of
{\it the same} description; here one would work with
{\it different levels of description};
the inner, ``ultimate microscopic" solver provides the basic detailed physics; and
it is ``coaxed" through the computational superstructure to also 
provide the relaxation
of the basic detailed physics to the ``coarse slow" system-level dynamics we want
to study.
That such a ``coarse slow" description exists, and the level at which it exists, is
a basic premise of the entire numerical skeleton we build around the microscopic
timestepper.
This basic premise requires validation during any
computation.

There are several levels at which the procedure can fail; some of them are
traditional numerical failures (inadequate coarse discrete approximations
in space and/or time) and the ways to deal with them on-line during
a computation through
refinement or adaptive meshing are parts of traditional scientific computation.
The more interesting and new feature, that has to be combined with these more
traditional numerical aspects, is the ``closure failure", i.e. the possibility
that equations do not close any more (as conditions change) at the level of
coarse description we have been using up to now.
A typical example would be that we need to write equations, in chemical kinetics,
not just in terms of densities, but of densities and pair probabilities.
Alternatively, in fluid mechanics, that we would need to write equations
for stresses as independent variables as the Deborah number grows,
and not just have them slaved to momentum gradients.
These considerations have been discussed to some extent in \cite{Alexei1}; we
believe that testing such ``constitutive failures" can be reasonably done ``on line",
and we discuss how to do it in \cite{Alexei2,Graham}.
We believe that the ability to detect,
test for, and possibly remedy such ``constitutive failures" is one of the
strong points of the computational framework we discuss.
We only mention here that ways to test for ``constitutive failures" have to
be integrated in the framework, along with tests for more
traditional discretization-based ``numerical failure".

To validate EF computations for any given problem
we always start in a parameter regime
where we {\it know} a valid macroscopic description explicitly,
and we reproduce its results.
We then perform homotopy/continuation towards the regime of interest in
parameters/operating conditions, constantly ``numerically" 
as well as ``constitutively"
validating the procedure.
As we gain more experience from applications, our knowledge of levels of
closure appropriate at different parameter regimes will give enough confidence
to avoid the overhead of constant validation.

Figure \ref{fig:oldAIChE_schem} shows our first schematic of the gaptooth 
scheme from the original
presentation \cite{AIChE2000}; the notation is slightly different than what
we have currently ``converged on" above, but the main ideas are clear.

Figure \ref{fig:oldAIChE_res} shows, from the same presentation, our 
first gaptooth integration
results for the FHN equation (this computation did not involve projective integration).
In those simulations the domain length was $L=20.0$ and the parameter,
$\epsilon$ was set to $\epsilon=0.5$. At this parameter value a steady
front is produced. The domain was discretized in 30 nodes, therefore
30 boxes were used, each of them having length, $h$ equal to  $1/4$ of the
internode distance, $H$ (i.e. $h=0.1724$). 
For each overall step,
starting from values of the density (zeroth moment for the LB-BGK code)
on a coarse grid we computed a coarse density  profile through interpolation
(taking advantage of smoothness); we also used this coarse profile to compute
temporary (slope) boundary conditions for the little boxes.
We lifted the density profile to a consistent LB profile
in each box, and then evolved the problem through LB-BGK
for one reporting horizon in each box, keeping the
temporary boundary conditions constant; the ``outer" scheme here is
an explicit finite difference scheme.
Implicit (iterated to self-consistency)
versions of this evolution may also be possible \cite{Gear1,GKT}.
After the problem has evolved for one time horizon in each box using the ``inner"
solver, we {\it restrict} the solution, 
computing the averages of the densities in each box.
We have thus created an ``inner solver-assisted" map from coarse grid
to coarse grid.
The procedure was then repeated.

We verified the effectiveness of this approach
in a nodal-FD mode (inner code: FD in (fine) space,
trapezoidal in (fine) time; nodal macroscopic field representation)
as a sanity check.
We then performed the computations in a nodal-LB code (we will call this an LB-BOX
code): inner LB in fine space and fine time, with lifting and with slope boundary
conditions; and outer nodal representation of the coarse fields.
Some of the results presented in \cite{AIChE2000} are 
shown in Fig.\ref{fig:oldAIChE_res} both
for ``inner FD" and ``inner LB" evolvers for comparison purposes.

In the simulations for the gaptooth procedure shown
in the following Figures, 101 boxes were used with $h/H=0.1$, 
i.e. the box length was $h=0.02$.
Figure \ref{fig:profbox1} shows comparisons of full LB and LB-BOX
profiles of the variable $u$ at various times in the
{\it oscillatory} regime of the FHN ($\epsilon=0.01$).
The solution
is a spatiotemporal limit cycle, and the figure shows a portion of it,
involving the palindromic movement of a sharp reaction front.
Figure \ref{fig:attrlbbox_sh} shows a projection of this spatiotemporal
long-term periodic attractor computed through full LB and through the
gaptooth LB-BOX approach.
The effect of the coarse time reporting horizon, $T_c=7.5 \times 10^{-4}$,
can be seen comparing with Fig. \ref{fig:attrlbbox_lo} where the
{\it long-term attractor} of the same scheme {\it but with a larger
coarse time horizon} ($T_c=2.5 \times 10^{-3}$) is shown.
In the latter case the communication between the
boxes, by constructing the new coarse profile and the ``new" box boundary
conditions for the next microscopic sumulation ``era", 
occurs a factor of three less frequently.
It is important in comparing these two figures to remember that what is shown
are {\it long term attractors}.
The short term itegration with the two different
reporting horizons will, for quite some macroscopic time, appear visually practically
the same at the resolution of these plots.

Comparable results have been obtained for
the diffusion equation and also for the Allen-Cahn equation \cite{AlCa}
of the form
\begin{eqnarray}
u_t=u_{xx}+ 2 (u-u^3),
\label{AC}
\end{eqnarray}
where the relaxation of the macroscale profile to the
exact steady state of the form $u=\tanh(x-x_0)$ was observed.
The
simulation used an ``inner FD'' scheme in each patch (with
different numbers of nodes in the short space of the patch and appropriate
time steps between each ``outer" reporting horizon).

There is one more issue that is important to discuss: the repeatedly imposed,
``macroscopically motivated" boundary conditions for the patches,
and the appropriate microscopic ensembles for imposing them.
This discussion will follow the next section on
``Patch Dynamics", since
patch boundary conditions are also important in that context.

\section {PATCH DYNAMICS: A COMPOSITION}

In Section III we were able to use ``large space, short time" computations,
embodied in calls to the microscopic ``large space" timestepper,
in order to perform coarse stability and bifurcation calculations (see Table 1).
Since the results of this procedure are invariant objects (including elements
of the $\omega$-limit set of the unavailable coarse equation) we can consider that
the coarse bifurcation results are ``large space, infinite time".
Correspondingly, projective integration as presented above allows
the use of ``large space, short time" computations (again embodied in
calls to the microscopic ``large space" time stepper), in order to perform numerical
integration over larger macro-time steps.
We could therefore summarize this as ``large space, short time" to
``large space, long time" computations (``long time" here is
to be contrasted with ``infinite", {\it i.e.} with bifurcation-level computations.

The ``gaptooth" time-stepper discussed in the previous section is a
``short space, short time" to ``large space, short time"
computational framework.
One could, in principle, use the gaptooth timestepper repeatedly, tiling
longer and longer intervals in time ``completely".
In the spirit, however, of our coarse projective integration discussion,
it is preferable that the gaptooth timestepper can be combined with
projective integrator templates to provide an EF framework bridging ``short space, short time"
simulations and ``large space, long time" evolution.
The gaptooth timestepper can also be combined with
coarse bifurcation calculations to provide a ``short space, short time" simulation
``large space, infinite time" computational framework.

A schematic summary of the
space-time features of the algorithms can be seen in Fig. \ref{fig:sp_time_sum}.
In the middle (a) we see that, in order to evolve the problem with the
``detailed microscopic" physics description, we must tile the space-time
area with extremely fine grids in both directions.
This would be practically computationally intractable.
The basic building block of our EF ``computational enabling technology" is the 
same detailed description, but in a ``little box" in space for a short
time (a small space-time region).
What enables the simulation to be done faster, if at all, is the basic
premise that {\it coarse behavior} exists -- that ``emergent" (for lack
of a better word), smooth in space and time at much larger scales, behavior
exists for the moments of our microscopically evolving distributions.
Projective integration is shown in caricature in (b), where we evolve in
large space but for short times, and then we {\it project}, and start
again (the lifting part has been suppressed).
The gaptooth scheme is shown in caricature in (c), where we evolve in small
boxes for a short time horizon, and then we stop, estimate coarse fields,
interpolate to compute new box boundary conditions (let the boxes talk
to each other) and start again.
Putting the two together is shown in (d) and -really- in (e): exploiting
smoothness in time we need only to integrate small times, and exploting
smoothness in space we need only to integrate in patches -- as a result
small space-time regions emerge from the overlap of the 
temporal and the spatial ``strips".

It is only {\it within} these ``space-time elements" 
(small space, short time) that the actual
microscopic evolution takes place -- the results from this are ``intelligently
postprocessed" by the identification part of the software, and ``passed on"
to the traditional, continuum numerical routines for number-crunching.
Since this way the closures for the unavailable macroscopic equations have been
computed in real time (``on demand"), the macro-solvers (the ``outer"
computational codes) do not realize that the information they process does not come
from a closed formula.

What is ``space" in this caricature is
{\it not necessarily physical space} -- it is whatever our evolving distribution is
distributed over, in addition to physical space.
For population dynamics an additional coarse dimension may be the cell size;
in the coarse integration of turbulence spectra, additional
coarse dimensions would involve wavenumber space.
The higher the dimensions of the problem we are evolving, the
higher the potential savings compared to ``full space" microscopic simulations.
Consider for example, agent based models arising in ecology or epidemiology,
where the required solution moments are distributed not only over space and time,
but also over several features of the populations in question (e.g. cell size, age, or
income); if we only solve in a fraction $h/H$ of each coarse dimension, the
savings grow as powers of this fraction.
It will be extremely interesting to explore and devise ``patch boundary
conditions" for such alternative descriptions of distributed systems.

Fig. \ref{fig:patchfull} contains a schematic
summary of the patch dynamics procedure which we now repeat in
words:

\begin{itemize}
\item[(a)] \underline{Coarse variable selection} (as in Section 2, but now the
variable $u(x)$ depends on ``coarse space"  $x$). We have chosen for
simplicity to consider only one space dimension. This defines
the restriction operator $\restrict$.

\item[(b)] \underline{Choice of lifting operator} $\lift$ (as in Section 2, but now we lift
entire profiles of $u(x)$ to profiles of $U(y)$, where $y$ is
microscopic space corresponding to the macroscopic space $x$). This
lifting involves therefore not only the variables, but the space
descriptions too. The basic idea is that a coarse point in $x$
corresponds to an interval (a ``box" or ``patch" in $y$).

\item[(c)] \underline{Prescribe a macroscopic initial profile} $u(x,t=0)$ - the
``coarse field". In particular, consider the values $u_i(t=0)$ at a
number of macro-mesh points (or, for that matter, finite element coefficients,
or spectral cofficients); the macroscopic profile comes then
from appropriate interpolation of these values of the coarse field.
Compute boundary conditions for the ``patches" from the 
coarse field.

\item[(d)] \underline{Lift} the ``mesh points" $x_i$ and the values $u_i(t=0)$ to
profiles $U_i(y_i)$ in microscopic domains (``patches") $y_i$
corresponding to the coarse mesh points $x_i$. These profiles
should be conditioned on the values $u_i$, and it is a good idea
that they are also conditioned on certain boundary conditions
motivated by the coarse field (e.g. be consistent with coarse
slopes at the boundaries of the microdomain that come from the
coarse field, see the discussion below).

\item[(e)] \underline{Evolve each of these ``micro-patches"} for a short time
based on the microscopic description, and through ensembles that
enforce the ``macroscopically inspired" boundary conditions - and thus
generate $U_i(y_i,T)$.

\item[(f)] \underline{Obtain the restriction} from each patch to coarse variables
$u_i(T) = {\cal M} U_i(y_i, T)$.

\item[(g)] \underline{Interpolate} between these to obtain the new coarse-field
$u(x,T)$ and, from it, new boundary conditions for each patch.
\end{itemize}

The steps above define the ``gaptooth scheme", that is, 
a scheme that computes in small domains
(the ``teeth") which periodically communicate over the gaps between them through
``coarse field inspired" boundary conditions. 
This provides us with
the ``small space, short time" to ``large space, short time" enabling
technology.
We now proceed by combining the gaptooth scheme with the Projective
Integrator ideas to

\begin{itemize}
\item[(h)] \underline{Repeat the process} (compute patch boundary
conditions, lift to patches,
evolve microscopically for a few steps) and then

\item[(i)] \underline{Advance coarse fields ``long" into the future} through
projective integration.
This first involves estimation of the time-derivatives for the coarse field
variables, using successively reported coarse fields.
A projective step then follows, advancing the coarse field into the future.

\item[(j)] \underline{Repeat the entire procedure} starting with the lifting (d)
above.
\end{itemize}
Microscopic simulations in each patch yield averaged quantities on a macroscopic
``sampling mesh."  These quantities are then interpolated to define the macroscopic fields.
The macroscopic time derivatives can be estimated at the nodes directly from the
detailed computational data in the patches.  Alternatively, they can be estimated from a
time sequence of the basis functional representation of the macroscopic fields. 

Results of ``patch dynamics" and ``patch bifurcation" computations
for our FHN equation based LB-example are now presented and discussed.
We already showed in Fig.\ref{fig:bifhop} the coarse bifurcation diagram of the
FHN obtained through a ``large space" LB coarse timestepper; the figure
also shows, for comparison, the same bifurcation diagram obtained with
a ``small space" LB coarse timestepper (a two-tier nodal-LB scheme that we
call LB-BOX).
Here, 101 lattice points were used for the LB timestepper and
correspondingly 101 boxes with $h/H=0.1$ (so the ``inner" solver
operates on only one tenth of the full spatial domain).
It is important to notice that, beyond the main features of the
diagram, the actual shapes of the computed spatially varying
steady states, both stable and unstable, are well captured.
Furthermore, their stability (embodied in the coarse linearization) is also
well captured.
This can be seen both for the LB and for the LB-BOX scheme.
The LB-BOX scheme clearly captures quantitatively the linearized stability
of the detailed problem at low wavenumbers (at the coarse level) as well
as the ``large space" coarse LB scheme does.
The combination of the gaptooth scheme with techniques like RPM allows us,
therefore, to perform coarse bifurcation computations using ``boxes" or
``patches" in space-
it provides a ``small space, short time to long space, infinite time" framework.

Figure \ref{fig:gearbox_tser} compares full LB with patch dynamics
(gaptooth combined with projective integration) results;
the algorithm now has several tiers.
There are two tiers in time: the inner integrator is
a ``gaptooth" LB, and the outer integrator is forward Euler on the coarse
profile (to be exact, on the macroscopic mesh density values).
There are, however, tiers
also in space: the ``coarse density profile" interpolated from the density
values at the coarse mesh points; the (auxiliary) coarse profile in each
box; and the Lattice-Boltzmann profile in each box, the one that we lift to,
and which we evolve using ``microscopic evolution physics".
In particular we call this a 100-100 ``inner'' 200 ``outer''
integrator, since the forward Euler ``jump'' is equivalent to 200
time steps of the inner integrator (see also \cite{GKT}). 
Fig.\ref{fig:gearbox_short} shows the corresponding short-term integration results, and
Fig.\ref{fig:gearbox_attr} shows the long-term periodic attractor, successfully captured
by the LB-BOX-Projective scheme.
These are, of course, preliminary results in a regime where the
``coarse PDE" is known, where a lot of variance reduction is not really
necessary, and where therefore we find a context for troubleshooting
the computational procedure and exploring  the numerical analysis
of the multi-tiered schemes.

This completes the ``simulation" part of EFM toolkit.
We now have, through the combination of the gaptooth scheme with
projective integrators and RPM what we call ``patch dynamics" 
and ``patch bifurcations" respectively:
a ``small space, short time" to ``large space, long/infinite time"
enabling methodology.
The accuracy and stability analysis of patch dynamics algorithms obviously
will rely on the corresponding analyses of the ``gaptooth" scheme and the
projective integrator schemes, and is the subject of current research.
This ``coarse simulation" technology can also be coupled with both
controller design and optimization algorithms.

\section{ON PATCH BOUNDARY CONDITIONS}

In the description, above, of the ``nodal-Exact" gaptooth scheme, we
directly introduced slope boundary conditions for the microscopic
computational domains (``teeth", ``boxes").
We will now briefly discuss
{\it how} one would repeatedly impose such boundary conditions in the
case of microscopic simulators, and then
{\it what} boundary conditions can be used for the little boxes.
The nomenclature is quickly summarized by the one-dimensional
sketch in Fig.\ref{fig:patch1d}.
The boundary conditions for the microscopic dynamics will be imposed using
buffer regions surrounding the patches;
converting the macroscale interpolant into a microscale boundary condition is
clearly a key algorithm in this framework.
The appropriate boundary conditions for the patches will, of course,
be problem specific. 

The macroscale average quantities are interpolated (e.g. through piecewise linear
or cubic Hermite interpolation) across the spatial gaps to give macroscopic
fields.
The approach is straightforward in higher-dimensional problems, when the patches are
centered at the gridpoints of a rectangular or cuboid grid.
The coarse, macroscale variables will be defined at the grid points and interpolated
through, for example, tensor product piecewise Hermite polynomials. 
These interpolations are fast and efficient on regular grids for high dimensions
(e.g. \cite{deBoor}).

If the ``inner solver" for the patches is based on a known partial differential equation,
then this equation acts as a guide in defining appropriate boundary conditions for
the patches.
However, even for known PDEs, in many cases it is not clear how to best choose the boundary conditions to
``tie" the inner solution to the macro solution.
If the macroscale field is sufficiently accurate, using it to overpose the BCs for a microscale
patch may yield acceptable results if the frequent communication between the scales regularizes
the solution.
That is, if the errors resulting from the temporarily frozen BC grow slowly, so that they remain
small and confined to the buffer regions, their effect will be minimized at the next reporting horizon.
Such a self-regularizing effect is reminiscent of AMR simulations \cite{BergerOliger}, where the coarse grid
solution is interpolated in time and used to overpose the boundary conditions for the
fine scale imbedded solution.

Figure \ref{fig:liju}, borrowed from \cite{LiJu2} shows an example of current practice in
imposing arbitrary field ``macroscopically inspired" boundary conditions to microscopic
simulations.
Such methodologies have resulted from extensive research in the coupling of
continuum with microscopic simulations, leading to current hybrid simulations
practice \cite{OConnel,HadjPat,LiJu2,Hadj,Garcia,Weinan1}. 
The so-called ``extended boundary conditions" (EBS) are applied in
buffer regions, which are not real, physical domains, and which
surround the actual computational ``patch".
It is the ``core region" 
over which we will eventually restrict in order to obtain coarse field information.
By acting away from the region of interest,
across a buffer zone where the microscopic dynamics are allowed to relax, these conditions
attempt to minimize disturbances to the microscopic dynamics in the core region.
Knowing that the microscopic distribution functions will relax within a few ``collision
lengths" dictates the size of the buffer regions.
Various methods have been proposed to impose the actual boundary conditions in the buffer
regions.
These approaches include using distribution functions \cite{Hadj,Garcia} when these are
known, constrained dynamics \cite{OConnel}, and feedback control (the Optimal
Particle Controller of \cite{LiJu2}).
Imposing a known distribution will, of course, be accurate; we hope that the theoretical
underpinnings of other approaches will gradually become more firm.
The ``dual volume grand canonical" ensembles also use buffer zones to impose
``chemical potential Dirichlet" boundary conditions.

The ingredients of the EBS implementation for continuum/MD fluid
simulations involve (a) A field estimator, based
on maximum likelihood inference; this corresponds to our restriction operator,
obtaining moment fields from particle realizations.
For Monte Carlo simulations we have also been using the cumulative probability
density function, projected on
the first few of a hierarchy of appropriate basis functions as the field estimator.
(b) Three distinct regions only the first of which is physical:
(the core region of interest, the action region of the OPC,
and a buffer zone between the two); a particular illustration is also shown in
the Figure); and
(c) A method for enforcing the macroscopic BC in the action region.

There is an overhead involved with imposing ``nontrivial" (e.g.
slope) boundary conditions to microscopic simulations, certainly in the MD case.
The conceptual steps will be the same for any microscopic simulator; the implementation
may be easier or more difficult depending on the nature of the simulator and on
the problem under study.
It might make sense to trade ``worse" boundary conditions for larger buffer zones.
In the case, for example, of the diffusion problem, instead of lifting to particles
over a small box, and imposing slope boundary conditions, we may lift consistently
with the coarse profile over a much larger box, but make no effort to impose
particular boundary conditions at the outer boundary of the buffer region (i.e.
let the particles ``spill out").
The inaccurate boundary conditions will slowly degrade the accuracy of the
quantities of interest in the patch; the reporting horizon is chosen so that the
computation in the core region remains accurate.
Imposing ``better" boundary conditions will permit smaller buffer region and/or
longer reporting horizons.

The effect of the boundary conditions on the quality of the computation
also depends on how often they are updated (compare, for example, the long
term attractors in Fig.\ref{fig:attrlbbox_sh} and Fig.\ref{fig:attrlbbox_lo}.
This can be analyzed in the
ideal case where an exact inner continuum solver is available (for example,
the ``inner exact" diffusion problem mentioned earlier).
Keeping the slope boundary conditions fixed for the exact solver, will eventually
result in an infinite (or zero) average in the box;
it is vital that the ``exact" simulation in each box is stopped periodically, a
restriction followed by interpolation to a new
coarse field made, and a new round of ``inner" simulations started.
The length of time over which the slope boundary conditions can be used in each
box without reinterpolation (without letting the boxes ``talk" to each other)
is clearly bounded by the accuracy and stability of the
``outer scheme"; for this problem gaptooth reduces to the FD Equ.(\ref{expl})
with well known stability and accuracy criteria.
For this example, there exists a direct connection between updating the BC in
the gaptooth scheme and the stability of an associated ``standard" finite
difference scheme.

When microscopic ensembles imposing macroscopically-motivated
boundary conditions are used, the quality of the computation will
be affected by the same factors as for an ``inner continuum" gaptooth
scheme, and, in addition, by how the particular macroscopic BC are
imposed on the microscale (the particular EBS ``ensemble").

In summary the quality of the overall patch dynamics scheme depends on the
interplay of these factors: (a) the choice of macroscopic boundary
conditions; (b) the size of the buffer region at the outer boundary of
which they will be imposed; (c) how often the BC are updated from the
macroscale (how patches ``talk to" each other) and (d) the details of how they are
imposed on the microscale.
The first three factors would also arise in analyzing the gaptooth scheme for
a continuum problem.
It is conceptually possible (by analogy to implicit projective integrators)
that implicit ``patch schemes" can also be constructed; the interplay between
``overall patch size" and ``time reporting horizon" will be different for
these.

For the diffusion equation, the slope boundary conditions are natural
(they govern the flux of physical material in and out of the patch).
In general, however, we can only guess at the number and type of boundary
conditions we should use, since we do not know what the coarse equations actually
are.
What we {\it do} know (or, at least, postulate)
is what the coarse variables are (at what level of
coarse description a deterministic equation exists and closes).
New challenges therefore arise for identification: from just the nature of
the coarse variables and the microscopic simulator, we must extract
enough information about the unavailable equation to determine the appropriate boundary
conditions.
This is an ongoing research subject, which also has a counterpart in the ``legacy
code" context: if we are given a ``black box" PDE solver, which we can initialize
at will, how can we use the solver to find out
how many, and what, boundary conditions to impose in a ``patch dynamics" context ?

A first step towards answering this question is made by observing that
appropriate BC for macroscopic evolution problems are often closely related to
the highest spatial derivative in the model equations.
To estimate the highest order spatial derivative in the right hand side
$P(\partial_x, x, u)$ for evolution PDEs of the form  $u_t=P(\partial_x, x, u)$
we developed the ``baby-bathwater" algorithm (BBA) \cite{LiJuYannis}.
In the BBA we evolve ensembles of initial conditions in a periodic domain
using a ``black box" simulation code (this can be the micrscopic solver,
or even a ``legacy timestepper" for an unknown PDE).
We construct ensembles of (spatially periodic)  initial conditions which at some point in space $x_0$,
share the same value; or share the same value and the same first spatial derivative value; the same
value and the same first and second spatial derivative values; and so on.
Let $\{u^M_i(0,x)\}_{i=1}^n$ denote an ensemble of $n$ initial conditions
such that $\partial_x^m u^M_i(0,x_0)=\partial_x^m u^M_j(0,x_0)$
for $i,j=1,\ldots, n$
and $m=0,\ldots, M$.
We note that if $P(\partial_x,x,u)$ only
depends on derivatives of order equal to or less than $M$, then so does
$\partial_t u^M_i(0,x_0)=\partial_t u^M_j(0,x_0)$ for $i,j=1,\ldots, n$.
Therefore, tracking the results of a short time simulation at the point $x_0$,
i.e. studying $\{u^M_i(t,x_0)\}_{i=1}^n$ with $t\ll 1$,
shows at what level we have thrown the ``baby"
(the highest {\it significant} spatial derivative) along with the ``bathwater" (the higher
order, non-significant ones).
We have already tested this algorithm
both at the ``legacy PDE" level and at the level of ``coarse" studies of
microscopic simulators \cite{LiJuYannis}.

If the microscale model is in divergence form, then the natural  boundary conditions would be in
terms of fluxes.  For example,  the boundary conditions for the heat equation ($u_t =
(Ku_x)_x$) would involve prescribing the flux ($Ku_x$) at the boundary. 
Similarly, for the KdV equation  ($u_t = (u^2 + u_{xx})_x$)
boundary conditions for the flux ($u^2+u_{xx}$) result in a well posed problem. 
Another consideration for these conservation laws is to retain global conservation of
the relevant quantities, including the material not being simulated between the patches.
Not keeping track of this material between the patches will result in only approximately
conservative algorithms. 
For the full calculation to  preserve global conservation, the material in the gaps
between the  patches must be accounted for. 
This can easily be done in one dimension by calculating auxiliary conserved variables
in the gaps between the patches; these
conserved variables in each gap are updated on every macroscopic timestep by continuously
accounting for the fluxes in and out of its neighboring patches.
These auxiliary variables play a key role in constructing the macroscale 
interpolant defining the patch boundary conditions. 
A conservative interpolant is  constructed to agree with the average
values of the material in each patch {\it and} the integral of the interpolant in the
gaps between the patches.
Conservative patch dynamics algorithms still need to be developed in higher
dimensions.
A formulation for conservation problems based on the estimation of fluxes, 
leading to a generalized Godunov scheme, is proposed in \cite{Weinan2}.

\section{SOME APPLICATIONS}

To demonstrate the scope of the techniques we have discussed, we
present here a brief anthology of results that we have obtained
doing ``coarse" numerics on problems with various ``inner timesteppers".
This collage, and
references to the corresponding publications, were selected to 
emphasize the broad scope of the problems that can be 
solved through the EFM computational approach.

Figure \ref{fig:bubbles} arises in the modeling of multiphase flows through LB-BGK
simulations. It is a study of the dynamics and instabilities of a two-dimensional
periodic array of gas bubbles rising, under the action of gravity, in a liquid.
Time-dependent simulations both before and after this ``coarse Hopf"
instability, leading to macroscopic bubble shape oscillations, are shown,
and the coarse bifurcation diagram shows both stable {\it and unstable} coarse
steady bubble shapes. 
This is collaborative research with Dr. Sankaranarayanan
and Prof. Sundaresan at Princeton \cite{Bubbles,BubblesPRL,Bubbles3,Bubbles4}

Fig.\ref{fig:MC} shows bifurcation diagrams for lattice gas models of surface
catalytic reactions (a caricature of CO oxidation) in the presence of adsorbate
interactions. Here the ``inner integrator" is a kinetic Monte Carlo code. One
deals with average surface concentrations (that is, coarse ODEs for expected coverages,
as opposed to coarse PDEs in the examples above). It is interesting to compare
the coarse bifurcation analysis obtained through our scheme with the
results of ``traditional" explicit
closures included in the graph (mean field as well as quasi-chemical approximations)
and with true KMC simulations (also included).
Clearly, at these strong value of adsorbate interactions, microsctructure has
started to develop on the surface; yet it is still possible (for a long enough
reporting horizon) to analyze unavailable closed equations in terms of only
the zeroth moments of the distribution (two coverages). This work is in
collaboration with Dr. A. Makeev of Moscow State University, and Profs. D. Maroudas
at UCSB and A. Panagiotopoulos at Princeton.

Figure \ref{fig:Graham} shows the bifurcation diagram for an order parameter arising
in the dynamics of nematic liquid crystals (more exactly, the Doi theory for
rigid rods in shear). Here the inner time-stepper is Brownian Dynamics,
the initial conditions are rod orientation distributions (over the unit sphere)
evolved through a
stochastic integrator,  and the bifurcation parameter corresponds to the density of
the rods. The Smoluchowski/Fokker-Planck equation for the evolution 
of the distribution has
been analyzed through both spherical harmonic and wavelet methods 
(\cite{Russo,Armstrong}).
This coarse bifurcation diagram (and the coarse eigenvalues computed in the
process) confirms what is known from the numerically integrated Fokker-Planck itself. 
We have used
this example to study the effect of closure conditioning on several moments,
and have confirmed that the ``next fastest" governing moment is approximately 500 times
faster than the order parameter $S$ \cite{Graham}.
This justifies one-mode closures in the literature, which
can be rationalized as a one-dimensional approximate inertial manifold for the
Smoluchowski equation. 
This work is in collaboration with Dr. C. Siettos and Prof. M. D.
Graham at Wisconsin (and aspect of it are also studied with Prof. R. Armstrong
and Dr. A. Gopinath at MIT).

Figure \ref{fig:Olof} shows a ``different" (non-microscopic) application of the EF
methodology: a computation of ``effective medium"
behavior using coarse timesteppers.
The example is taken from our recent paper \cite{Olof}; it is a study
of the dynamics of pulses in a reaction-diffusion problem, traveling
in a one-dimensional periodic medium (consisting of alternating ``stripes" of
different reactivity).
Assuming that an effective equation for a complicated medium exists, but is not
available in closed form, we can construct an approximate timestepper for it by
giving a coarse initial condition to an ensemble of realizations of the detailed medium,
solving (for short times) the detailed equations, and averaging the result.
In general, for a random medium, one would have to construct the ensemble of
medium realizations through Monte-Carlo sampling; but in this example the medium
is periodic, and so the different realizations simply correspond to shifts of the
initial condition over the unit cell. 
The ``effective equation" version of the
coarse time-stepper is shown above in caricature, and the ``effective" bifurcation
diagram, showing a ``coarse Hopf" bifurcation of a ``coarsely traveling" wave
to a ``modulated coarsely traveling" wave is shown below.
We believe that, at the appropriate limit, the time-stepper constructed through this
procedure will approach the time-stepper arising from the discretization of the
corresponding homogenized equations.

Figure \ref{fig:control} is a sample of an ``alternative" computational task - neither
integration nor steady/bifurcation computation. 
It shows the coarse control of the
expected value of a kinetic Monte Carlo simulation for a caricature of a catalytic
surface reaction \cite{Alexei1,Siettos_CC}.
Here the unstable coarse (expected) steady states, their linearization and
bifurcations have been computed off-line using the coarse time-stepper.
The identified linearization and
coarse derivatives with respect to control parameters around an unstable
steady state were used to design both an
observer (a Kalman filter) and a (pole placement) controller for the system.
The bifurcation parameter (corresponding to the gas phase
pressure of a reactant) is used as the actuator in a regime where the coarse
dynamics are oscillatory.
The ``coarse linear controller" is clearly capable of
stabilizing the (unstable) coarse expected stationary state of the KMC simulation. 
This work is in
collaboration with Dr. C. Siettos, A. Armaou and Prof. A. Makeev;
work on the coarse control of distributed microscopic systems is also
underway.

We are currently exploring, through collaborations, several problems
involving several types of ``inner timesteppers": with Prof. Katsoulakis
we study coarse optimization and the computation of
``coarse rare" events using timesteppers. 
With Prof. Bourlioux 
we explore turbulent combustion using
coarse front evolution timesteppers. 
With Prof. Tannenbaum we study ``coarse" computationns of
microscopic edge detection algorithms;
With Prof. Hadjiconstantinou and Drs. Alexander and Garcia
we study coarse DSMC fluid simulation.
Finite temperature
MD stress induced crystal transformations is studied 
with Profs. Maroudas and Li Ju, and pattern formation in
granular flows is studied with Professors Swinney and Swift.

The last application we include is very simple, but has, we believe,
important implications. Fig.\ref{fig:selfsim} shows the computation of a {\it coarse
self similar solution} -- the decaying self-similar solution of the diffusion
equation--, computed through a microscopic timestepper (a Monte Carlo random
walk simulator). 
We have recently proposed the use of a template based
dynamic renormalization approach to compute self-similar solutions for
evolutionary PDEs \cite{PNAS2,ClancyYannis}.
This template-based approach was
motivated by the work of Rowley and
Marsden for problems with translational invariance 
and traveling solutions \cite{marsden00}.
The solution is rescaled either
continuously, or discretely (after some time) based on minimizing its distance
from a (more or less aritrary) template function (see also \cite{Goldenfeld}). 
Here we combine this idea
with the ``lift-run-restrict" approach as follows:  we start with a random walker
density profile (a coarse initial condition); 
for convenience, the coarse variable will be 
the cumulative distribution function (percent total walkers to the left of a
given location in space).
We then ``lift" it to random walker location, and let the
random walkers evolve for some time. 
We interpolate to cumulative distribution
function (using appropriate orthogonal polynomials) and rescale the solution
based on template minimization.
The procedure is then repeated giving a
discrete-time renormalization evolution, which converges quickly (the solution
is stable).
Successive coarse portraits are used to extract the 
relevant self-similarity exponents.
Finally, we find the same solution not through successsive
coarse timestepping, but through a contraction mapping:
a secant-based quasi-Newton method, for the profile that is invariant under
finite time evolution and template-based rescaling - see the bottom panel
in the Figure.
This work \cite{CSS} is in collaboration 
with Prof. Aronson and Dr. Betelu at Minnesota,
and we hope that it will allow the ``coarse" study of many phenomena that
involve self-similarity (``from cells to stars", \cite{Brenner})
when microscopic physics descriptions exist but the macroscopic equations
are not available.
It is conceivable that even asymptotically self-similar
solutions might be treatable in some cases. 
We have recently located an
interesting bifurcation (the onset of dynamic self-similarity from steady
state solutions) in the context of the focusing Nonlinear Schroedinger (NLS)
equation \cite{NLS}. 
It is conceivable that similar bifurcations, involving the onset
of dynamic coarse self-similarity from coarse steady states (e.g. in molecular
dynamics equilibrium simulations) might share some features with this instability
and offer a useful computational perspective of the transitions involved.

\section {CONCLUSIONS, DISCUSSION, OUTLOOK}

We have described an equation-free  framework for the computer-aided analysis of
multiscale problems, in which the physics are known at a microscopic level,
but where the questions asked about the expected behavior are at a much higher,
macroscopic, coarse level.
This is a ``closure on demand", system identification based framework, which
provides a bridge between microscopic/stochastic simulators and macroscopic 
scientific computation and numerical analysis.
The basic idea is to use short, intelligent ``bursts" of appropriately initialized
microscopic simulations (the ``inner" code), and process
their results through system identification techniques to 
estimate quantities (residuals, time derivatives,
right-hand-sides, matrix-vector products, actions of coarse slow Jacobians 
and Hessians) that,
if a macroscopic model was available, would simply be evaluated from
closed formulas.
The quantities thus estimated are ``passed" to an ``outer level" code,
usually a traditional macroscopic computational code that performs
integration, controller design, steady state calculation or optimization tasks.
This outer code proceeds to perform its task ``without realizing" that the
model it is studying is not available in closed form at the macroscopic level.
We showed also that these short bursts of microscopic simulation may, under
some cicrumstances, be performed for small ``patches" of computational space;
these patches repeatedly communicate with each other, as the computation
evolves, through macroscopically inspired boundary conditions (``patch dynamics").

In our discussion, the main tool was the ``coarse timestepper", that evolves
fields of the macroscopic variables through microscopic simulation.
If the timesteppers are ``legacy simulators" (one of the original motivations of
time-stepper based enabling algorithms, like the adaptive condensation of Jarausch
and Mackens, the RPM of Shroff and Keller, but also the Arnoldi based stability
computations of Christodoulou and Scriven \cite{KostasCh})
we are in the realm of ``numerical
analysis of legacy codes".
EF-type algorithms can significantly 
accelerate the convergence of large scale industrial simulators to
fixed points (e.g. periodic steady states for pressure swing
adsorption or reverse flow reactor simulations) as well the stability
analysis of these fixed points (e.g. \cite{Siettos_PSA}).

This paper is focused on situations 
where the macroscopic model is not available in closed
form, and one needs to use microscopic simulation over the entire 
spatial domain, or over ``patches" distributed across the domain.
As we briefly discussed, however, it is also conceptually straightforward to
construct coarse timesteppers using existing {\it hybrid} simulation frameworks.
The map from current to future coarse fields is obtained through the combination
of the lift-evolve-restrict approach in the ``molecular regions" and continuum
solvers in the continuum regions.
If the ``molecular regions" are large, and the macroscopic expected
solution is smooth enough across these regions, one can conceivably
evolve the microscopic simulation only in a grid of patches along this region;
one then has a hybrid continuum-``patch" overall coarse timestepper.

On a different note, a connection between coarse timesteppers and what
is termed ``gray-box identification" is best discussed considering
an ``unavailable macroscopic equation" of the form of Equ.(\ref{tim}).
The coarse timestepper estimates the result of integrating the entire
right-hand-side of the equation for time $T$.
Frequently, we may know (e.g. from irreversible thermodynamics, or from
experimental experience) a lot about
the form of several terms in the RHS; and only miss a particular phenomenological
term, such as an effective viscosity, to have a useful closure.
In such cases, microscopic simulations are typically done to extract a ``law"
for this phenomenological coefficient.
For engineering purposes, discrete-time data obtained through the coarse
timestepper can be ``fed to" discrete time gray-box identification schemes, 
with built-in partial knowledge of the RHS of the macroscopic equation; only
the unknown parts of the postulated equation will then be identified (fitted).
A particular framework for doing this, based on recurrent artificial neural networks
templated on numerical integration schemes can be found in \cite{NN1,NN2}.
We are exploring the ability of this framework to practically close microsopic
rheological simulations in collaboration with Prof. Oettinger.

While we discussed mostly parabolic problems in presenting coarse integration
results, clearly coarse bifurcation algorithms like RPM provide the solutions
to elliptic problems, and -as we briefly discussed in the section above- hyperbolic
problems also fit naturally within this framework.

Most of the ``mathematics assisted" enhancement we expect from these algorithms
comes from the estimation of coarse derivatives (in time, in space, with respect
to parameters or with respect to the coarse variables themselves) through calls
to the microscopic timestepper.
Variance reduction through sheer number of copies, but also through ``the best"
estimation we can have, is thus vital in establishing computational savings
\cite{Spall}; its importance cannot be overestimated.

In most of our exposition we assumed that the coarse variables of
choice would be low order moments of the distributions evolved through the
microscopic timesteppers.
There will, however, exist problems where ``intelligently chosen" phase field variables
will provide a much ``leaner" set of equations than moments (in the same sense that good
basis functions can provide a much more parsimonious description of a PDE discretization).
We believe that the discovery (through statistical algorithms that search for nonlinear
correlations across data fields) of ``good coarse variables" from which to lift is
one of the requests this technology will have from modern computer science. 
The fact
that we need not come up with explicit formulas for the time
derivatives in terms of the variables generates 
a lot of  freedom in allowing the study of systems currently intractable
(because good closures could not be explicitly computed, and ``full" microscopic
simulations were too expensive to perform over relevant space-time scales).

A simple case in point is the attempt to create timesteppers for expected energy
spectra of DNS or LES simulations of the Navier Stokes equations by (a) conditioning
NS initial condition ensembles on a given spectrum; (b) evolving the ensemble
through DNS or LES for some time and (c) restricting back down to spectra.
Aspects of this problem are being explored 
in collaboration with Profs. Martin, Yakhot, Jolly and Foias.
The list of problems that can be attempted through this methodology (with no
guarantee of solution, of course; the premise of ``coarse behavior" has to be
satisfied, and it cannot be {\it a priori} established) is vast.
In the previous section we provided an anthology of systems we are working on
currently; we have promising initial results on Monte Carlo models of
bacterial chemotaxis (with Dr. Sima Setayeshgar and Prof. Hans Othmer)
and in the derivation of translationally invariant effective equations for
discrete systems (e.g. lattices of coupled neurons) in which we study the effect
of front pinning and an analogy to the Portevin-le Chatelier effect (with K. Lust,
J. Moeller and D. Srolovitz).
Agent based models in ecology, epidemiology and evolutionary biology (the timesteppers
do not have to be in physical time, they can involve mutations in a population)
are also a possibility, which we are exploring with Prof. Levin;
another possibility might be the coarse integration of 
unavailable equations for the expected backbone shapes of oligopeptides.
While most of the problems we have suggested so far come from nonequilibrium
statistical mechanics, equation-free dynamic approaches
may play a role in studying {\it equilibrium phenomena}
throgh dynamics (see recent work
in \cite{Escobedo,Kofke,Maginn1,Maginn2} as well as in \cite{Jarzynski}).
It is also conceivable that appropriate limits of quantum problems, 
such as the weak coupling limit
of the $N$-particle Schroedinger, which in the semi-classical limit
yields the 1-body Vlasov equation \cite{Bardos}, can
be studied through ``coarse timesteppers".
In fact the Vlasov equation (as well as the Fokker-Planck-Vlasov)
has analytically known equilibrium states; one could try to retrieve
them using time-steppers and compare to the explicit analytic results.
Another possibility is the semiclassical limit of a single Schroedinger equation,
which, using the Wigner transform, gives rise to a Vlasov equation
Further ad-hoc closures yield a system of conservation laws
with a dispersive correction \cite{JinLi}; coarse timesteppers
could sidestep (and/or validate) the derivation of such closures.

The availability of continuum equations underlies many of the methods
developed and the results obtained in the modeling of physical systems.
These continuum equations are, in many cases, only caricatures (sometimes
excellent caricatures) of ``better but dirtier" descriptions of the physics, 
such as those provided by microscopic/stochastic models. 
Equation-Free approaches hold the promise of applying our best 
computational/ mathematical tools (whose development was targeted at
continuum evolution equations) to the ``dirty" description directly, 
circumventing the necessity of passing through a caricature.
The transition between analytical ``paper and pencil" solution methods 
(based on perturbation theory
and special functions) to computer-aided analysis (based on large scale
linear algebra and PDE discretization techniques) discussed in \cite{Brown,Scriven}
serves, in some sense, as a prototype for a new transition:
from equation-based computer-aided analysis to equation-free computer aided analysis.
In the first transition, quantities that would be computed by paper and
pencil if the {\it solution} was explicitly available (e.g. eigenvalues
of a linearization) were computed numerically.
In the second transition, quantities that would be computed numerically if
the {\it equation} was explicitly available (e.g., again, eigenvalues of
a {\it coarse} linearization) are estimated numerically through 
appropriate calls to microscopic/stochastic simulators.

Beyond the type of systems that one can attempt to approach through this
computational enabling technology, the type of tasks that are affected go
beyond coarse simulation, steady state calculation and bifurcation analysis.
We illustrated ``coarse control" in the previous section, and timesteppers
also naturally fit in a ``coarse optimization" framework, where optimal control
problems (such as those arising in rare events) can be solved without ever
having to derive explicit models in terms of ``coarse coordinates" that
parametrize the transitions in question. Just guessing what a set of sufficient
such coordinates might be, would be enough.
The ability to find coarse self-similar solutions by combining
coarse timesteppers with renormalization flow approaches for PDEs
opens up the playing field even more.
It would be interesting to see, as experience is gathered and
developments are made, how much of the conceptual promise of these
techniques becomes reality.

{\bf Acknowledgements}.  The work reported in this paper spans several
years and involves interactions with several people, whom we would like
to acknowledge. Original discussions with Herb Keller about RPM (at the
IMA in Minneapolis) and with Jim Evans at Iowa State about his hybrid
MC simulators of surface reactions played an important role. Discussions
over the years with Profs. J. Keller, K. Lust, D. Maroudas, D. McLaughlin,
A. Majda, L. Ju, S. Yip, R. Armstrong, A. Panagiotopoulos, 
M. Graham, R. Kapral, H.-C. Oettinger, V. Yakhot, C. Foias, E. Titi, 
P. Constantin, M. Aizenman, J. Burns, F. Alexander, N. Hadjiconstantinou, A. Makeev,
C. Siettos, C. Pantelides, C. Jacobsen, and A. Armaou have affected this work. 
In the course of the last two years we have discussed this research
direction with our colleagues, Professors Engquist and E at PACM in Princeton.
In the companion paper \cite{Weinan2} they present an alternative formulation which, 
in the time-dependent case, corresponds to a generalized Godunov scheme.
The role of AFOSR (Dynamics and Control, Drs. Jacobs and King),
to whom this work was proposed in 1999, and which they supported
through the years, has been crucial. Partial support by NSF through
a KDI and an ITR grant, by DARPA, by a
Humboldt Forschungspreis to I.G.K, and Los Alamos National Lab
(through LDRD-DR-2001501) are also gratefully acknowledged.

\newpage

\begin{figure}
\centerline{\psfig{file=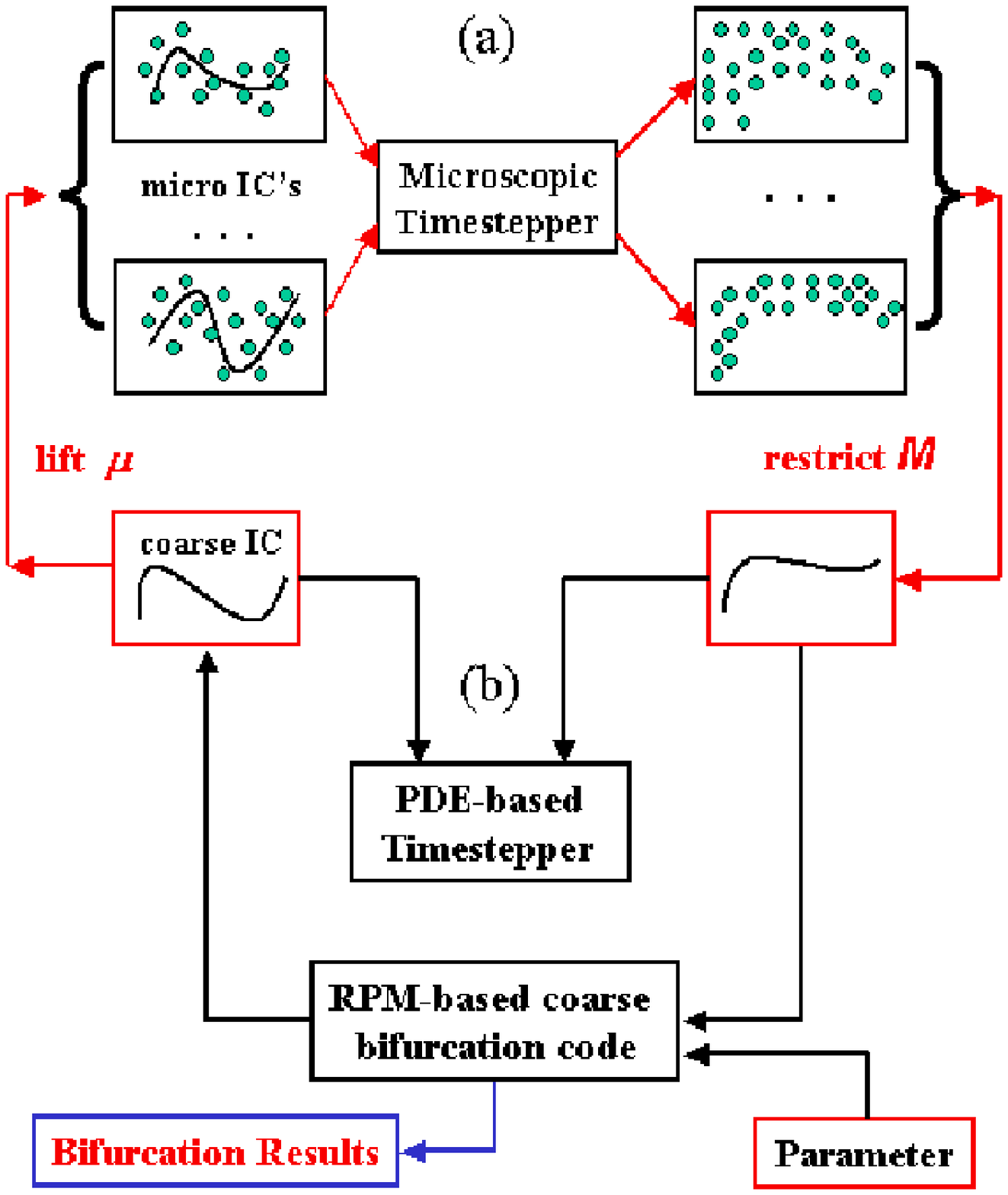,width=12.5cm}}
\vspace{0.1in}
\caption{Schematic description of time-stepper based bifurcation analysis,
fine and coarse. Notice the lifting of a macroscopic initial condition to
an ensemble of consistent microscopic ones, as well as the restriction of
the microscopic integration results back to the macroscopic (usually
moments-based) description.}
\label{fig:RPM}
\end{figure}

\begin{figure}
\centerline{\psfig{file=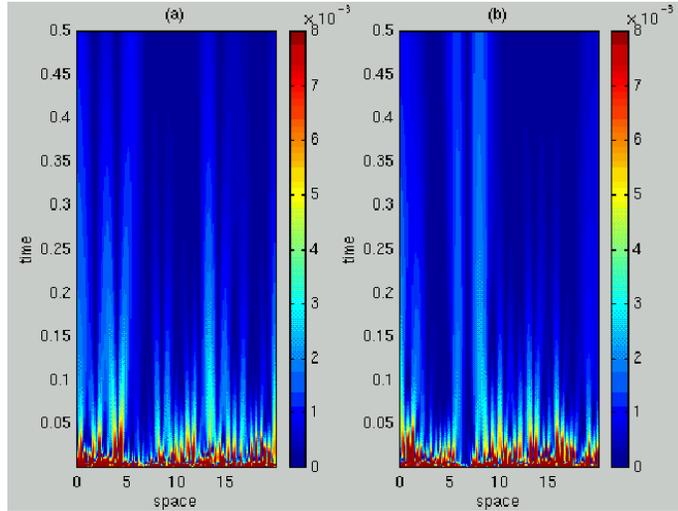,width=9cm}}
\vspace{0.1in}
\caption{Decay (``healing") of the lifting error: comparison between the
evolving restrictions of a ``mature" LB-BGK trajectory, and two distinct liftings
of its restriction: (a) a local equilibrium fine lifting and (b) a
more ``coarse" random lifting. Healing has visibly taken place
at times much shorter than the timestepper reporting horizon (here 0.5).}
\label{fig:heal}
\end{figure}

\begin{figure}
\centerline{\psfig{file=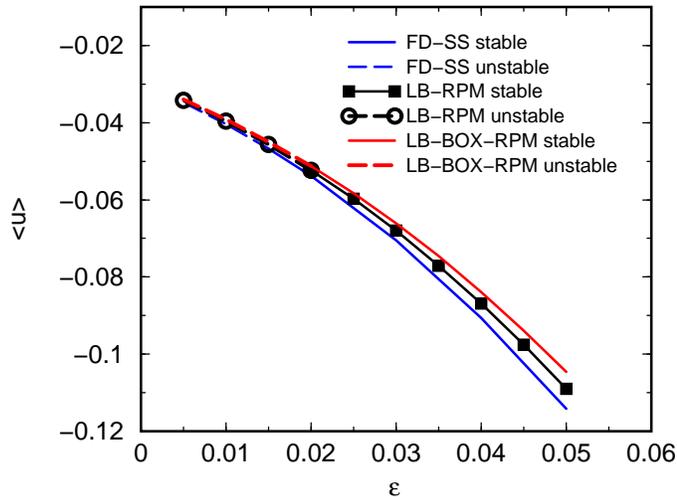,width=9cm}}
\vspace{0.1in}
\caption{FHN bifurcation diagram near a Hopf bifurcation point computed with a steady Finite Difference
(FD-SS) code (blue line), our LB-RPM code (black line) and our LB-BOX gaptooth scheme combined
with RPM, called LB-BOX-RPM (red line). Stable (unstable) steady states:
solid (broken) lines.}
\label{fig:bifhop}
\end{figure}

\begin{figure}
\centerline{\psfig{file=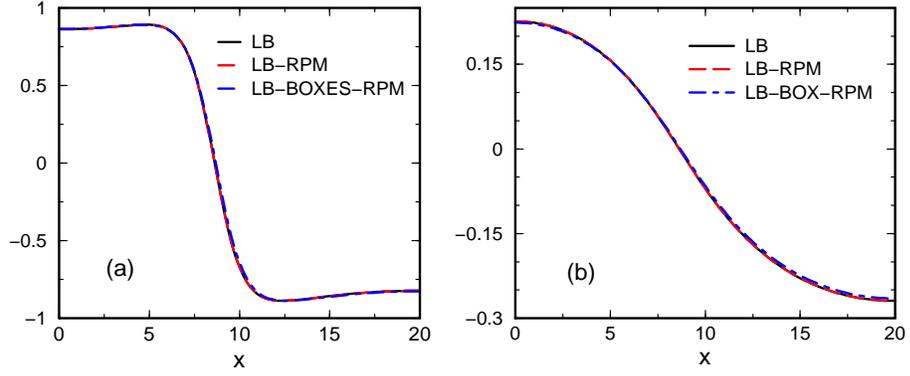,width=12cm}}
\vspace{0.1in}
\caption{Comparison of the coarse steady state spatial profiles of (a) $u$ and (b) $v$ at
$\epsilon=0.05$ (a stable steady state) computed with direct LB simulations
(black solid lines,
LB combined with RPM (LB-RPM) simulations (red broken lines) and
our LB-BOX gaptooth scheme combined with RPM (LB-BOX-RPM -blue broken lines).}
\label{fig:uvss}
\end{figure}

\begin{figure}
\centerline{\psfig{file=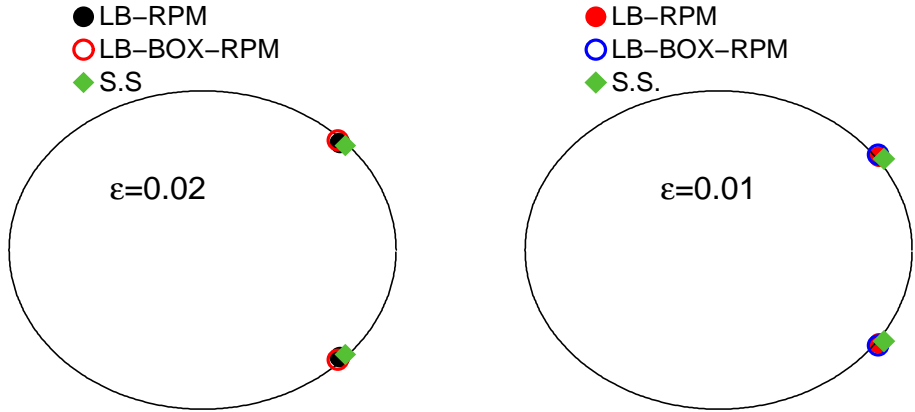,width=12cm}}
\vspace{0.1in}
\caption{Leading coarse eigenvalues before ($\epsilon=0.02$) and after ($\epsilon=0.01$) the
Hopf point, obtained upon convergence of LB-RPM (filled black circles) and LB-BOX-RPM (open red circles).
The exponentials of the eigenvalues (for the same reporting horizon)
of a steady FD code (based on the PDE) are also depicted for comparison 
(green diamonds, SS).}
\label{fig:eigval}
\end{figure}

\begin{figure}
\centerline{\psfig{file=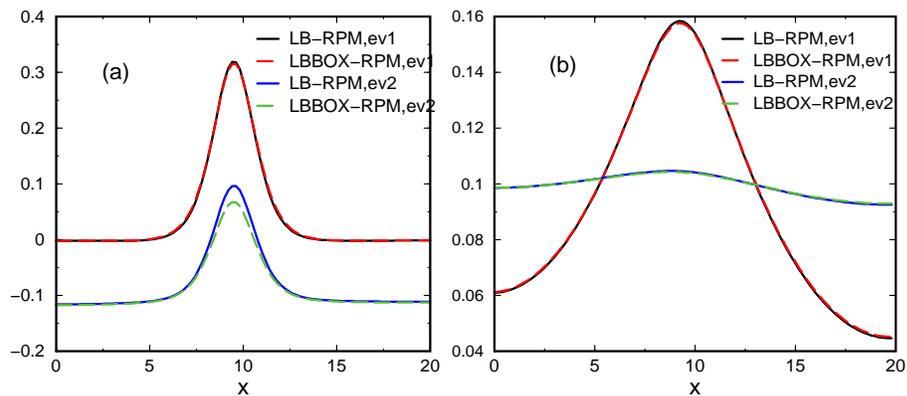,width=12cm}}
\vspace{0.1in}
\caption{Comparison of (a) $u$ and (b) $v$ components of the
critical coarse eigenvectors (ev1 and ev2)
at $\epsilon=0.01$ obtained upon convergence of the LB-RPM and LB-BOX-RPM computations.}
\label{fig:eigvec}
\end{figure}

\begin{figure}
\centerline{\psfig{file=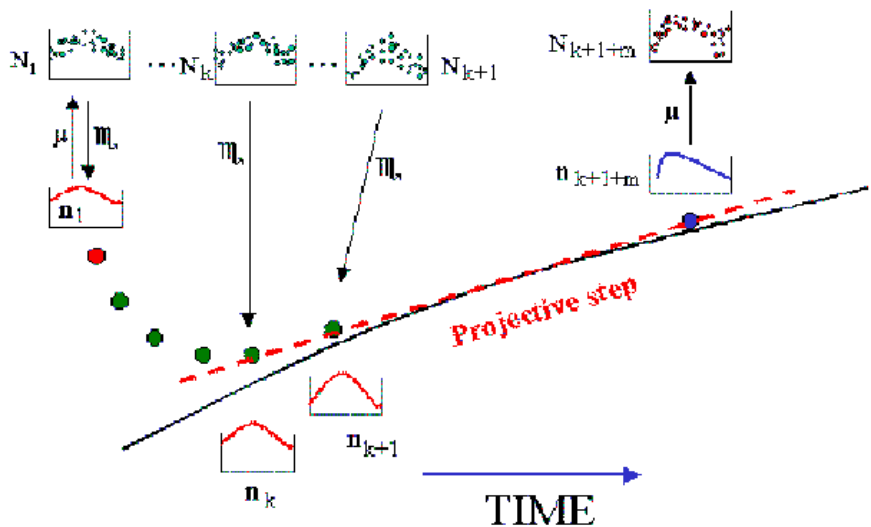,width=12cm,clip=t}}
\vspace{0.1in}
\caption{Schematic representation of an explicit projective integrator. Notice
the lifting, the evolution with successive restrictions, the estimation of
the ``coarse time derivative", the projection step, followed be a new lifting.}
\label{fig:project}
\end{figure}

\begin{figure}
\centerline{\psfig{file=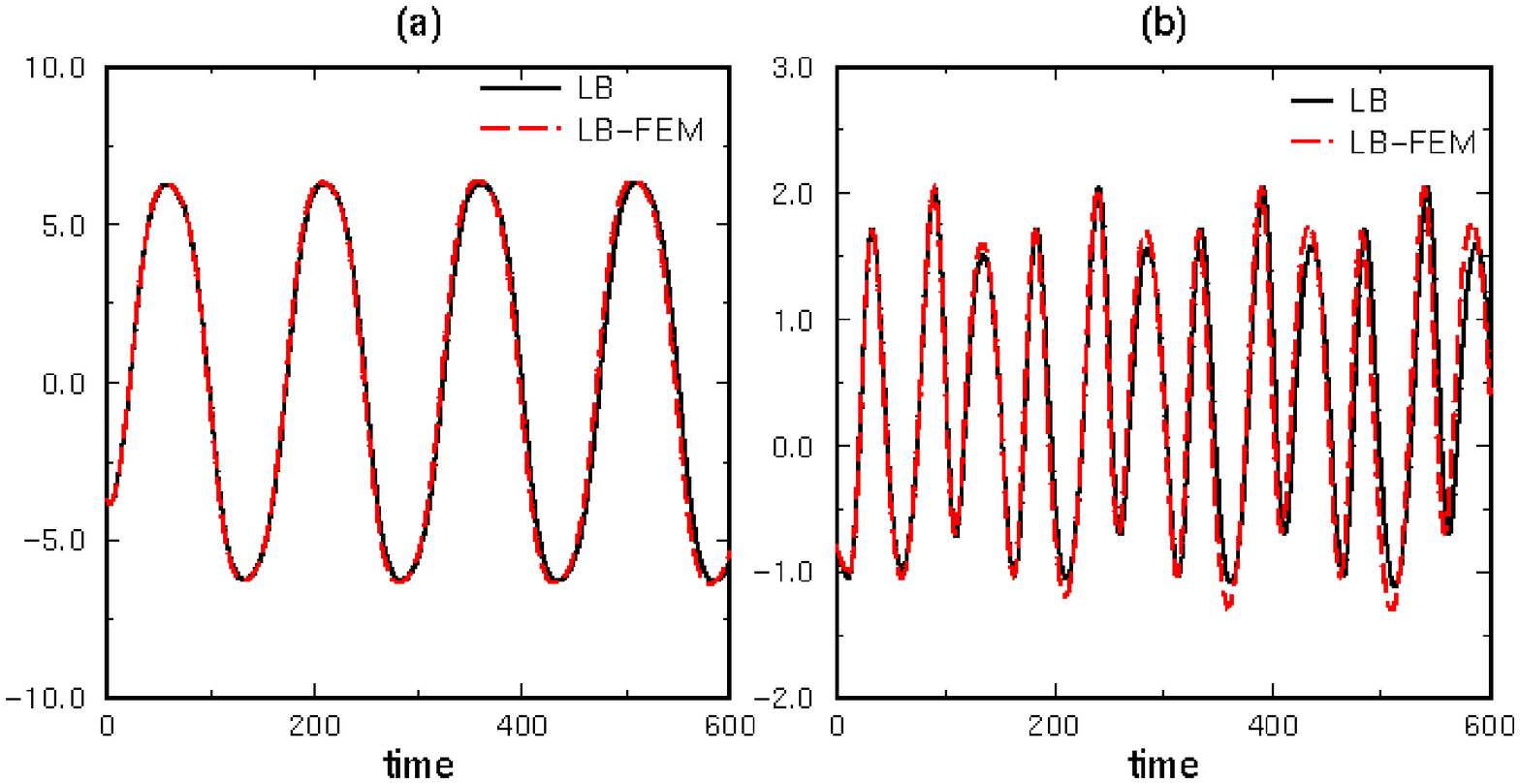,width=10.9cm}}
\centerline{\psfig{file=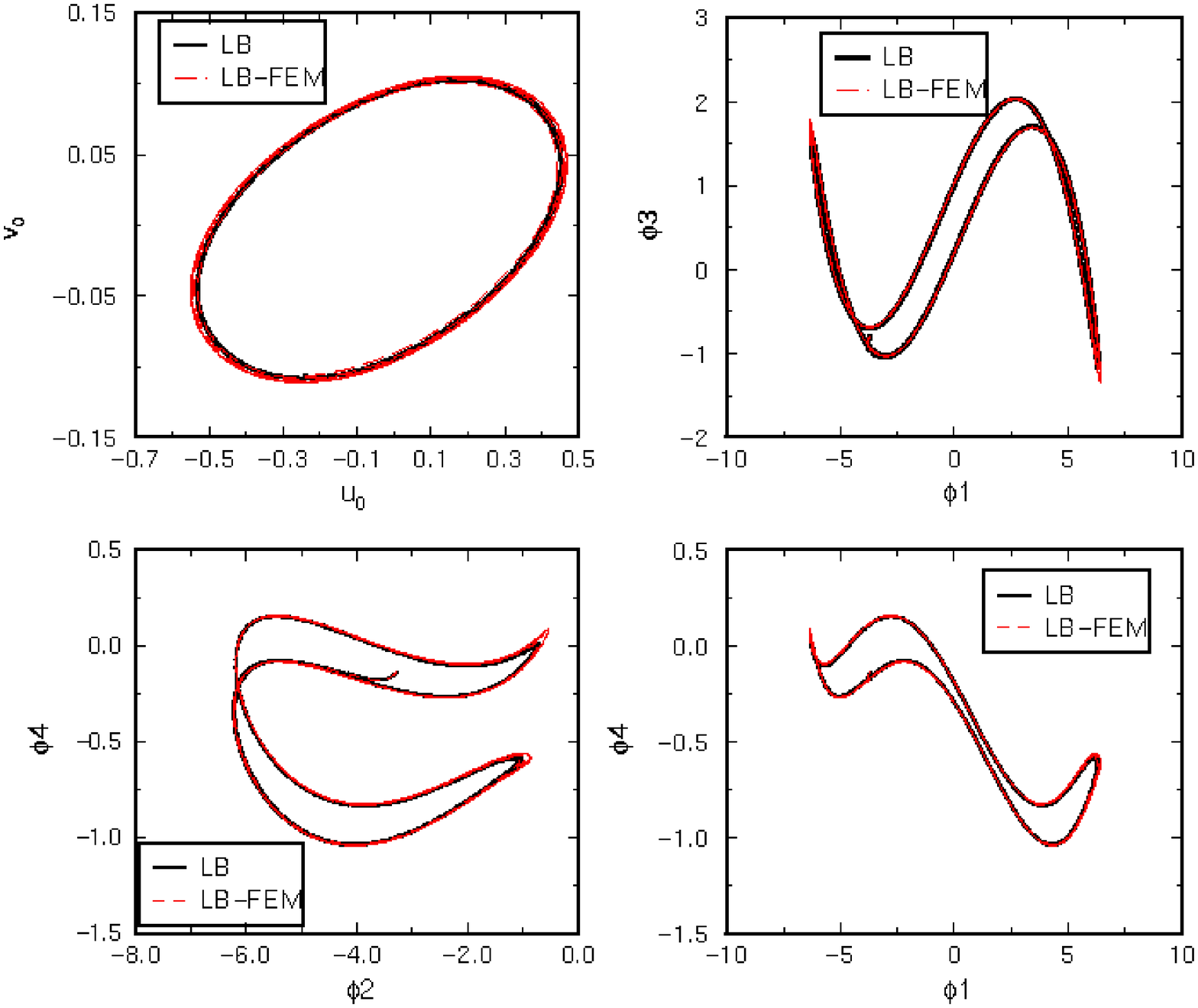,width=10.9cm}}
\vspace{0.1in}
\caption{A comparison of FHN dynamics computed through (1)
an LB-BGK code and  (2) a 100-50-350 inner-LB - outer
FEM (explicit Euler) projective integrator (from \cite{GKT}).
The comparisons are performed by projection
on subspaces spanned by a few empirical orthogonal eigenfunctions
(EOFs or PODs, see text).
Top two panels: time series for (1, black) and (2, red) starting
from the same initial conditions.
Bottom four panels: various phase space projections (in POD-mode space)
of the attractors of the LB-BGK (1, black)
and the 100-50-350 LB-FEM projective scheme (2,red).}
\label{fig:lbfem}
\end{figure}

\begin{figure}
\centerline{\psfig{file=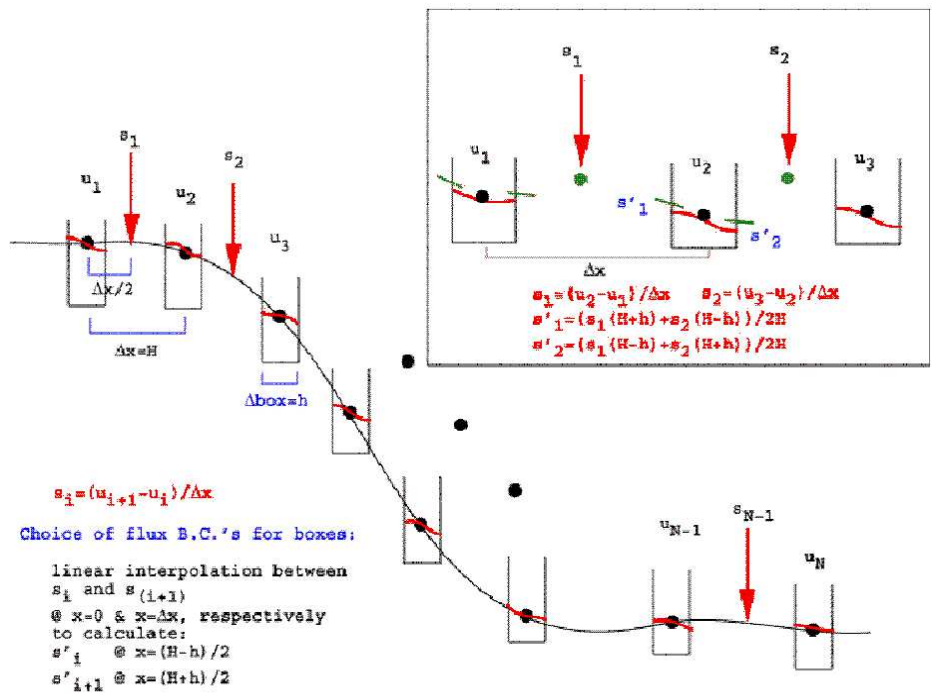,width=12.5cm}}
\vspace{0.1in}
\caption{An original schematic description of the gaptooth scheme (from
the 2000 AIChE Area 10 plenary lecture, \cite{AIChE2000}); the notation is slightly different
from the one in the paper. Notice the macro-mesh and the smoothly interpolated
macro solution, the ``teeth" surrounding the macro mesh, and the interpolated
boundary conditions at the outer edges of the ``teeth" or ``patches". For this
continuum description the buffer zones are not depicted.}
\label{fig:oldAIChE_schem}
\end{figure}

\begin{figure}
\centerline{\psfig{file=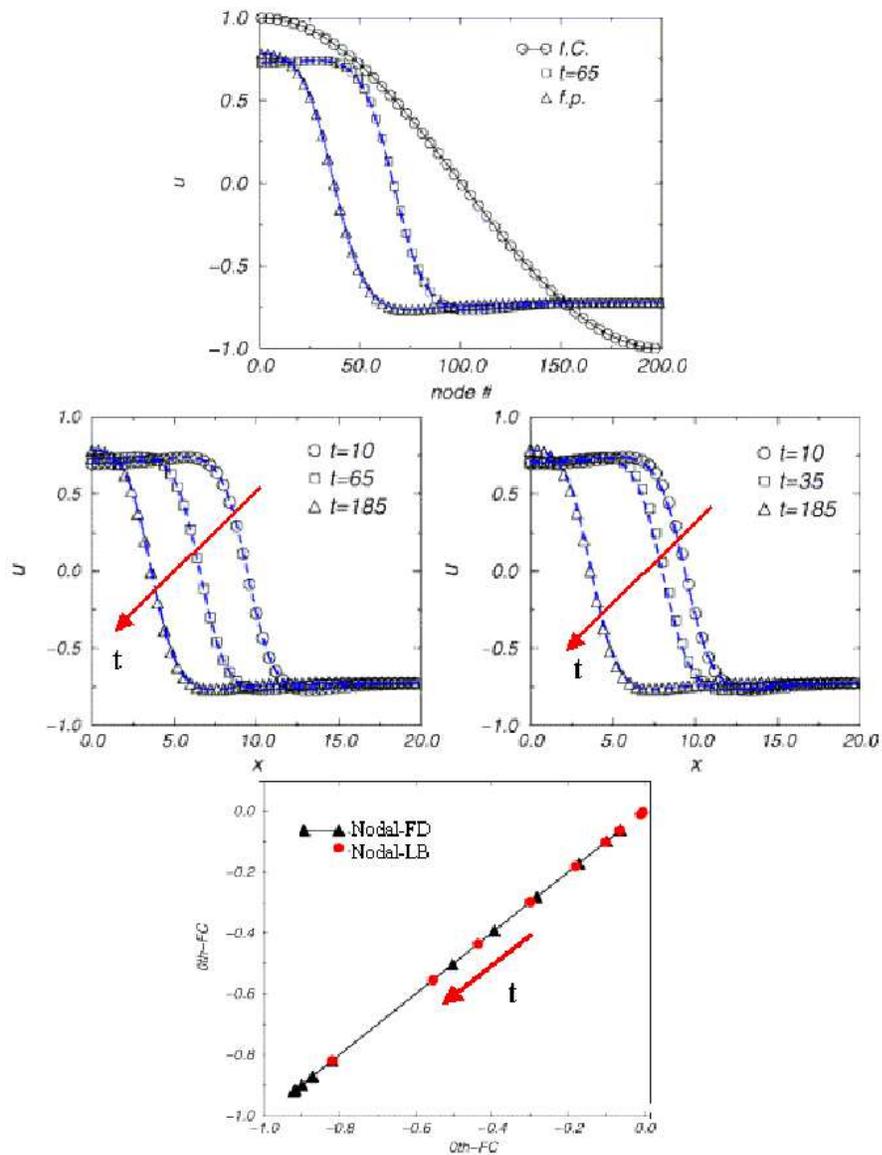,width=12cm}}
\vspace{0.1in}
\caption{``Inner LB" gaptooth integration results for the FHN equation (symbols)
compared to full LB simulations (curves) for various initial conditions. The last
figure shows a phase portrait for this simulation for two different gaptooth schemes:
an ``inner LB" (FD-LB) and an ``inner FD" (FD-FD) continuum scheme.}
\label{fig:oldAIChE_res}
\end{figure}

\begin{figure}
\centerline{\psfig{file=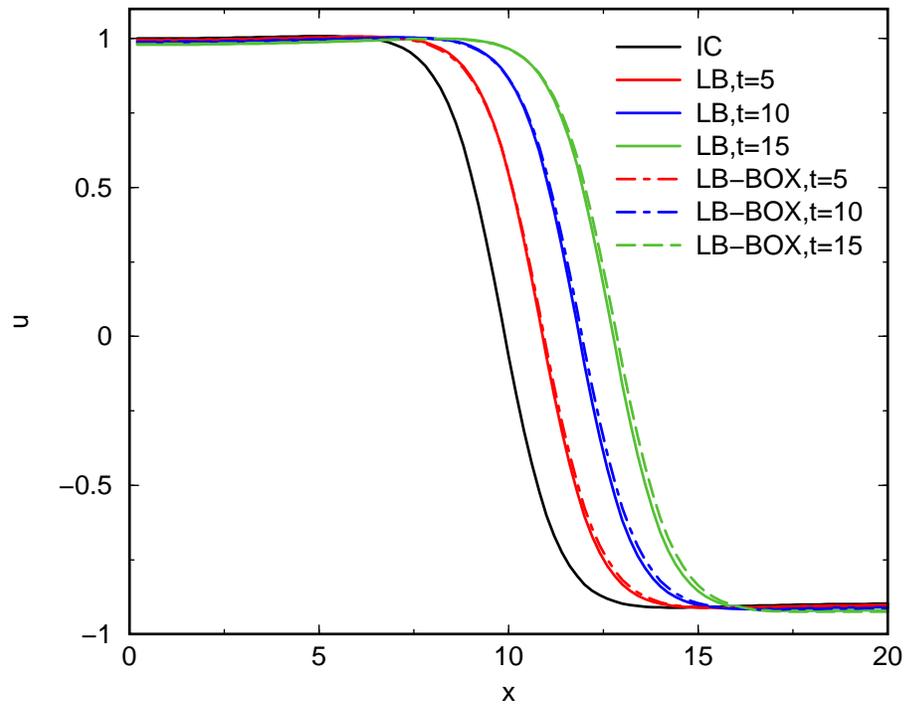,width=12cm}}
\vspace{0.1in}
\caption{Comparisons of LB and LB-BOX spatial profiles of $u$
at various times in the unstable (oscillatory) FHN region ($\epsilon=0.01$)
showing a palindromic movement of the reaction front.}
\label{fig:profbox1}
\end{figure}

\begin{figure}
\centerline{\psfig{file=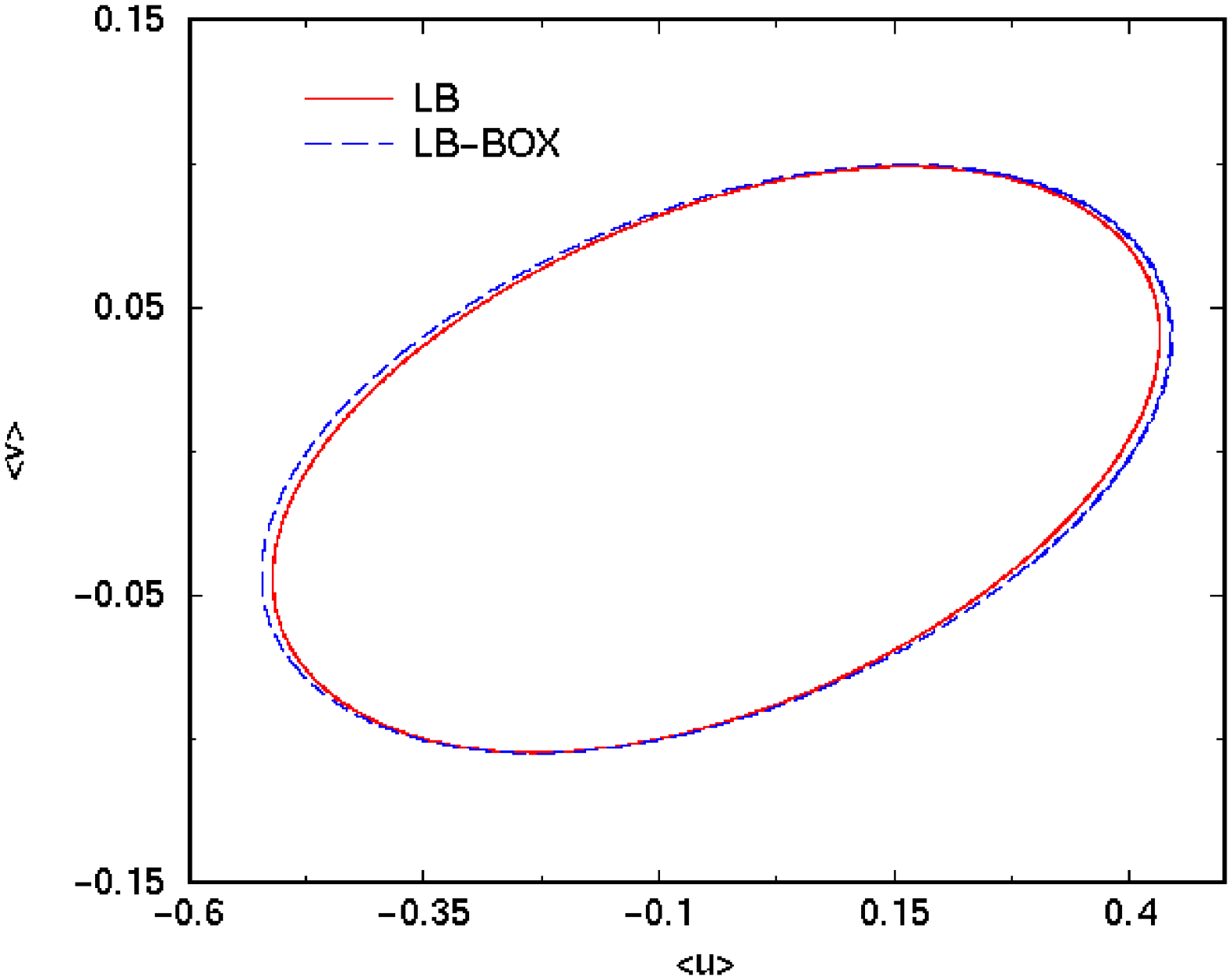,width=9.8cm}}
\vspace{0.1in}
\caption{Comparison of phase space projections of the LB (black)
and LB-BOX (red) long-term attractors. Reporting horizon for communication between
boxes $T_c=7.5 \times 10^{-4}$.}
\label{fig:attrlbbox_sh}
\end{figure}

\begin{figure}
\centerline{\psfig{file=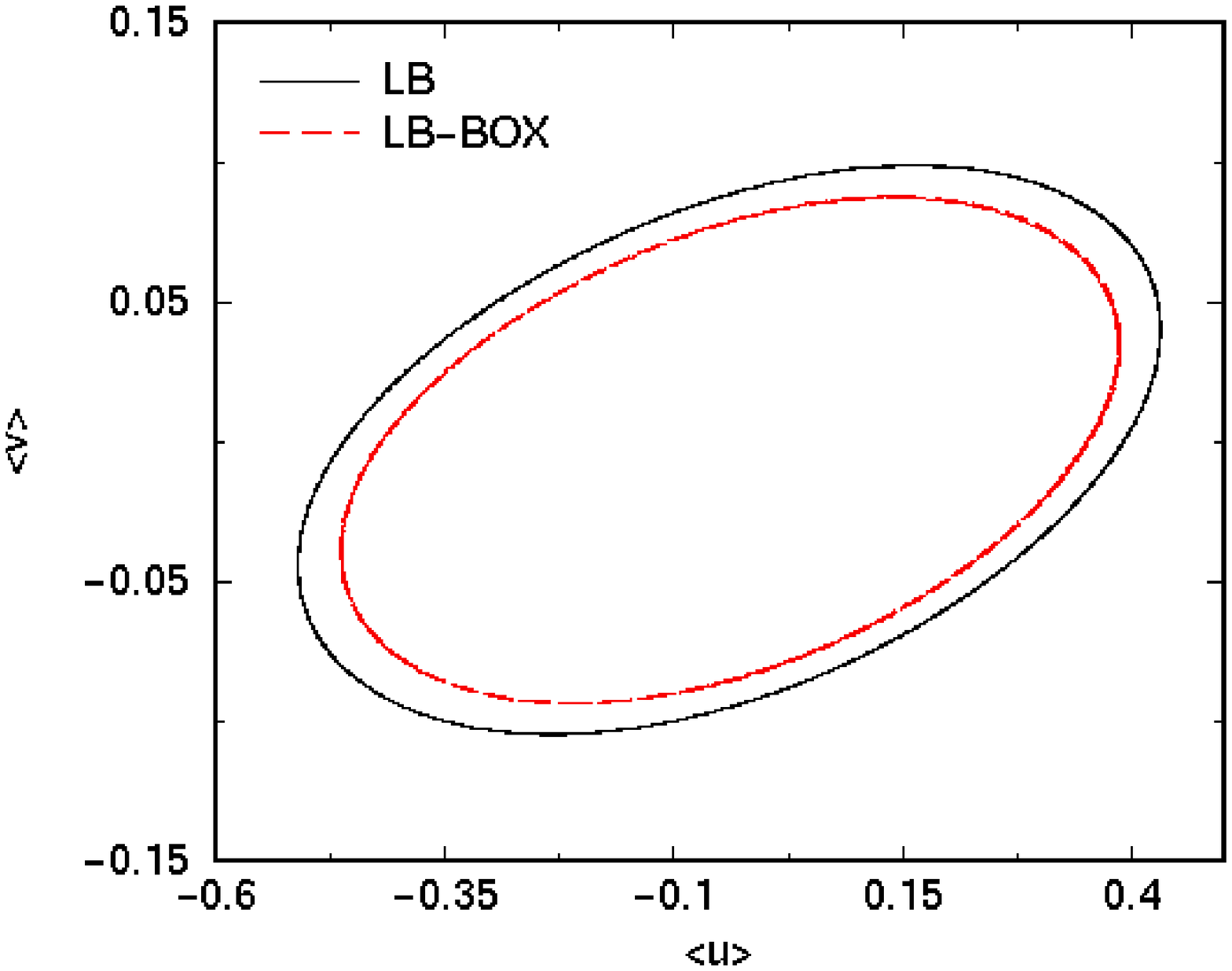,width=9.8cm}}
\vspace{0.1in}
\caption{Comparison of phase space projections of the LB (black)
and LB-BOX (red) long-term attractors. Reporting horizon for communication between
boxes $T_c=2.5 \times 10^{-3}$.}
\label{fig:attrlbbox_lo}
\end{figure}

\begin{figure}
\centerline{\psfig{file=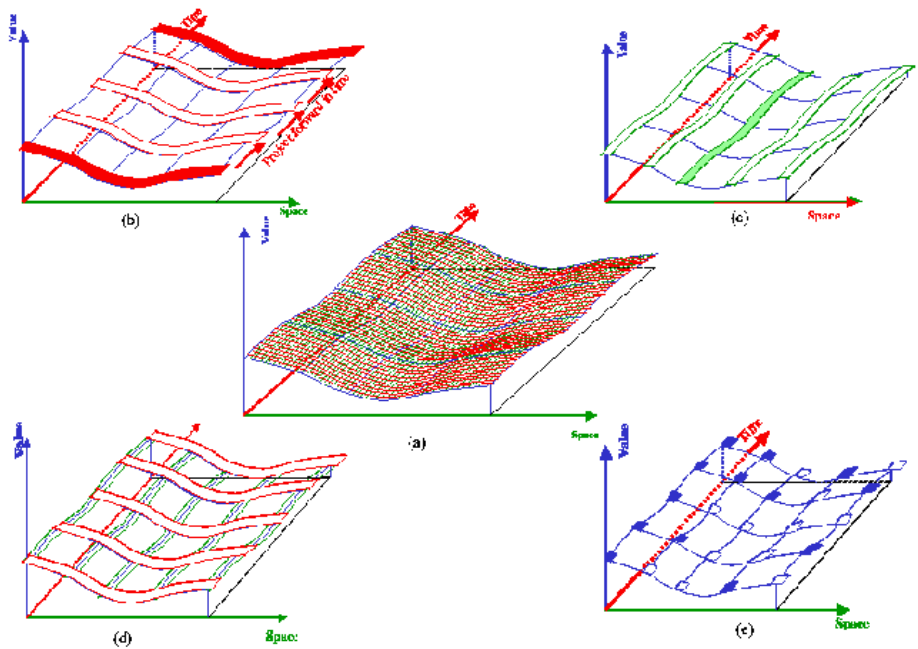,width=12.5cm,clip=t}}
\vspace{0.1in}
\caption{Schematics of coarse integration and patch dynamics techniques.
Full microscopic simulation (a) requires impractically fine discretizations.
Coarse integration (b) simulates in large space for short times, and then projects
to longer times. The gaptooth scheme (c) simulates in short spacial domains and
periodically reinterpolates between the domains. Combining the gaptooth scheme with
coarse integration (d) is better seen in (e): microscopic simulations are
performed over short space
domains (patches) and for relatively short times (sparse space time ``elements"). 
Interpolation
in coarse space and projection in coarse time is used to advance the macroscopic
quantities.}
\label{fig:sp_time_sum}
\end{figure}

\begin{figure}
\centerline{\psfig{file=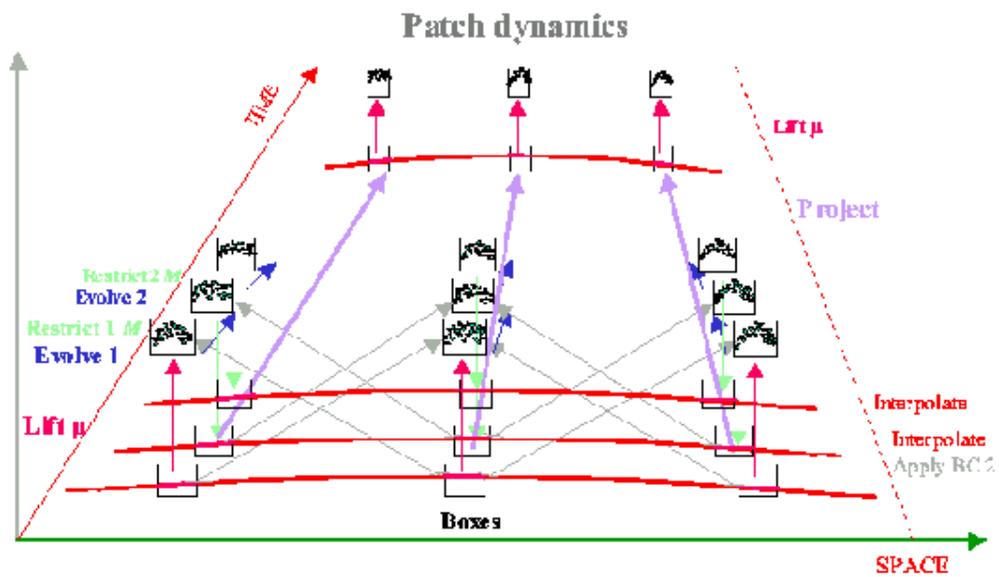,width=14cm,clip=t}}
\vspace{0.1in}
\caption{A schematic of the full patch dynamics simulation cycle, including
lifting in patches, computation and imposition of patch BC, microscopic
evolution, repeated restrictions and BC recomputations as necessary, and (after
sufficient time has elapse so that the coarse time-derivatives can be estimated,
a coarse projective step, followed by a new lifting.}
\label{fig:patchfull}
\end{figure}

\begin{figure}
\centerline{\psfig{file=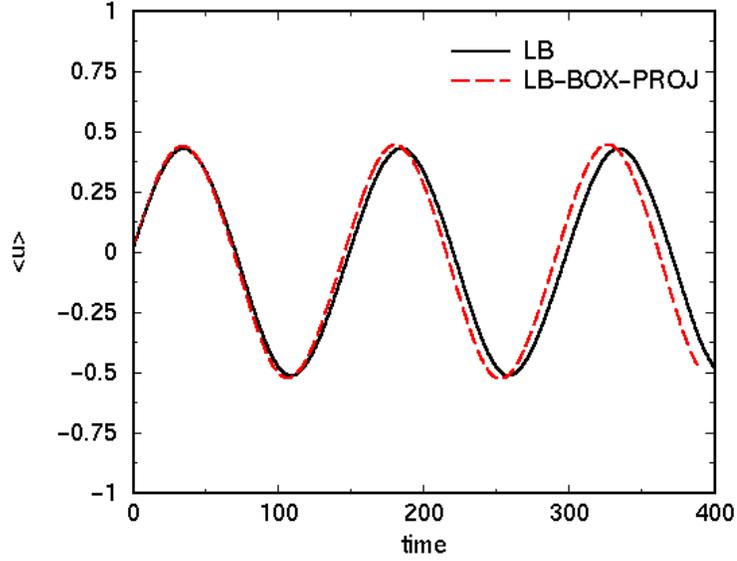,width=10cm}}
\vspace{0.1in}
\caption{Computed time-series comparison of the oscillatory FHN dynamics ($\epsilon=0.01$)
computed by a full LB code (solid black line)
and by a 100-100-200 LB-BOX projective scheme (broken red line).}
\label{fig:gearbox_tser}
\end{figure}

\begin{figure}
\centerline{\psfig{file=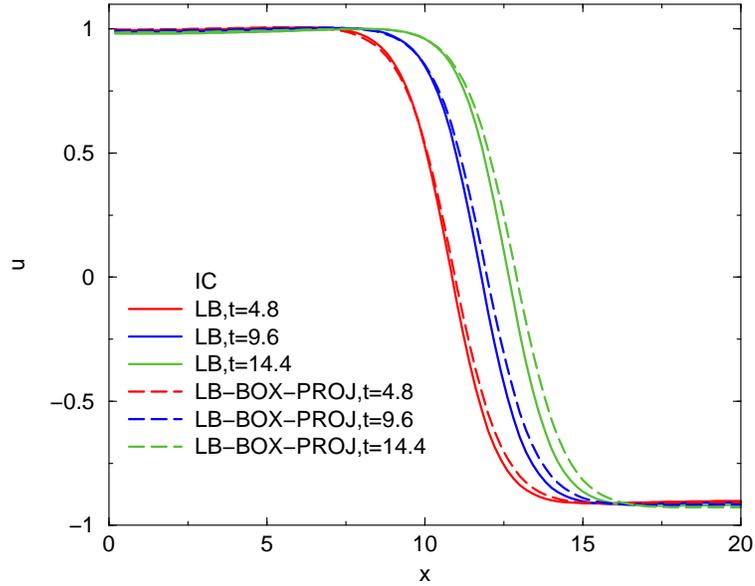,width=10cm}}
\vspace{0.1in}
\caption{Comparison of spatial profiles of $u$
at various times in the unstable (oscillatory) FHN region ($\epsilon=0.01$)
computed by a full LB code (solid lines) and by a 100-100-200 LB-BOX 
projective code
(broken lines). The boxes cover 10\% of the macroscopic computational
domain.}
\label{fig:gearbox_short}
\end{figure}

\begin{figure}
\centerline{\psfig{file=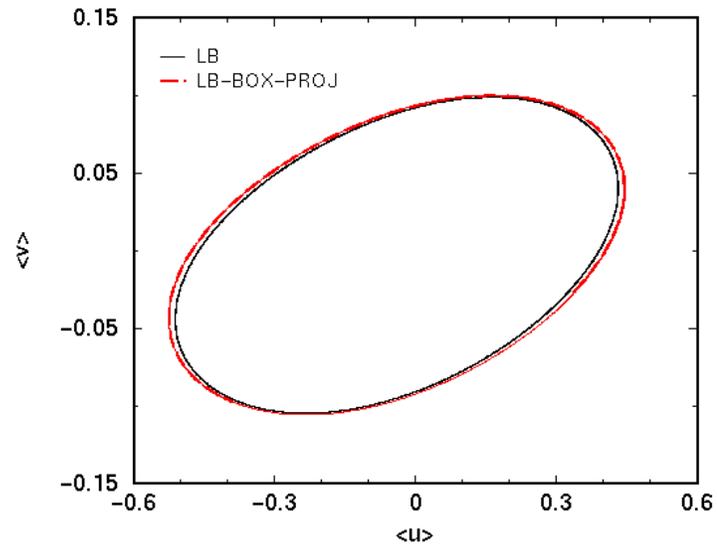,width=9.8cm}}
\vspace{0.1in}
\caption{Comparison of phase space projections of the long term
attractors of a full LB (black line)
and the 100-100-200 LB-BOX projective (red line) attractors.
Reporting time for communication between boxes $T_c=7.5 \times 10^{-4}$.}
\label{fig:gearbox_attr}
\end{figure}
\clearpage

\begin{figure}
\centerline{\psfig{file=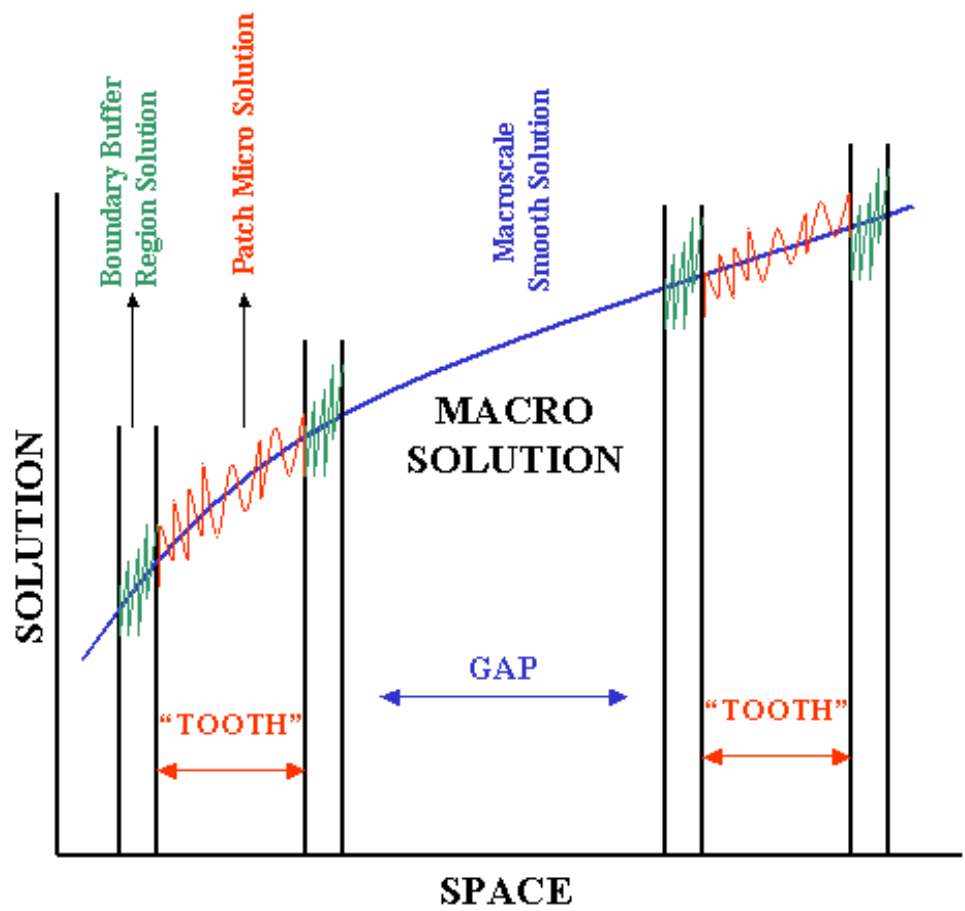,width=14cm,clip=t}}
\vspace{0.1in}
\caption{A schematic summary of the gaptooth/patch dynamics nomenclature in 1D.}
\label{fig:patch1d}
\end{figure}
\clearpage

\begin{figure}
\centerline{\psfig{file=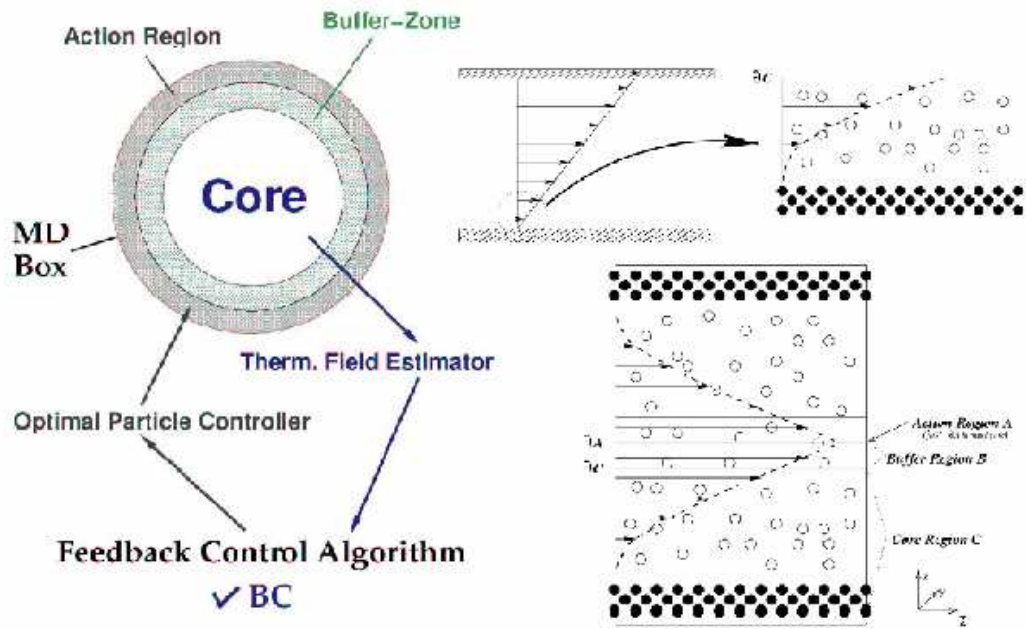,width=15cm}}
\vspace{0.1in}
\caption{A schematic of the ``extended boundary conditions" from \cite{LiJu2}
and a fluid flow application.}
\label{fig:liju}
\end{figure}

\begin{figure}
\centerline{\psfig{file=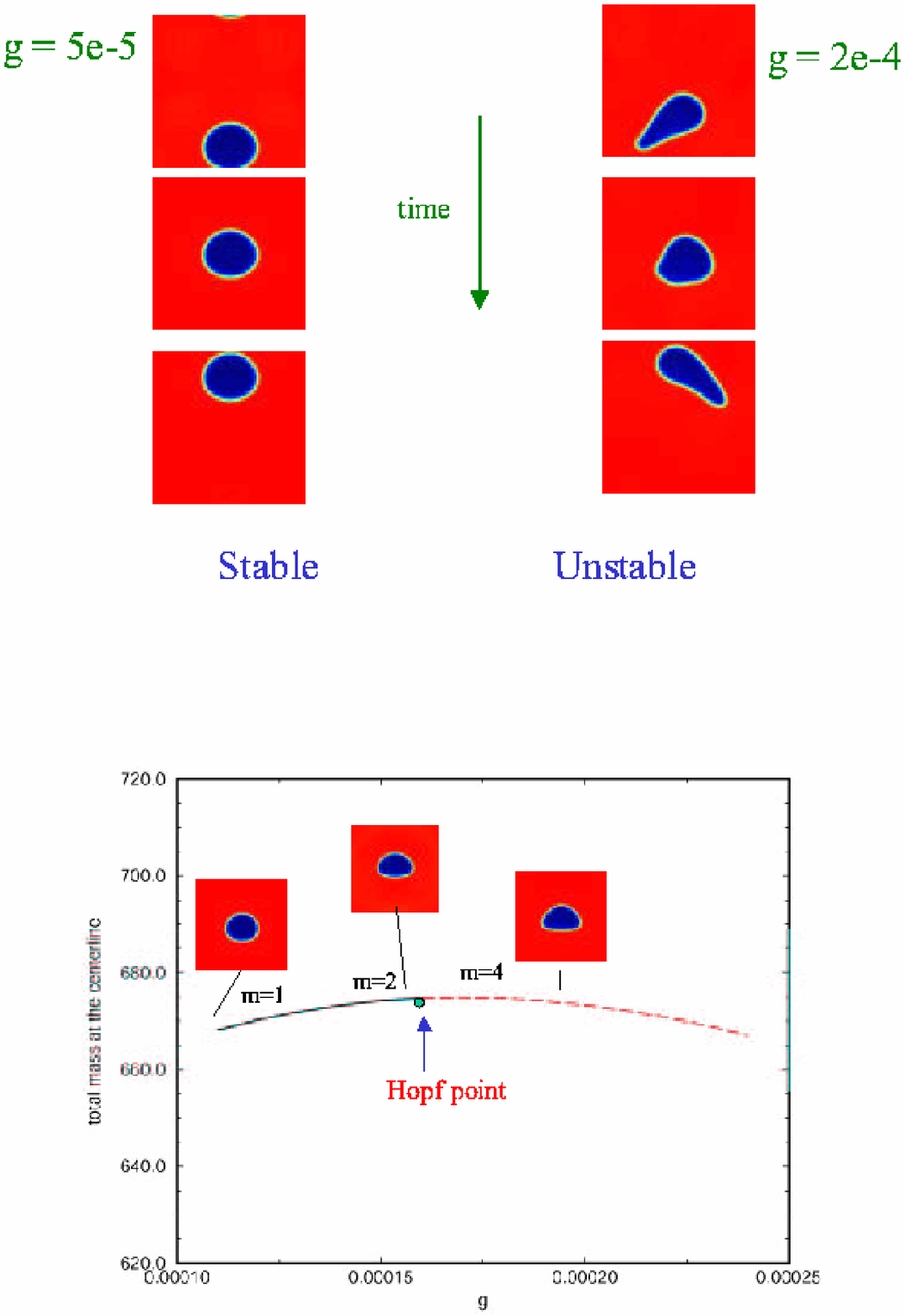,width=13cm,clip=t}}
\caption{Coarse bifurcation diagram of a rising bubble flow instability (bottom) obtained
with an inner LB-BGK simulator. The top panels show simulations of a steadily
rising and an oscillating bubble before and after the bifurcation 
(from \cite{BubblesPRL}). Solid(resp. broken): coarse stable(resp. unstable)
steady rising bubbles. $m$ is the dimension of the slow coarse subspace 
adaptively identified by RPM).
}
\label{fig:bubbles}
\end{figure}

\begin{figure}
\centerline{\psfig{file=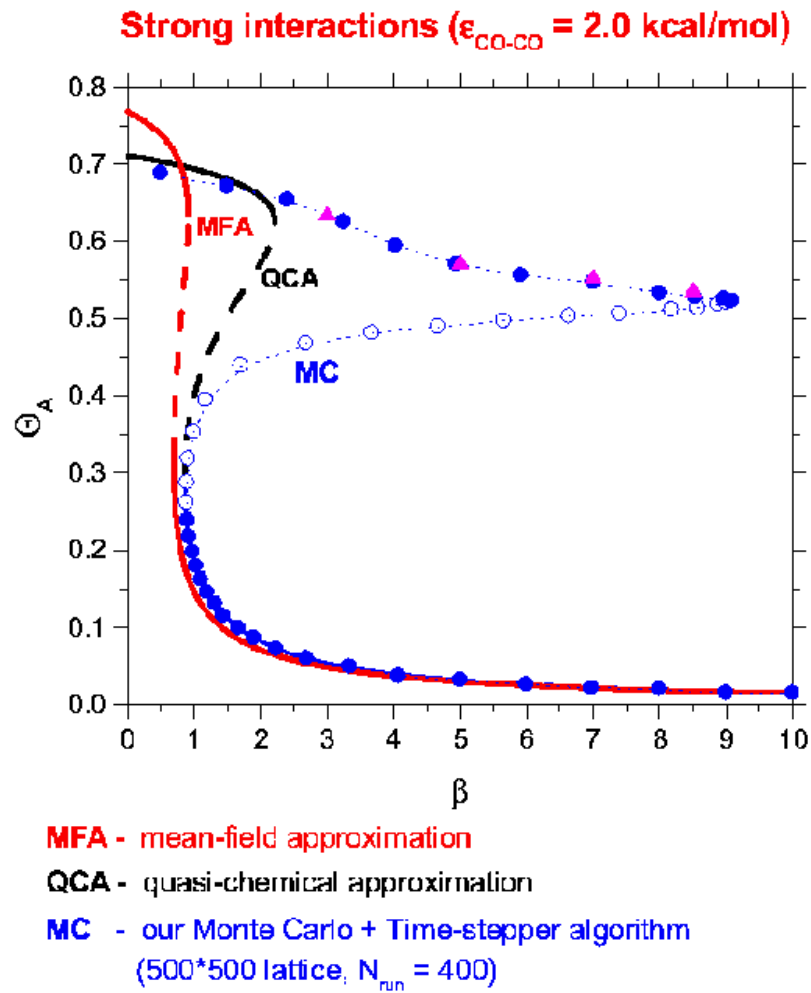,width=11cm,clip=t}}
\vspace{0.1in}
\caption{Coarse bifurcation diagram of a Kinetic Monte Carlo simulation of
a model surface reaction (CO oxidation with repulsive adsorbate interactions).
The diagram compares the bifurcation diagram obtained trough the coarse KMC
timestepper with the mean field and the quasichemical bifurcation diagrams; the
triangles are long-term KMC simulations (from \cite{Alexei2}).}
\label{fig:MC}
\end{figure}

\begin{figure}
\centerline{\psfig{file=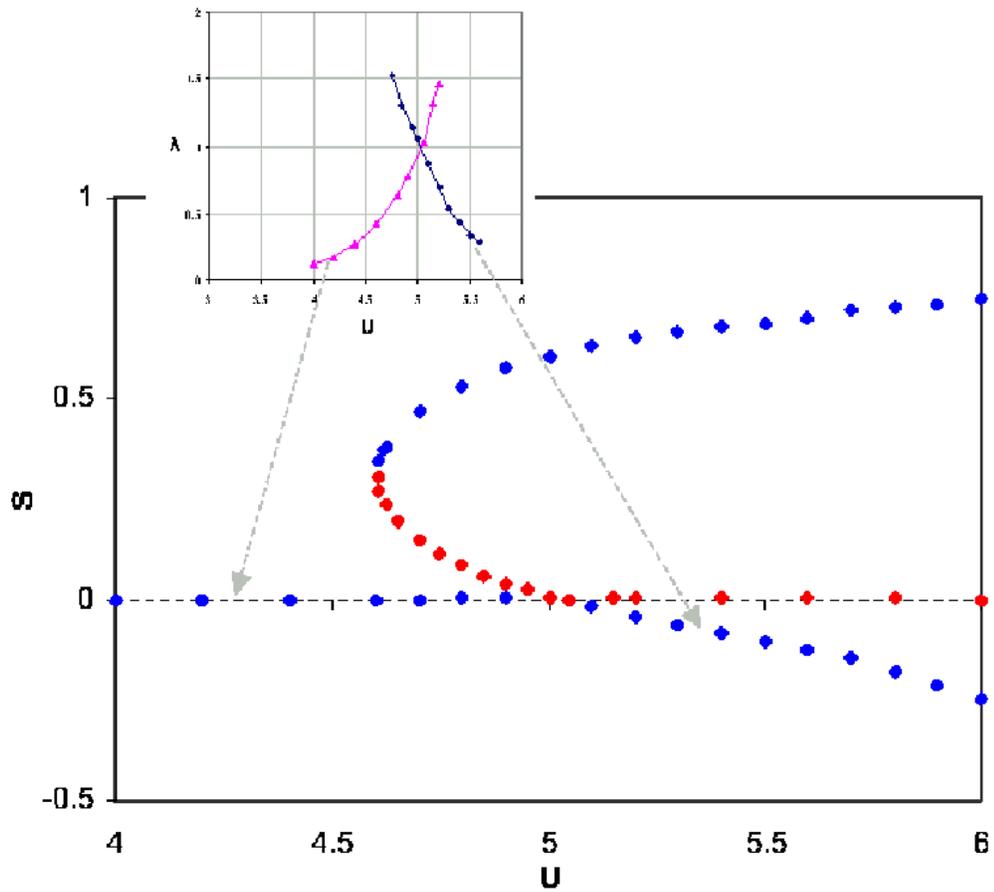,width=14cm,clip=t}}
\vspace{0.1in}
\caption{Coarse bifurcation diagram of a Brownian Dynamics simulation arising in
the rigid rod model of nematic liquid crystals. Orientation distributions of rigid rods
conditioned on one or two of their moments are used to construct the coarse timestepper
with an inner stochastic Euler integrator. Stable and unstable branches of the order
parameter S as a function of the rod density U are marked by color, and the corresponding
coarse eigenvalues are also indicated (from \cite{Graham}).}
\label{fig:Graham}
\end{figure}

\begin{figure}
\centerline{\psfig{file=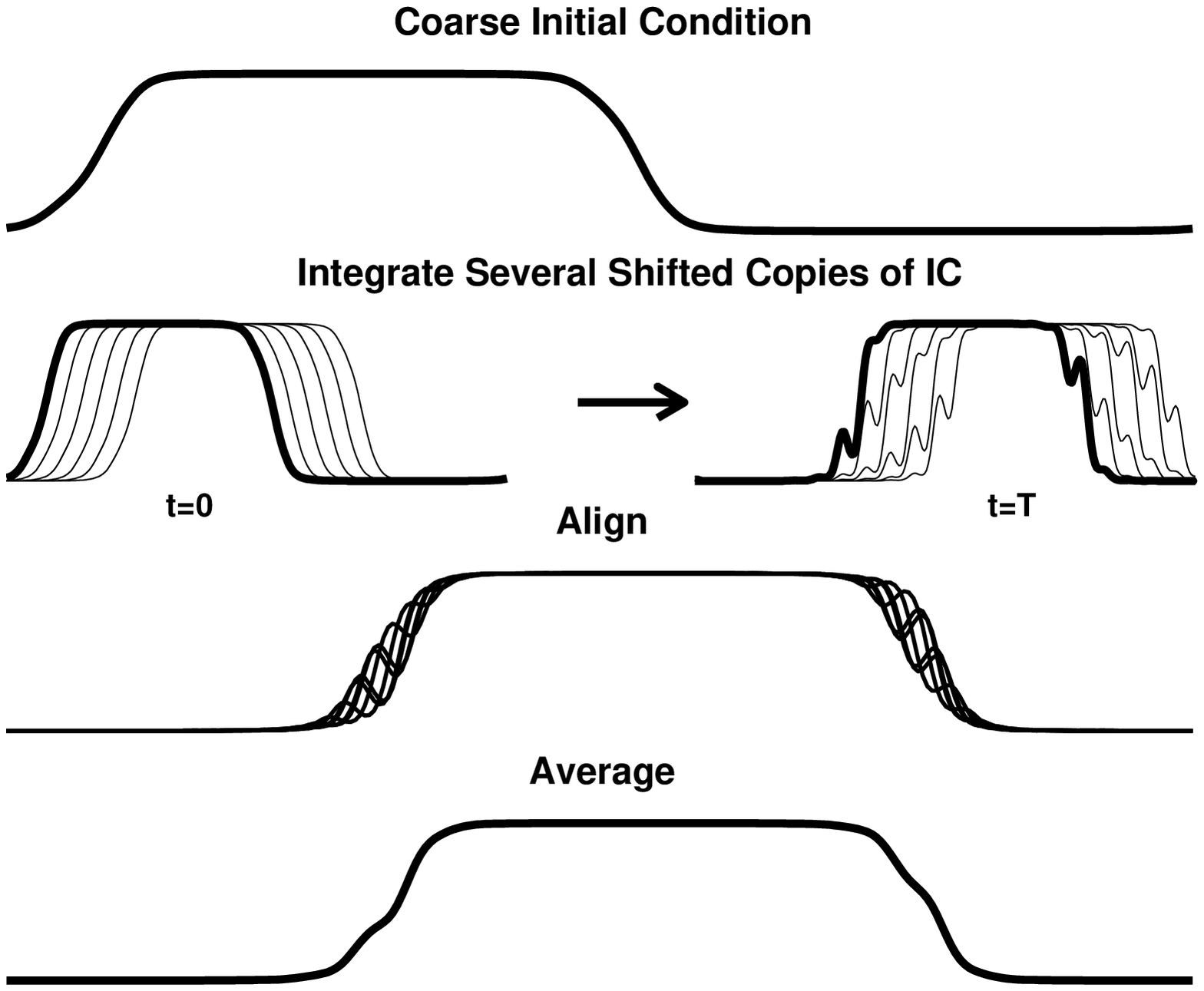,width=10.5cm}}
\vspace{0.4in}
\centerline{\psfig{file=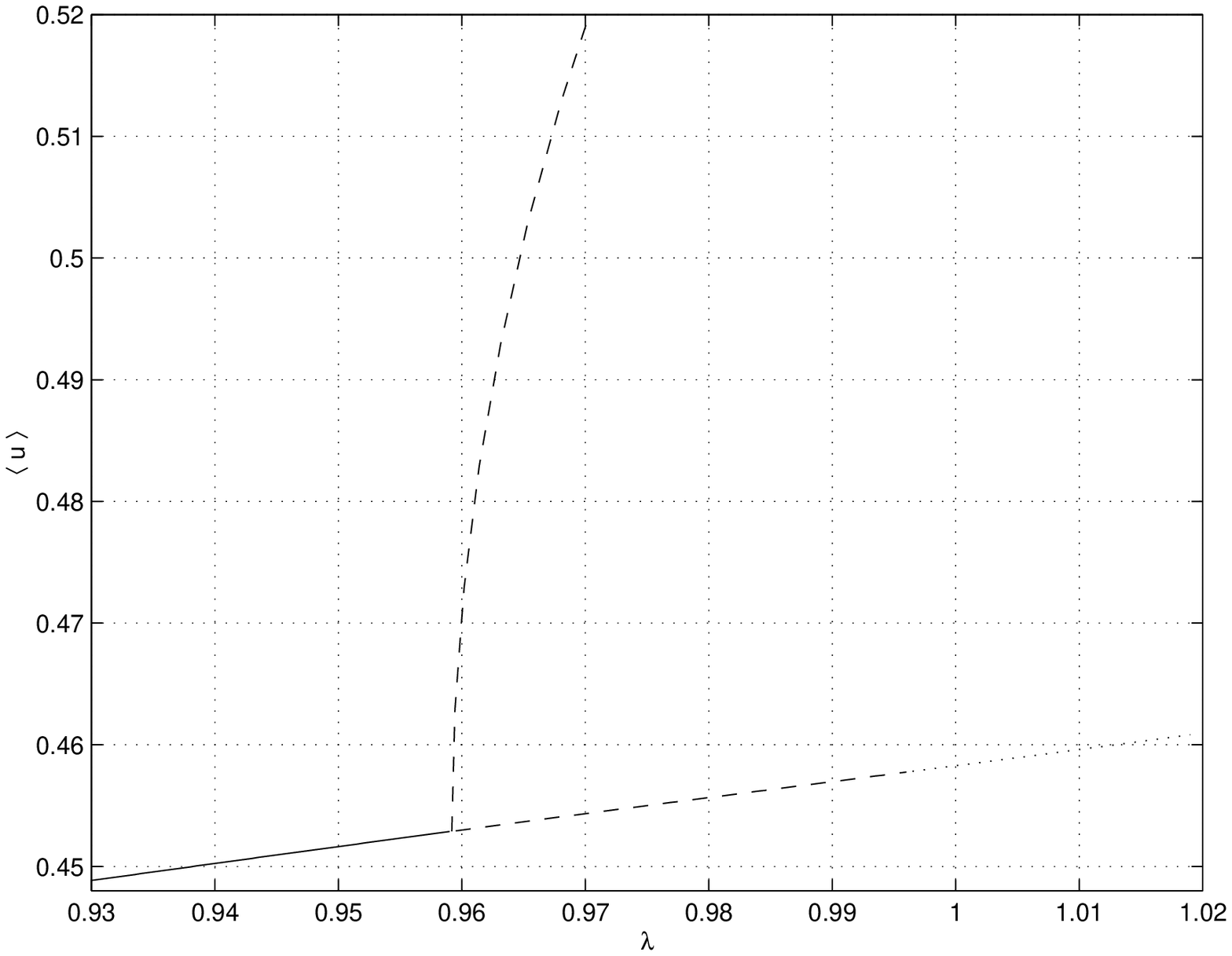,width=10.5cm}}
\vspace{0.1in}
\caption{Schematic of the coarse time-stepper in a homogenization/effective equation analysis
framework from \cite{Olof}. The ``effective bifurcation diagram" for a traveling reaction
pulse through a periodic medium as a function of the medium periodicity is shown at the bottom,
indicating a coarse instability and the birth of a modulated coarse traveling pulse.}
\label{fig:Olof}
\end{figure}

\begin{figure}
\centerline{\psfig{file=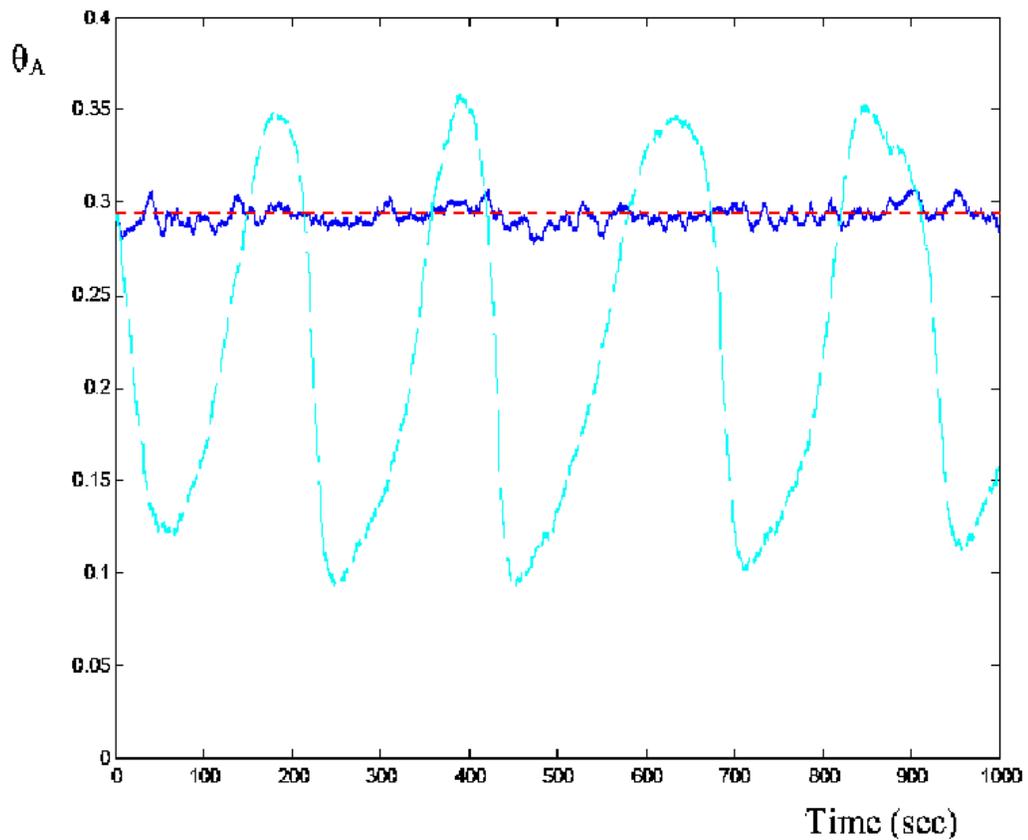,width=14cm,clip=t}}
\vspace{0.1in}
\caption{Coarse control of a KMC simulation of a surface reaction model \cite{Siettos_CC}; the (coarse)
unstable steady state to be stabilized was located, 
the estimator (a Kalman filter) and the controller
(pole placement) constructed based on Newton-Raphson
results obtained exploiting the coarse timestepper. The open loop
behavior is a noisy oscillation (light blue), while the closed loop (using the bifurcation parameter as the
actuator) clearly shows stabilization of the coarse steady state.}
\label{fig:control}
\end{figure}

\begin{figure}
\centerline{\psfig{file=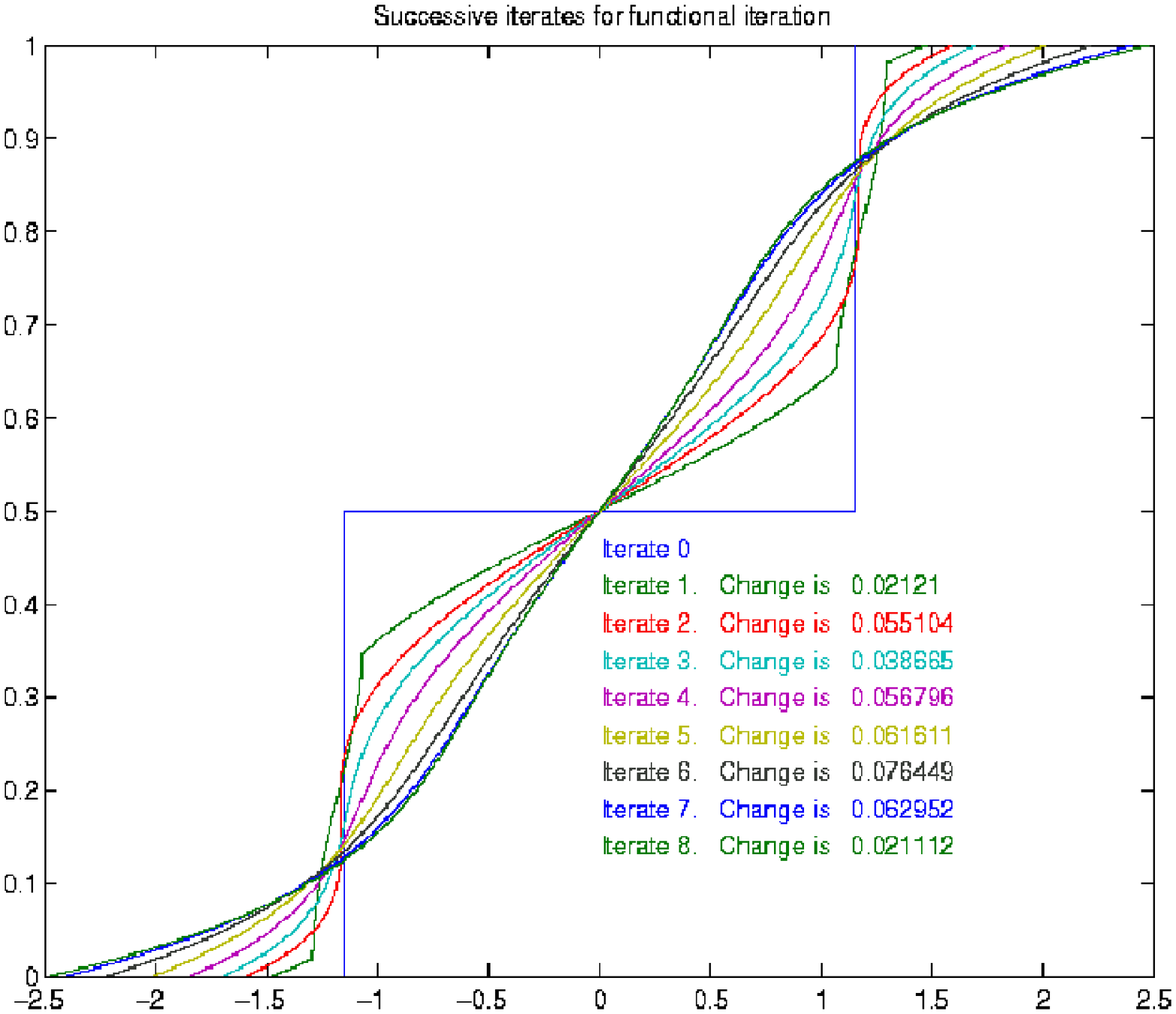,width=9cm}}
\vspace{0.1in}
\centerline{\psfig{file=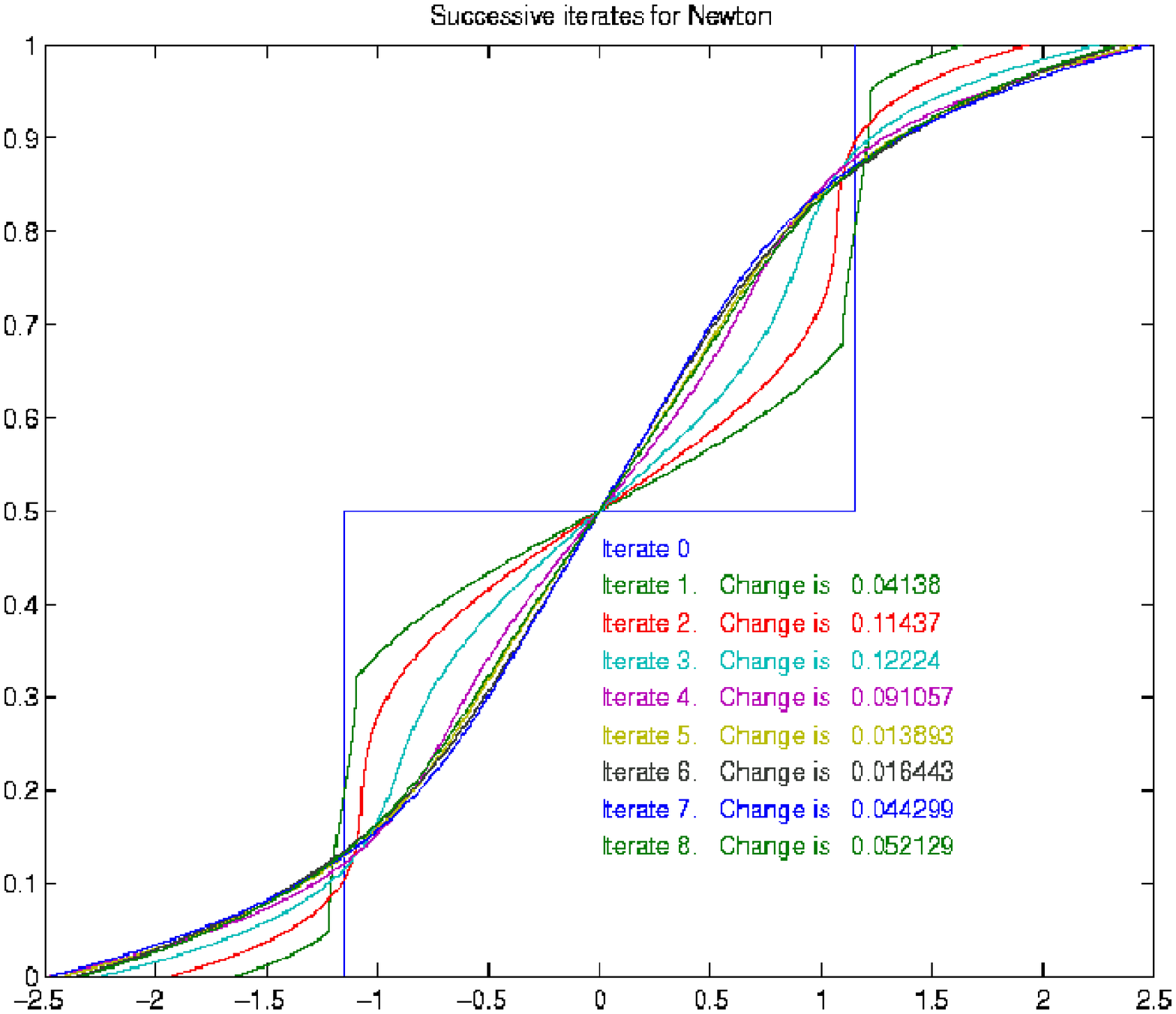,width=9cm}}
\vspace{0.1in}
\caption{Coarse renormalization flow and coarse self-similar solution calculations
based on KMC random walk simulations. The top panel shows the results of coarse
renormalization flow projective integration for the one-dimensional 
diffusion equation, while
the bottom panel shows iterations of a fixed point algorithm converging
on the same coarse self-similar solution. The evolution of the rescaled cumulative probability density
is plotted at various times above, and at various iterations of the fixed point 
scheme below. The restriction operator was the cumulative distribution function.}
\label{fig:selfsim}
\end{figure}

\end{document}